\documentclass[12pt,preprint]{aastex}

\shortauthors{Carpenter \etal}
\shorttitle{FEPS Survey of Solar-type Stars}

\newcommand{\ts}{\thinspace}
\newcommand{\simless}{\mathbin{\lower 3pt\hbox
     {$\rlap{\raise 5pt\hbox{$\char'074$}}\mathchar"7218$}}}
\newcommand{\simgreat}{\mathbin{\lower 3pt\hbox
     {$\rlap{\raise 5pt\hbox{$\char'076$}}\mathchar"7218$}}}

\newcommand{\about}    {$\sim$\ts}
\newcommand{\aboutless}{$\simless$\ts}
\newcommand{\aboutmore}{$\simgreat$\ts}

\newcommand{\RI}{$R_{8/3.6}$}
\newcommand{\RII}{$R_{16/8}$}
\newcommand{\RM}{$R_{24/8}$}
\newcommand{\RMM}{$R_{70/8}$}

\newcommand{\Spitzer}{{\it Spitzer}}
\newcommand{\IRAS}{{\it IRAS}}
\newcommand{\ISO}{{\it ISO}}
\newcommand{\msun}{\ts M$_\odot$}
\newcommand{\lsun}{\ts L$_\odot$}
\newcommand{\etal}{\ts et~al.}
\newcommand{\zody}{\ts $\cal Z$}

\begin{document}

\title{\hfill Formation and Evolution of Planetary Systems (FEPS):\hfill\break
       Properties of Debris Dust around Solar-type Stars}

\author{John M. Carpenter\altaffilmark{1}}
\author{Jeroen Bouwman\altaffilmark{2}} 
\author{Eric E. Mamajek\altaffilmark{3}} 
\author{Michael R. Meyer\altaffilmark{4}} 
\author{Lynne A. Hillenbrand\altaffilmark{1}} 
\author{Dana E. Backman\altaffilmark{5}}
\author{Thomas Henning\altaffilmark{2}}
\author{Dean C. Hines\altaffilmark{6}} 
\author{David Hollenbach\altaffilmark{7}} 
\author{Jinyoung Serena Kim\altaffilmark{4}} 
\author{Amaya Moro-Martin\altaffilmark{8}}
\author{Ilaria Pascucci\altaffilmark{7}} 
\author{Murray D. Silverstone\altaffilmark{9}}
\author{John R. Stauffer\altaffilmark{10}} 
\author{Sebastian Wolf\altaffilmark{2}} 
\altaffiltext{1}{Department of Astronomy, California Institute of Technology,
       Mail Code 105-24, 1200 East California Boulevard, Pasadena, CA 91125.}
\altaffiltext{2}{Max-Planck-Institut f\"ur Astronomie, D-69117 Heidelberg, Germany.}
\altaffiltext{3}{Harvard--Smithsonian Center for Astrophysics; current address: Department of Physics \& Astronomy, University of Rochester, Rochester NY 14627}
\altaffiltext{4}{Steward Observatory, The University of Arizona, 933 North Cherry Avenue, Tucson, AZ 85721.}
\altaffiltext{5}{SOFIA/SETI Institute}
\altaffiltext{6}{Space Science Institute, 4750 Walnut Street, Suite 205,
   Boulder, CO 80301}
\altaffiltext{7}{NASA-Ames Research Center}
\altaffiltext{8}{Department of Astrophysical Science, Princeton University}
\altaffiltext{9}{Eureka Scientific, Inc., 113 Castlefern Dr., Cary NC 25713}
\altaffiltext{10}{Spitzer Science Center, California Institute of Technology, Mail Code 314-6, 1200 East California Boulevard, Pasadena, CA 91125}

\begin{abstract}

We present \Spitzer\ photometric (IRAC and MIPS) and spectroscopic (IRS low
resolution) observations for 314 stars in the Formation and Evolution of
Planetary Systems (FEPS) Legacy program. These data are used to investigate the
properties and evolution of circumstellar dust around solar-type stars spanning
ages from approximately 3~Myr to 3~Gyr. We identify 46 sources that exhibit 
excess infrared
emission above the stellar photosphere at 24\micron, and 21 sources with
excesses at 70\micron. Five sources with an infrared excess have 
characteristics of optically thick primordial disks, while the remaining 
sources have properties akin to debris systems. The fraction of systems
exhibiting a 24\micron\ excess greater than 10.2\% above the photosphere is
15\% for ages $<$ 300~Myr and declines to 2.7\% for older ages. 
The upper envelope to the 70\micron\ fractional luminosity appears to 
decline over a similar age range. The characteristic temperature of the debris
inferred from the IRS spectra range between 60 and 180~K, with evidence for the
presence of cooler dust to account for the strength of the 70\micron\ excess 
emission. No strong correlation is found between dust temperature and 
stellar age. Comparison of the observational data with disk models containing
a power-law distribution of silicate grains suggest that the typical inner disk
radius is \aboutmore 10~AU. Although the interpretation is not unique, the lack
of excess emission shortwards of 16\micron\ and the relatively flat
distribution of the 24\micron\ excess for ages \aboutless 300~Myr is consistent
with steady-state collisional models.

\end{abstract}


\section{Introduction}

The {\it IRAS} mission left a lasting legacy with the surprise discovery of
dust surrounding Vega and other main-sequence stars \citep{Aumann84,Aumann85}.
The orbital lifetime of the dust, limited by collisional and radiative
processes in a gas poor environment, is orders of magnitude shorter than the 
stellar age, and implies that the dust must have been created recently
\citep{Backman93}. The model currently favored to explain the presence of dust
in these systems postulates that planets gravitationally stir a population of
planetesimals, which subsequently collide and fragment into a cascade of
smaller debris \citep{Williams94}. In support of this conjecture, high-angular
resolution images have indeed shown that the dust is frequently distributed in
disk- or ring-like geometries
\citep{Smith84,Koerner98,Holland98,Kalas04,Schneider06}.

The presumed causal relationship between debris dust and planetary systems has
motivated many studies to investigate the frequency and properties of debris
disks \citep[][see also \citealp{Lagrange00} and references therein]
{Aumann91,Mannings98,Moor06,Rhee07}. Analysis of \IRAS\ data indicates that
\about 15\% of nearby AFGK type stars are surrounded by a debris disk with a
dust-to-photosphere luminosity ratio exceeding that of the Vega debris disk
\citep{Backman87,Plets99}. Subsequent observations with \ISO\ demonstrated that
the frequency of debris disks declines with age
\citep{Habing99,Habing01,Spangler01}. \citet{Decin03} emphasized, however, that
a large range of dust properties are present at any given stellar age, which
suggests that planetesimals with a wide range of properties exist around 
main-sequence stars.

As an extension of the \IRAS\ and \ISO\ heritage, the high photometric
precision and extraordinary sensitivity of the \Spitzer\ Space Telescope
\citep{Werner04} have enabled a comprehensive characterization on how the
debris disk phenomenon correlates with stellar mass and age.
\citet{Beichman05b} obtained the first secure demonstration of debris dust
around extrasolar planetary systems \citep[see also][]{MoroMartin07a} to
support the basic tenet that debris and planets are connected.
\citet{Beichman06b} showed that fractional luminosity from dust in the 1~AU 
region around solar-type stars is rare at levels more than 1400 times the 
brightness of the current zodiacal cloud,
although spectacular examples of warm (\about 300~K) dust have been
discovered \citep{Song05,Beichman05a,Rhee08}. \citet{Rieke05} found that 
\about 47\%
of A-type stars younger than 90~Myr have a 24\micron\ excess more than 25\%
above the photosphere, with a rapid decline in both the frequency and magnitude
of the excess toward older ages. Lower mass FGK type stars exhibit a similar
decline in the warm excess fraction with age \citep{Siegler07}, although the
persistence time may be longer for 70\micron\ excess emission
\citep{Su06,Bryden06}. For stars older than 600~Myr, the presence of 
70\micron\ excess shows no clear dependence on spectral type for AFGK stars
\citep{Trilling08}. Few debris disks have been detected around M dwarfs with
Spitzer \citep{Beichman06a,Gautier07}; it remains uncertain however whether
this indicates a real difference in disk frequency compared to higher mass
stars, or simply reflects the greater difficulty in detecting dust around low
luminosity stars \citep[see, e.g.,][]{Cieza08}.

The improved observational data have coincided with increasingly sophisticated
models to understand the connection between debris dust and the presence of
planets. \citet{Liou99} demonstrated how debris dust can be trapped in
resonances with orbiting planets to produce the asymmetric and clumpy structure
observed in some debris disks \citep[see
also][]{Wyatt03,MoroMartin03,MoroMartin05,Krivov07}. In a series of papers,
\citet{Kenyon01,Kenyon02,Kenyon04,Kenyon05} investigated the link between the
collisional growth of planets
\citep[see][]{Safronov69,Greenberg84,Wetherill89,Spaute91} and the subsequent
production of debris in the collisional cascade. They demonstrated that as
planet formation proceeds from the inner disk outwards and gravitationally
stirs the planetesimals, debris dust will be produced at sufficient levels to
be detected with current instrumentation. \citet{Dominik03} suggest that the
rapid decline in the observed debris emission at an age of a few hundred
million years is a consequence of the collisional depletion of the planetesimal
belts. \citet{Wyatt07a} extended their model and confirmed the basic findings,
and also identified several luminous debris systems which are best explained
by a recent collision between planetesimals that produced copious amounts of
dust. 

To date, the observational data and the application of models is most complete
for A-stars which are sufficiently bright to be detected to large distances by
\IRAS\ and \ISO. With the sensitivity of \Spitzer, extensive data sets can now
be collected for solar-type stars that may provide insights on the evolutionary
history of our own Solar System. Toward this goal, the Formation and Evolution
of Planetary Systems (FEPS) \Spitzer\ Legacy Program \citep{Meyer06} has
conducted a photometric and spectroscopic survey of 328 solar-type stars
spanning ages between \about 3~Myr and 3~Gyr. Previous FEPS studies 
have analyzed subsets of the FEPS data to address specific issues regarding 
debris evolution, including 
(1) the evolution of terrestrial temperature debris around stars younger 
    than 30~Myr \citep{Silverstone06},
(2) the temporal evolution of warm 24\micron\ excess 
    \citep{Stauffer05,Meyer08}, 
(3) the identification and properties of cool Kuiper-Belt analogs 
    \citep{Meyer04,Kim05,Hillenbrand08},
(4) analysis of debris disk properties around known extra-solar 
    planetary systems \citep{MoroMartin07a,MoroMartin07b},
(5) dust composition for a sample of optically thick accretion disks
    \citep{Bouwman08},
and 
(6) the gas dissipation timescales in debris disks 
    \citep{Hollenbach05,Pascucci06,Pascucci07}.
In this contribution, we present the most comprehensive analysis to date of 
photometry and low resolution spectra in the FEPS survey to investigate the
evolution of debris disks around solar-type stars. The broad goal of our
investigation is to quantify the incidence of the debris phenomenon around
solar-analogs, infer the properties of the implied planetesimal belts around
these systems, and understand the similarities and differences of these
presumed planetary systems with the Solar System.

\citet[hereafter Paper~I]{Carpenter08} describe the observation and data 
reduction procedures adopted for this study. A summary of the
FEPS sample and important aspects of the FEPS data reduction are summarized in
\S\ref{data}. In \S\ref{excess}, we analyze each of the \Spitzer\ data products
to identify stars that have an infrared excess between 3.6 and 70\micron. We
use these results to trace the temporal evolution of debris properties
(\S\ref{evolution}), and to constrain the location of the planetesimal belts
(\S\ref{models}). We then compare these properties to the Solar System debris
disk (\S\ref{comparison}) before summarizing our results (\S\ref{summary}).

\section{Stellar Sample and Observations}
\label{data}

This paper analyzes data for 314 stars that were selected for the FEPS program 
without regard to the presence or absence of a known infrared excess from 
pre-\Spitzer\ observations. We exclude 14 other FEPS stars with suspected
excesses from \IRAS\ or \ISO\ that were observed explicitly for a gas-detection 
experiment \citep{Hollenbach05,Pascucci06,Pascucci07}. \citet{Meyer06}
describe the sample properties in detail and only a summary is presented here. 
The stars span spectral types between K7 and F5 and ages between approximately
3~Myr and 
3~Gyr. The stellar masses range from 0.7 to 2.2\msun, with 90\% of the stars
having masses less than 1.4\msun. Stellar ages were estimated by a variety of 
methods. In brief, for the youngest clusters/associations, the ages were based
on fits to pre-main sequence isochrones; for older clusters, the ages are
primarily based on isochronal fits to the upper main sequence turnoff and/or
the ``lithium-depletion boundary" method \citep{Rebolo92}. Field stars were 
assigned ages based
on stellar activity indicators (e.g. X-ray, rotation, CaII H\&K), whose decline
in strength with age has been calibrated with respect to open clusters of known
age \citep{Mamajek08}.

The sample breakdown between field stars and associations is provided in 
Table~\ref{tbl:sample_associations} and by stellar age in
Table~\ref{tbl:sample_age}. Stellar ages are distributed approximately in 
uniform logarithmic intervals with between 34 and 60 stars for every factor of 
three in age. For ages less than 30~Myr, the sample is weighted toward stars 
found in associations, and stars older than 1~Gyr consist entirely of field 
stars because of the lack of nearby, old clusters. Younger stars tend to have 
later spectral types since solar-mass stars increase in temperature as they 
evolve from the pre-main-sequence phase to the main-sequence. 

FEPS obtained IRAC 3.6, 4.5, and 8\micron\ 
images\footnote{The FEPS IRAC observations were obtained in sub-array mode,
where a subsection of the full-array is read out to enable observations of 
bright stars. Sub-array observations obtains images in the four IRAC band 
separately. IRAC 5.8\micron\ observations were not obtained for the majority 
of the FEPS targets (see Paper~I).}, MIPS~24 and 70\micron\
images, and IRS low resolution spectra between 8 and 35\micron\ (SL1, LL1, and
LL2 orders) for most of the sample, with observations for a few stars
supplemented from the \Spitzer\ archive. Five (HD~80606, HD~139813, HII~2881,
HIP~42491, and RX~J1544.0-3311) of the 314 sources do not have an IRS
spectrum since the peak-up observations selected the incorrect star. 
In a few sources, portions of the IRS spectra appear corrupted and were 
discarded. The flux densities in the LL spectra for ScoPMS~52 are \about 60\% 
higher than the MIPS~24\micron\ flux density, and the IRS spectrum is likely 
contaminated by a source 18\arcsec\ away that has an order of magnitude 
higher flux density. The flux density in the SL1 spectrum for HD~13974 is 
2.6$\times$ lower than expected, and the extracted LL1 and LL2 spectra for 
R45 have negative flux densities. Also, the IRS spectra for two sources 
(HII~1015 and HE~699) were discarded because of poor signal-to-noise. 
Finally, since Paper~I was published, we have reanalyzed the spectrum for the 
star 1RXS~J051111.1+281353 using the S15 data products from the \Spitzer\
Science Center.

The signal to noise ratio on the stellar photosphere is \aboutmore 30 for both
the IRAC and MIPS~24\micron\ photometry. At 70\micron, the photosphere was
detected above the 3$\sigma$ noise level for only one source. The
signal-to-noise ratio of the spectra was assessed by computing synthetic
broad-band fluxes at 16\micron\ using a square-wave response function between
15 and 17\micron, at 24\micron\ using the MIPS~24\micron\ instrumental response
function\footnote{http://ssc.spitzer.caltech.edu/mips/spectral\_response.html},
and at 32\micron\ using a square-wave response function between 30 and
34\micron. In the 16\micron\ bandpass, the signal-to-noise ratio of the IRS
spectra is $\ge$ 20 for 305 sources. At 24\micron, the signal-to-noise ratio is
$\ge 10$ for 302 stars and $\ge 5$ in 305 sources. At 32\micron, the
signal-to-noise ratio is $\ge 5$ for 264 sources and $\ge 3$ for 292 stars.
A complete description of the data reduction procedures is presented in
Paper~I.

\section{Identifying Sources with Infrared Excesses}
\label{excess}

In this section, we analyze the photometric and spectroscopic data to identify
sources that exhibit infrared emission diagnostic of circumstellar dust. Each
\Spitzer\ instrument is sensitive to dust emitting over a range of
dust temperatures, and a given source will not necessarily exhibit detectable
infrared excesses in all instruments. In general terms, IRAC photometry
is most sensitive to hot dust located in the terrestrial planet zone (as
defined by our Solar System), MIPS~24\micron\ to warm dust in the
gas-giant formation region, and MIPS~70\micron\ to cold Kuiper Belt analogs.
IRS spectra probe regions similar to that by IRAC and MIPS~24\micron. We first
analyze the data for each instrument individually, and then synthesize the
results in \S\ref{excess_sum} to identify a reliable sample of excess sources.
In subsequent sections, we analyze these data to infer disk properties.

\subsection{IRAC}
\label{excess:irac}

\begin{figure}
\begin{center}
\includegraphics[angle=0,scale=0.7]{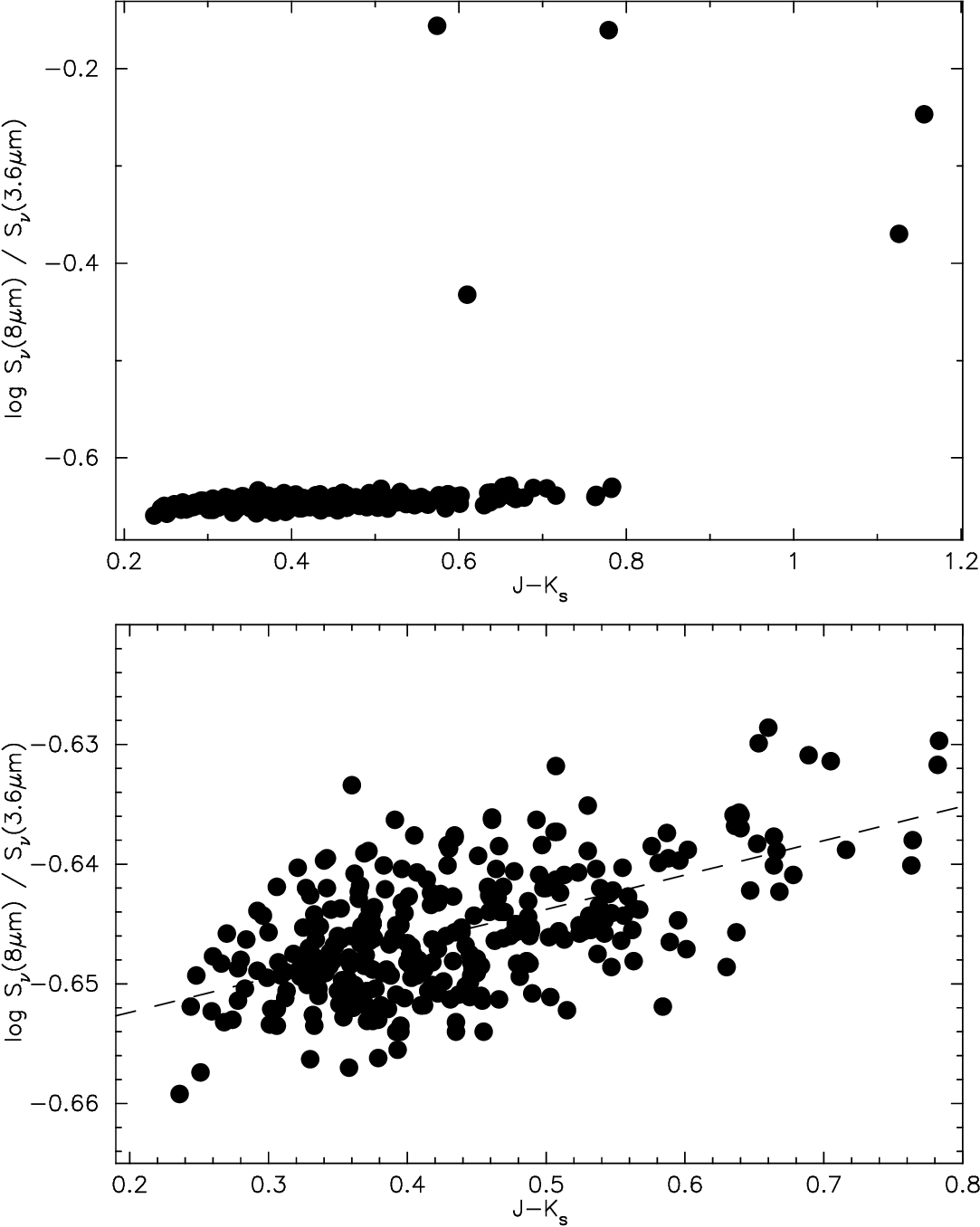}
\caption{
  \label{fig:irac_ccd}
  Plot of dereddened 2MASS $J-K_{\rm s}$ color versus the
  8\micron\ to 3.6\micron\ flux ratio (\RI) for 309 stars in the FEPS program. 
  Five FEPS stars were omitted that do not have high-quality \citep[PHQUAL=A;
  see][]{Cutri03} 2MASS $J$ or $K_{\rm s}$ photometry. The top panel shows 
  data for all stars, and the bottom panel excludes five stars with flux ratio 
  greater than 0.25 that have infrared excesses from optically thick disks 
  \citep{Silverstone06}. The dashed line in the bottom panel represents a
  linear fit to the data [log(\RI) = $0.029\,(J-K_{\rm s}) - 0.658$].}
\end{center}
\end{figure}

In the top panel of Figure~\ref{fig:irac_ccd}, we present the 8\micron\ to
3.6\micron\ flux ratio ($\equiv$ \RI) as a function of $J-K_{\rm s}$ color for
the FEPS sample after dereddening the photometry using the extinction
corrections derived in Paper~I and the reddening law compiled by
\citet{Mathis90}. The median visual extinction for the sample is 0~mag with 
a maximum of 1.8~mag.
Sources with large values of \RI\ contain a possible infrared
excess at 8\micron. Since the youngest stars in our sample are \about 3~Myr
old, and the inner disk, as traced by photometric observations at $\lambda <
3.5$\micron, dissipates in half of solar-type stars by an age of 3~Myr
\citep{Haisch01}, we anticipate that the $J-K_{\rm s}$ color traces the stellar
photosphere for most stars in our sample. This expectation is confirmed from
inspection of a $J-H$ vs. $H-K_{\rm s}$ diagram which shows that only two stars
(RX~J1111.7$-$7620 and RX~J1842.9$-$3532) exhibit a $K_{\rm s}$-band excess
detectable by this technique \citep[see, e.g.,][for a discussion of the merits
and limitations of this diagram]{Meyer97}. In Figure~\ref{fig:irac_ccd}, most
sources lie along a tight locus of points, while five sources
(RX~J1111.7$-$7620, RX~J1842.9$-$3532, RX~J1852.3$-$3700, PDS~66,
 $[$PZ99$]$~J161411.0$-$230536) have values of \RI\ well above the locus and
have a clear 8\micron\ excess. Four of these sources have ages $< 10$~Myr, and
the fifth (PDS~66) is a member of the Lower Centaurus Crux association with an
age of \about 12~Myr \citep{Preibisch08}. \citet{Silverstone06} showed that
these five sources have infrared excesses over a broad range of wavelengths 
and other properties characteristic of optically thick, primordial, 
circumstellar accretion disks. 

In the bottom panel of Figure~\ref{fig:irac_ccd} we present the same
diagram after removing the five sources with strong 8\micron\
excesses to emphasize the colors for the majority of stars. The dashed line
indicates the best-fit linear relation to the trend between $J-K_{\rm s}$ 
color and log~\RI. We assume that the trend represents intrinsic variation in 
the photospheric value of \RI\ over the spectral type range in the FEPS sample. 
The dispersion about the best fit line is $\sigma({\rm log}$\RI) =
0.0043, or $\sigma($\RI)/\RI = 1.0\%, while the expected dispersion from the
observational uncertainties is 0.0044. The
maximum outlier with a positive apparent excess is 3.3$\sigma$ (HD~77407), 
and we expect \about 1 outlier more than 3$\sigma$ from the mean based on the 
sample size. The IRS spectrum for HD~77407 shows no evidence for an 8\micron\
excess (\S\ref{excess:irs}), and this source was one of two objects where the 
IRAC photometry was contaminated by a latent image (see Paper~I). 
We conclude that outside of the five sources with strong 8\micron\
excesses characteristic of optically thick disks, no individual source
shows conclusive evidence of a weak ($3\sigma$ limit of 3\% above the 
photosphere) 8\micron\ excess indicative of optically thin dust.

\subsection{IRS low resolution spectra}
\label{excess:irs}

To quantify the presence of an infrared excess in the IRS spectra, we
determined if the observed spectra are better fitted by a model photosphere
alone, or by the sum of a model photosphere and a modified blackbody that
represents circumstellar dust emission. The photospheric component was derived
by fitting Kurucz synthetic spectra to optical and near-infrared photometry
between 0.5 and 2.2\micron. The stellar effective temperature and visual
extinction were free parameters in the fits, while the surface gravity and
metallicity were fixed (see Paper~I for details of the fitting procedure). 

These model spectra cannot be compared directly with the IRS spectra to infer
the presence of an infrared excess for two reasons. First, the mean flux
density of the model often differs from the observed spectra, which may reflect
either uncertainties in the model fit or calibration uncertainties in the
observed spectrum. Second, in some spectra an offset is present between the SL1
and LL2 IRS orders that is manifested as an abrupt jump in the flux density at
a wavelength of 14.2\micron. To correct for these offsets, the best fit Kurucz
model to the broad-band photometry was renormalized to the IRS spectrum. The
renormalization included a term to account for an overall flux density offset
between the model and the observed spectrum, and a second term to account for
the offset between the SL1 and LL2 orders. 

The variance between the Kurucz model fit (including the flux offset terms) and
the observed IRS spectra was computed between 12 and 35\micron. The same IRS
spectrum was then fitted with a Kurucz model plus modified blackbody [i.e.
$S_\nu \propto \nu^{3+\beta} / (e^{h\nu/(kT_{\rm d})} - 1)$] that represents
thermal emission from dust grains. The free parameters for the modified
blackbody are the dust temperature ($T_{\rm d}$) and the solid angle of dust
grains, which is proportional to the total cross-sectional surface area if the
grains are at a single temperature. The appropriate value of $\beta$ depends on
the grain properties that contribute emission in the IRS wavelengths. In
practice, the IRS spectra probe the Wien tail of the dust emission (see
discussion below) and do not place meaningful constraints on $\beta$.
Therefore, we assume $\beta=0.8$ to conform with the typical value inferred
from submillimeter observations of debris disks \citep{Williams06}. Adopting a
blackbody function ($\beta=0$) produces warmer dust temperatures but does not
otherwise alter the results of our analysis. The variance from the Kurucz
spectrum fit alone and that from the Kurucz spectra plus modified blackbody
were compared by computing the $F$-test statistic \citep[$\equiv
p$;][]{Press02}. If $p \ll 1$, we conclude that the Kurucz spectrum alone is a
poor fit to the IRS data.

\citet{Protassov02} emphasized that the probability distribution from the
$F$-test is not formally valid for this application since the second model 
adds a modified blackbody component that is not present in the first model. We
conducted Monte Carlo simulations to establish the correct probability
distribution where we took the best fit Kurucz model spectrum, introduced a
random offset to the SL1 order, and added wavelength-dependent random noise to
the spectrum. We then repeated the $F$-test analysis for this synthetic 
spectrum. The empirical probability distribution was derived from 8,000 
simulated spectra. Simulations were run using the noise characteristics of 
a relatively faint star (MML~32, signal-to-noise ratio of 4 in the 32\micron\ 
bandpass), and a bright star (vB~1, signal-to-noise of 13). Whereas we 
expected 80 of the 8,000 simulated spectra to have a probability $\le 0.01$ 
from random noise, the $F$-test yielded 76 for the MML~32 simulation and 86 
for vB~1. Similarly, we expect 8 source to have a probability $\le 0.001$, 
and the simulation yielded 7 and 10 for MML~32 and vB~1 respectively.
We conclude that the $F$-test, while not formally valid for this
application, nonetheless provides a reasonably accurate probability
distribution.

\begin{figure}
\begin{center}
\includegraphics[angle=0,scale=0.7]{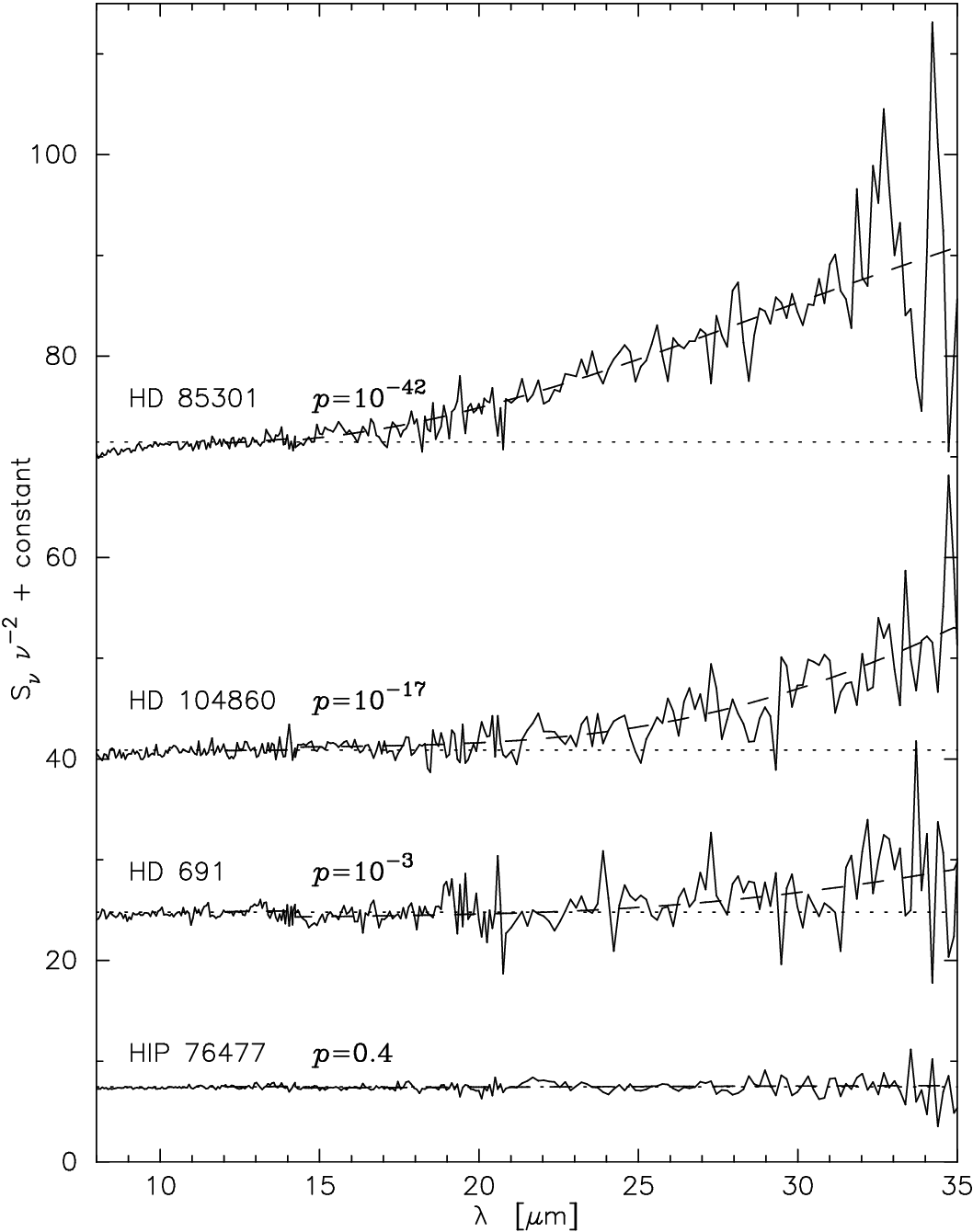}
\caption{
  \label{fig:spectra_example}
  Observed IRS low resolution spectra for four stars in the FEPS sample to
  illustrate how infrared excesses in the spectra were identified. The
  dotted line indicates the average value of $S_\nu\ \nu^{-2}$ between 12 and
  14\micron, and the dashed curve is the best fit Kurucz model (including flux
  offset terms; see text) plus modified blackbody. The value of $p$ is the
  $F$-test statistic that compares the variance from a Kurucz-model only fit 
  to the IRS spectra and a Kurucz-model plus modified blackbody. The lower the 
  value of $p$, the less likely the Kurucz model alone is a good fit to the IRS 
  spectrum. HIP~76477 is an example of a star where the IRS spectrum is 
  consistent with a stellar photosphere. The other 3 sources have emission 
  that departs from the stellar photosphere at longer wavelengths. 
}
\end{center}
\end{figure}

In Figure~\ref{fig:spectra_example}, we present IRS low resolution spectra for
four sources to illustrate the fitting results. For the star HIP~76477, we
derived $p=0.4$ and conclude that the IRS spectrum is consistent with 
photospheric emission. The spectrum indeed shows that the $S_\nu\ \nu^{-2}$
spectrum is roughly constant versus wavelength, which is expected for these
sources in the absence of dust since the emission is approximately in the
Rayleigh-Jeans limit at these wavelengths. The other three sources in
Figure~\ref{fig:spectra_example} have $p < 10^{-3}$, which indicates that the
spectra are poorly represented by Kurucz models. For these sources the observed
emission systematically exceeds a constant $S_\nu\ \nu^{-2}$ spectrum at the
longer IRS wavelengths and is consistent with the presence of emission from
dust grains.

In principle we can select a reliable list of sources with probable IRS 
excesses based solely on the $F$-test statistic. In practice, several sources 
have a low spectral intensity at long wavelengths (\aboutmore 30\micron) 
relative to the Kurucz model and are parameterized by a negative solid angle in
the modified blackbody fits. These spectra clearly do not result from dust 
emission, and likely indicate errors in the spectral extraction or low 
signal-to-noise in the IRS spectra at the longer wavelengths. We imposed the 
following criteria then to select sources with candidate IRS excesses:
({\it i}) the signal-to-noise ratio in a synthetic bandpass between 
    30 and 34\micron\ is $\ge 3$;
({\it ii}) the probability from the $F$-test is $p \le 0.003$,
and
({\it iii}) the solid angle of dust emission is $> 0$.
In total, 71 sources satisfied these criteria. However, whereas we expected 
one source detected at 32\micron\ to have such negative solid angles for 
$p \le 0.003$, the analysis yielded seven such sources. 
In \S\ref{excess_sum}, we combine the IRS and MIPS data to select a reliable 
sample of infrared excess sources that exhibit infrared excesses in both 
instruments.

\begin{figure}
\begin{center}
\includegraphics[angle=-90,scale=0.9]{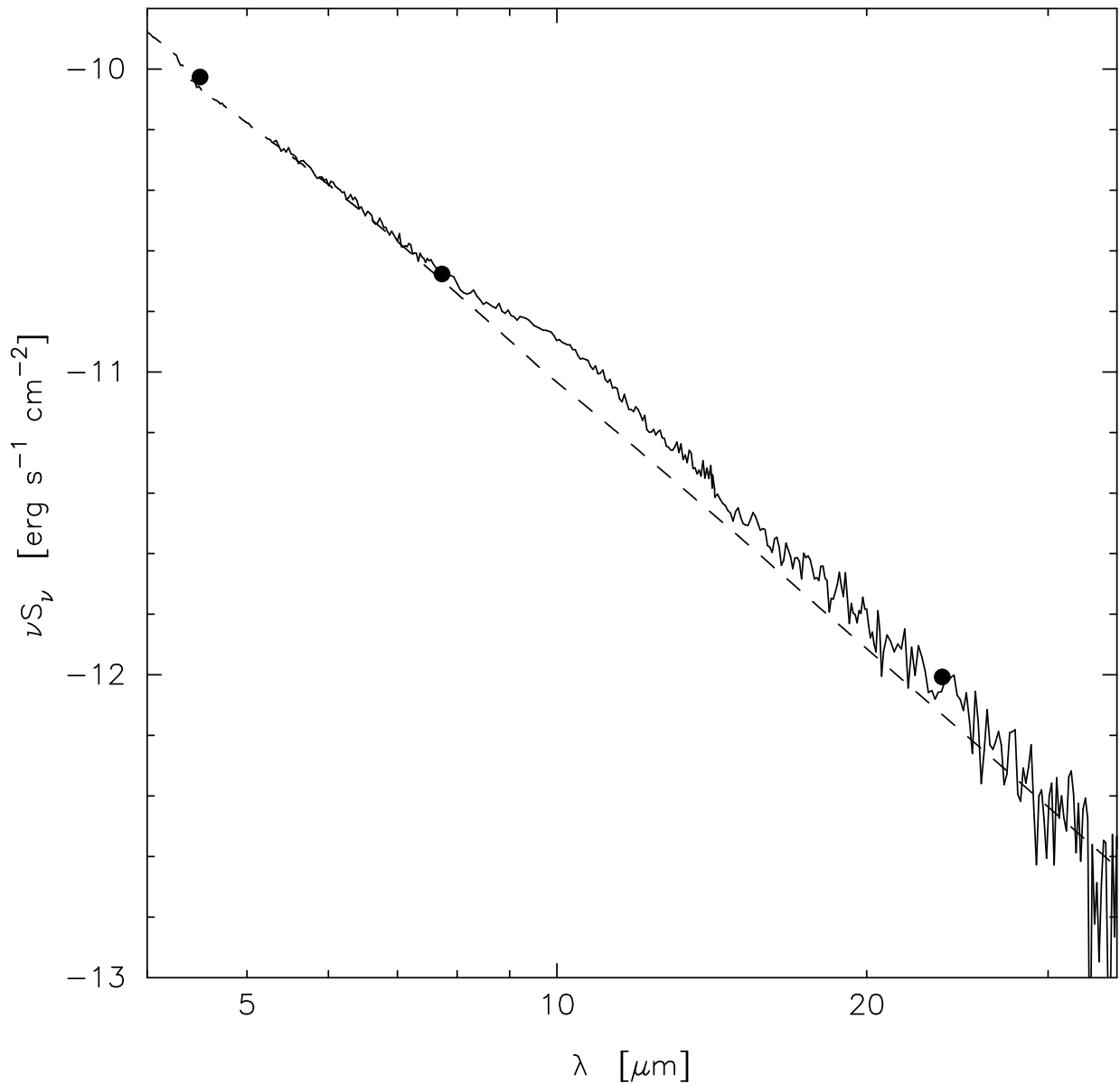}
\caption{
  \label{fig:sed_1rxs}
  Spectral energy distribution for 1RXS~J051111.1$+$281353 between
  5 and 35\micron. Solid curve is the IRS low resolution spectrum,
  dashed curved is the Kurucz synthetic spectrum normalized to
  optical and near-infrared photometry (see Paper~I), and solid
  circles represent IRAC and MIPS broad band photometry. Comparison
  of the IRS and model spectra suggest the presence of an infrared
  excess between 8 and 28\micron.
}
\end{center}
\end{figure}

From visual inspection of the IRS spectra, the star 1RXS~J051111.1$+$281353
had a spectral shape that was not amenable to the above analysis. In 
Figure~\ref{fig:sed_1rxs}, we present the spectral energy distribution between 
5 and 35\micron\ for this star, including the IRS spectrum (solid curve), IRAC 
and MIPS photometry (filled circles), and a Kurucz synthetic spectrum (dashed
curve) normalized to optical and near-infrared photometry (see Paper~I). The
IRS spectrum shows an apparent excess above the stellar photosphere between 9
and 28\micron, with perhaps a 10\micron\ silicate emission feature. The shape
of the excess emission is distinct from the other FEPS sources where the excess
emission increases toward longer wavelengths, and suggests that the excess
emission in 1RXS~J051111.1$+$281353 originates primarily from warm dust grains.
The star HD~72905 shows similar characteristics \citep{Beichman06b}.
The excess was not revealed by fitting a modified blackbody to the IRS spectrum
since this analysis allowed for an normalization constant, which 
removed the low-level infrared excess. A more detailed analysis of the 
spectrum is forthcoming. In the following analysis, we include 
1RXS~J051111.1$+$281353 as an IRS-verified excess to provide a total of
72 stars with candidate IRS excesses.

\subsection{IRS 16\micron}
\label{excess:16}

While analysis of the IRS spectra identified sources with infrared excesses
between 12 and 35\micron, in later sections it will be useful to quantify the
infrared excess at an intermediate wavelength between the IRAC 8\micron\ and
MIPS~24\micron\ bandpasses. For this exercise we used the 16\micron\ fluxes
computed from the IRS spectra (see \S\ref{data}). Since the 8\micron\ flux
density is mostly photospheric in origin (\S\ref{excess:irac}), we use the 16
to 8\micron\ flux density ratio ($\equiv$ \RII) to identify any stars with
16\micron\ excesses. After excluding the 5 known optically thick disks with
IRAC excesses, the median value of \RII\ for the FEPS observations is 0.244
with a dispersion about the median of $\sigma($\RII) = 5.4\%, which we adopt as
the photospheric value and minimum uncertainty in \RII, respectively. Of the
314 sources in the sample, only the 5 stars that are surrounded by optically
thick disks have a 16\micron\ excess more than 3$\sigma$(\RII) = 16.2\% above
the photosphere. The star 1RXS~J051111.1$+$281353 also has a 16\micron\
excess (see Fig.~\ref{fig:sed_1rxs}), but at a lower level.

\subsection{MIPS 24\micron}
\label{excess:mips24}

The methods adopted in the literature to identify
24\micron\ excesses include measuring how much the observed 24\micron\ flux 
density exceeds a model stellar spectrum normalized at optical and 
near-infrared wavelengths \citep[e.g.][]{Bryden06}, and identifying 
sources with anomalously red $K-[24]$ colors \citep[e.g.][]{Siegler07}. After 
considering these approaches, we used the 24 to 8\micron\ flux density 
ratio ($\equiv$ \RM) which empirically resulted in the most sensitive search
for sources with 24\micron\ excesses. \citet{Meyer08} adopted a similar
approach to identify 24\micron\ excesses in the FEPS sample, but used a
single detection threshold over all brightness levels to identify excess 
sources. We extend their analysis by using the final FEPS data processing (see 
Paper~I) and adopting separate detection thresholds for bright and faint 
sources.

\begin{figure}
\begin{center}
\includegraphics[angle=-90,scale=0.9]{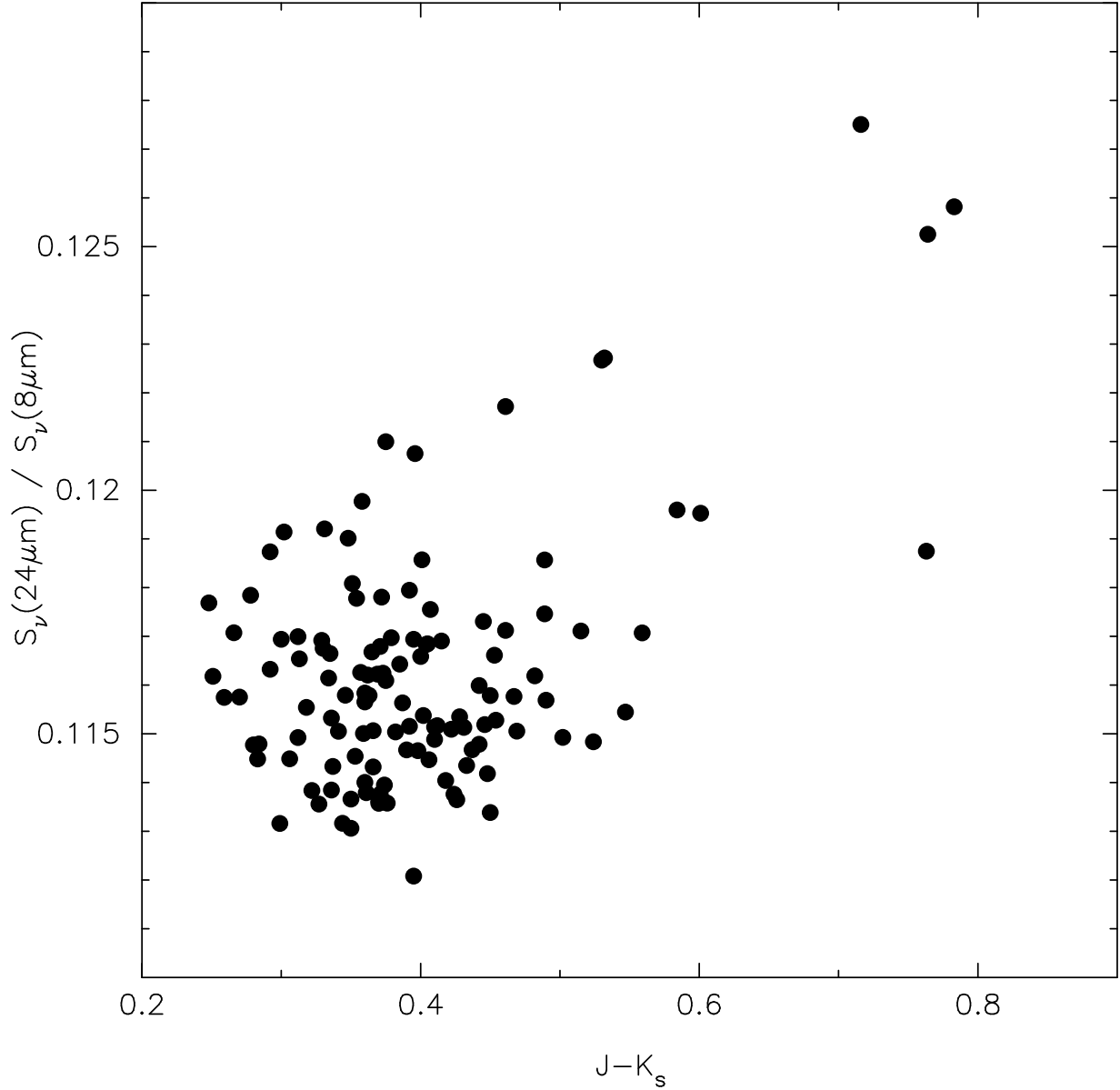}
\caption{
  \label{fig:r24_jk}
  24\micron\ to 8\micron\ flux density ratio (\RM) as a function of 
  $J-K_{\rm s}$ color for stars with $S_\nu(8\micron) \ge 100$~mJy that do not
  have an IRS excess ($p > 0.003$). An apparent trend exists in that stars 
  with the reddest $J-K_{\rm s}$ colors tend to have large value of \RM. Photometry 
  was dereddened using the extinction values listed in Paper~I.
}
\end{center}
\end{figure}

In Figure~\ref{fig:r24_jk}, we show the dependence of \RM\ on the dereddened 
2MASS $J-K_{\rm s}$ color, a proxy for spectral type, for stars brighter than 
$S_\nu(8\micron) = 100$~mJy that do not have an apparent IRS excess
($p > 0.003$). These criteria were adopted to select high signal to 
noise photometry consistent with photospheric emission.
An apparent trend exists in that stars with the largest value of \RM\ also 
have red $J-K_{\rm s}$ colors. All four stars with $J-K_{\rm s} > 0.7$ have 
spectral types between K0 and K5, consistent with an increasing value of \RM\
toward later spectral types. A similar trend is observed in $K$-band where
the $K-[24]$ color is \about 0.0~mag over the spectral type range F2 to K4, 
and becomes redder toward later spectral types \citep{Beichman06a}. However, 
we cannot confidently derive the 
functional dependence on \RM\ on $J-K_{\rm s}$ since the trend depends entirely 
on the four reddest stars. For simplicity, we assume a 
photospheric value of \RM\ = 0.116 for $J-K < 0.7$, and \RM\ = 0.125 
for redder stars, which are the median \RM\ colors in the respective 
$J-K_{\rm s}$ color range of stars without IRS excesses.

\begin{figure}
\begin{center}
\includegraphics[angle=-90,scale=0.9]{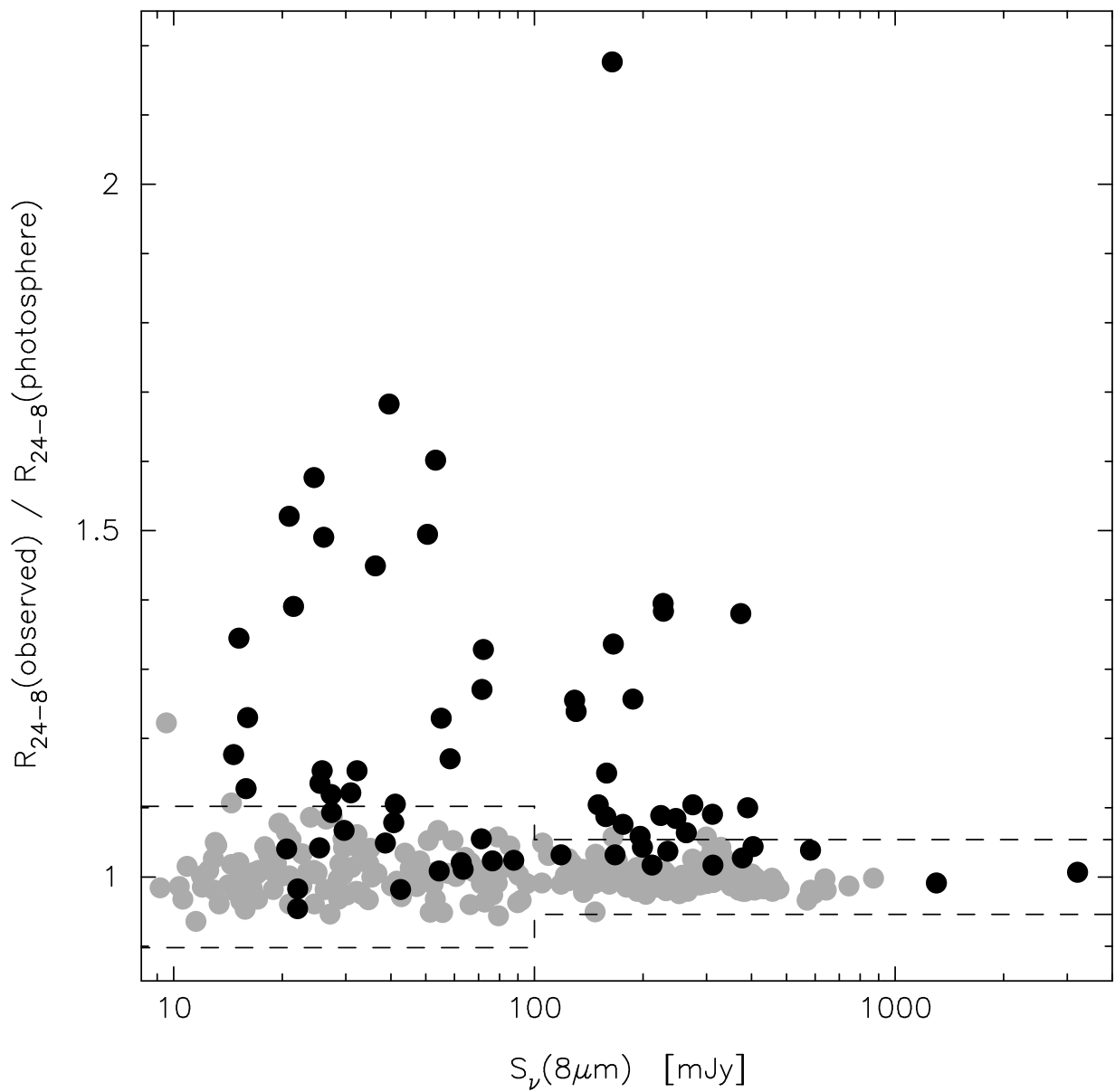}
\caption{
  \label{fig:r24_flux_irs}
  Observed 24\micron\ to 8\micron\ flux density ratio normalized by the 
  photospheric value as a function of the observed 8\micron\ flux density.
  The dashed lines show the 3$\sigma$ limits used to identify sources with 
  MIPS~24\micron\ excesses, which is a 10.2\% excess for stars fainter than 
  $S_\nu(8\micron) = 100~{\rm mJy}$ and 5.4\% for brighter stars. Black 
  circles represent sources that exhibit an apparent excess in the IRS 
  spectra independent of the MIPS~24\micron\ photometry. Five sources with 
  optically thick disks (see Fig.~\ref{fig:irac_ccd}) are offscale on this plot.
}
\end{center}
\end{figure}

In Figure~\ref{fig:r24_flux_irs}, we show the observed \RM\ ratio, normalized
by the adopted photospheric value, as a function of the 8\micron\ flux density.
Since the 8\micron\ emission is mostly photospheric in origin (see
\S\ref{excess:irac}), sources with a large value of \RM\ are candidate
24\micron\ excess sources. To determine the minimum detectable 24\micron\
excess, we consider the empirical scatter in \RM\ for sources without an IRS
excess ($p > 0.003$) as shown by the gray circles in
Figure~\ref{fig:r24_flux_irs}. The value of the normalized \RM\ appears nearly
constant at a mean value of 1.0, but the RMS about the mean increases for
sources fainter than $S_\nu(8\micron) \sim 100~{\rm mJy}$. The computed RMS
after rejecting one outlier point with normalized \RM\ $ > 1.2$ (W79;
see \S\ref{excess_sum}) is 1.8\% and 3.4\%
for sources brighter and fainter than 100~mJy respectively. By comparison, the
median uncertainty in \RM\ for the two brightness ranges is 1.1\% and 1.3\%,
which is smaller than the observed scatter in \RM. We adopt a minimum
uncertainty in \RM\ of 3.4\% for stars fainter than $S_\nu(8\micron)=100~{\rm
mJy}$ and 1.8\% for brighter sources. These minimum uncertainties imply a
3$\sigma$ detection limit for a 24\micron\ excess above the photosphere of
10.2\% for faint stars and 5.4\% for bright sources. The detection limits
for a 24\micron\ excess are indicated by the dashed lines in
Figure~\ref{fig:r24_flux_irs}. Of the 314 stars in the sample, 50 have a
value of \RM\ more than 3$\sigma$ above the photosphere. 

\subsection{MIPS 70\micron}
\label{excess:mips70}

Photometric excesses in the MIPS~70\micron\ band were identified from
comparison of the measured flux densities with the expected photospheric
contribution. 
The 70\micron\ photospheric flux density was estimated from the photospheric
value of \RM\ described in \S\ref{excess:mips24}, and further assuming that the
intrinsic photospheric $[24]-[70]$ color is 0~mag. We adopt flux densities for
a zero magnitude star of 7.14~Jy and 0.775~Jy for the 24\micron\ and 70\micron\
band respectively as reported on the MIPS calibration web
pages\footnote{http://ssc.spitzer.caltech.edu/mips/calib/} as of April 30,
2007. \citet{Hillenbrand08} also identified sources with 70\micron\ excesses in
the FEPS data, but they used the Kurucz synthetic spectra normalized to optical
and near-infrared broad-band photometry to estimate the 70\micron\ photospheric
emission.

\begin{figure}
\begin{center}
\includegraphics[angle=-90,scale=0.9]{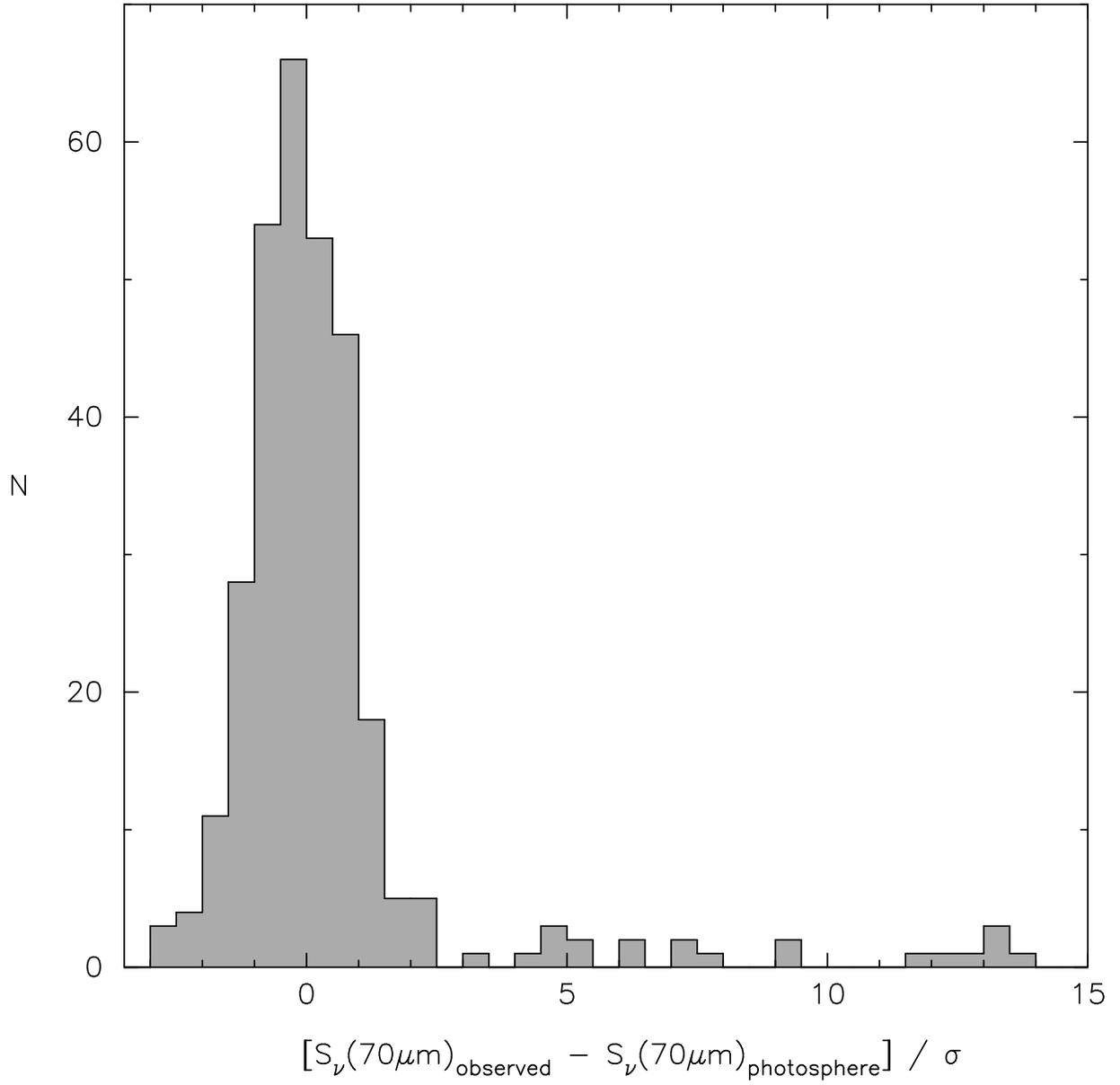}
\caption{
  \label{fig:mips70_excess}
  Histogram of the signal to noise ratio of the observed 70\micron\ flux
  density above the expected stellar photospheric value. The photospheric
  contribution was estimated from the observed 8\micron\ flux density and an
  assumed 8\micron\ to 70\micron\ flux density ratio. 
}
\end{center}
\end{figure}

Of the 314 sources, 70\micron\ emission was detected toward 22 sources with a
signal to noise ratio $\ge 3$. To establish which sources exhibit a
70\micron\ excess from circumstellar dust, we present in
Figure~\ref{fig:mips70_excess} a histogram of the signal-to-noise ratio of the
observed 70\micron\ emission above the stellar photosphere. The signal to noise
of the excess is defined as $(S_\nu(70\micron)_{\rm observed} -
S_\nu(70\micron)_{\rm photosphere}) / \sigma$, where $\sigma$ is the 
uncertainty in the difference between the observed and
photospheric flux densities including calibration uncertainties. The signal to
noise is centered near zero with a tail toward positive values that may
indicate 70\micron\ excess sources. Of the 22 sources detected at 70\micron,
one source (HD~13974) has 70\micron\ emission consistent with the photosphere,
and 21 sources have a 70\micron\ excess above the stellar photosphere with a
signal to noise ratio greater than three. All 21 sources were also identified
as 70\micron\ excess sources by \citet{Hillenbrand08}. 

\subsection{Synthesis}
\label{excess_sum}

We now combine results from the individual instruments to select a list 
of sources with infrared excesses that will be analyzed in the remainder
of this paper. Since the MIPS~24\micron\ bandpass is encompassed by the IRS 
spectral coverage, any 24\micron\ photometric excess should be verifiable with 
IRS assuming comparable sensitivity. As discussed in \S\ref{excess:irs},
approximately 7 of the candidate IRS sources could be spurious. The 
expected false detection rate for the MIPS~24\micron\ excesses is \aboutless
1 since no sources have observed \RM\ values more than 3$\sigma$ below the 
adopted photospheric value (see Fig.~\ref{fig:r24_flux_irs}). Given the 
detection rate of 3$\sigma$ MIPS~24\micron\ excesses (16\%), the expected
number of sources with a valid MIPS~24\micron\ and a false IRS excesses is 
\about 1. The expected number of sources with a false MIPS~24\micron\ and 
false IRS excess is \about 0.02. This is consistent with the fact that none of 
the 7 IRS sources with negative solid angles have a 24\micron\ photometric 
excess. By requiring both an IRS and MIPS~24\micron\ excess, we aim
to create a more reliable sample of IR excesses.

In Figure~\ref{fig:r24_flux_irs}, the 72 sources with apparent IRS excesses 
(see \S\ref{excess:irs}) are marked as black circles on the color-flux diagram 
used to 
identify MIPS~24\micron\ excesses. Of the 50 sources with a $\ge 3\sigma$
photometric excess at 24\micron, 46 are also identified with an IRS excess.
Conversely, 26 sources have an IRS excess but not a 24\micron\ photometric
excess.

The 4 sources with MIPS~24\micron\ excesses that are not verified
spectroscopically are R45, V343~Nor, and V383~Lac, and W79. R45 has a 10.2\%
excess with a SNR of 3.0 and is at the limit to define a 24\micron\ excess. R45
is located in bright nebulosity and the extracted IRS spectrum has negative
flux densities at the longer wavelengths, perhaps because of poor background
subtraction. The apparent 24\micron\ photometric excesses in V343~Nor and
V383~Lac are 6\% above the photosphere. These two stars have K0 spectral types
with $J-K_{\rm s}$ colors of 0.53~mag, and therefore we adopted a low value for
the intrinsic \RM. Since the photospheric value of \RM\ appears to increase
toward later spectral types (see Fig.~\ref{fig:r24_jk}), the apparent
photometric excess could be explained if we have underestimated the intrinsic
value of \RM\ by more than 0.6\%. This is possible given our simplistic
treatment on how \RM\ varies with $J-K_{\rm s}$ color. The star W79 has a
photometric 24\micron\ excesses of \about 22\% above the photosphere at a
signal-to-noise ratio of \about 6.5. We estimate that the apparent 24\micron\
excess should have been detected in the IRS spectra between 24 and 34\micron\
at $\ge$ 4.4$\sigma$ for dust temperature $\le$ 100~K. The 24\micron\ image for
W79 contains extended and structured cirrus emission that complicates
background subtraction and could conceivably create an apparent excess. Given
the discrepant results between the IRS spectra and MIPS~24\micron\ for W79,
V343~Nor, and V383~Lac, and that R45 is at the limit to identify a photometric
excess but does not have a confirming spectrum, we do not consider these four
sources to have a MIPS~24\micron\ excess in the remainder of this paper. 

\begin{figure}
\begin{center}
\includegraphics[angle=0,scale=0.8]{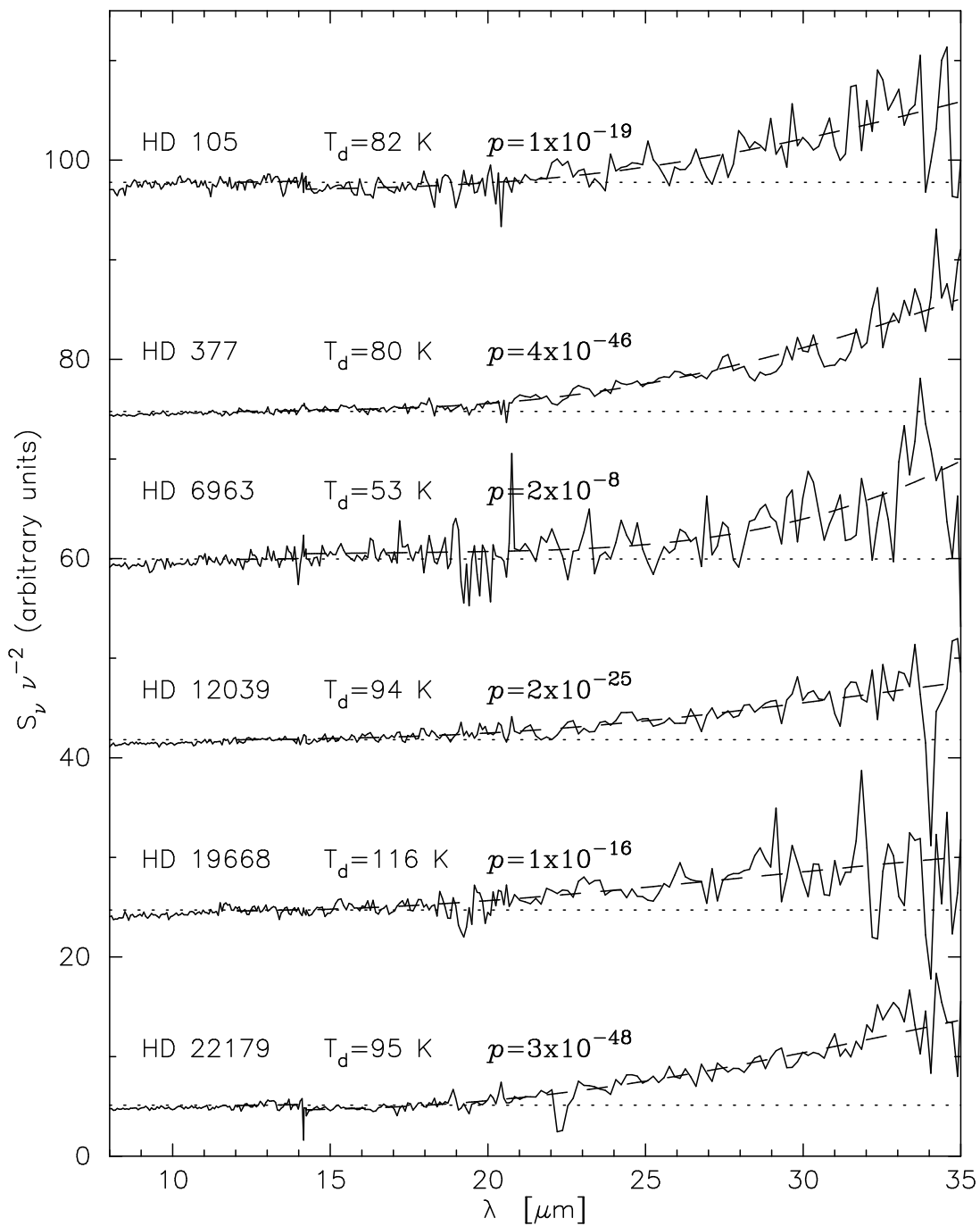}
\caption{
  \label{fig:mips24irs}
  IRS low resolution spectra for 40 FEPS sources that have both a
  24\micron\ photometric excess and an IRS spectroscopic excess.
  Spectra for five additional sources with optically thick 
  disks are presented in \citet{Bouwman08} and are not shown here.
  The dashed and dotted curves and the variable $p$ have the same meaning
  described in Figure~\ref{fig:spectra_example}. The dust temperature ($T_d$) 
  inferred from the modified-blackbody fit ($\beta=0.8$) are indicated 
  for each source.
}
\end{center}
\end{figure}

We thus have 46 sources with both a 24\micron\ and IRS excess. Spectra for 40
of these sources are presented in Figure~\ref{fig:mips24irs}; spectra for 5
sources appear in \citet{Bouwman08} in a study of optically thick disks in the
FEPS sample, and 1RXS~J051111.1$+$281353 is presented in
Figure~\ref{fig:sed_1rxs}. The detection of an infrared excess in both MIPS and
IRS does not ensure the dust emission is associated with the star, since
contamination by interstellar cirrus and galaxies will affect both
measurements. The expected extragalactic contamination can be assessed from
the observed extragalactic counts as a function of 24\micron\ flux density
from \citet{Papovich04}. For each star, we computed the probability that at
least one galaxy is present within the FWHM size (6\arcsec) of the
MIPS~24\micron\ point response function that will produce a $\ge 3\sigma$
photometric excess at 24\micron. The photospheric 24\micron\ fluxes were
computed from the observed 8\micron\ flux density and the adopted photospheric
values of \RM\ (see \S\ref{excess:mips24}). We find that \about
1 FEPS source could be contaminated by an extragalactic source bright enough
to produce an apparent 24\micron\ excess.

In Paper~I, we considered the positional coincidence of the
MIPS~24\micron\ sources with the stellar coordinates to search for potential
contaminants. For this sample of 314 stars, the largest astrometric offset
between a MIPS 24\micron\ excess source and the 2MASS stellar position is
1.3\arcsec\ for HD~201219, which is a 2.5$\sigma$ deviation based on the
observed dispersion in the coordinate offsets. (Two sources without 24\micron\
excesses had larger angular offsets.) The 70\micron\ detection of
HD~201219 is offset by 3.4\arcsec\ from the stellar position. We consider
this source to have an excess, but the photometry should be viewed with some
caution. We conclude that while the photometry for \about 1 FEPS source may be 
contaminated by a galaxy, there are no clear instances of contamination to 
the 24\micron\ photometry\footnote{\citet{Stauffer05} concluded that the
measured 24\micron\ excesses toward HII~152 and HII~250 were likely caused
by background galaxies based on the positional mismatches between 2MASS and
the 24\micron\ \Spitzer\ image from pipeline version S10.5. However, the 
positional differences in the \Spitzer\ data reduction pipeline version
S13 were not significant (see Paper~I) and we consider these apparent 
excesses to be real.}.

\begin{figure}
\begin{center}
\includegraphics[angle=0,scale=0.9]{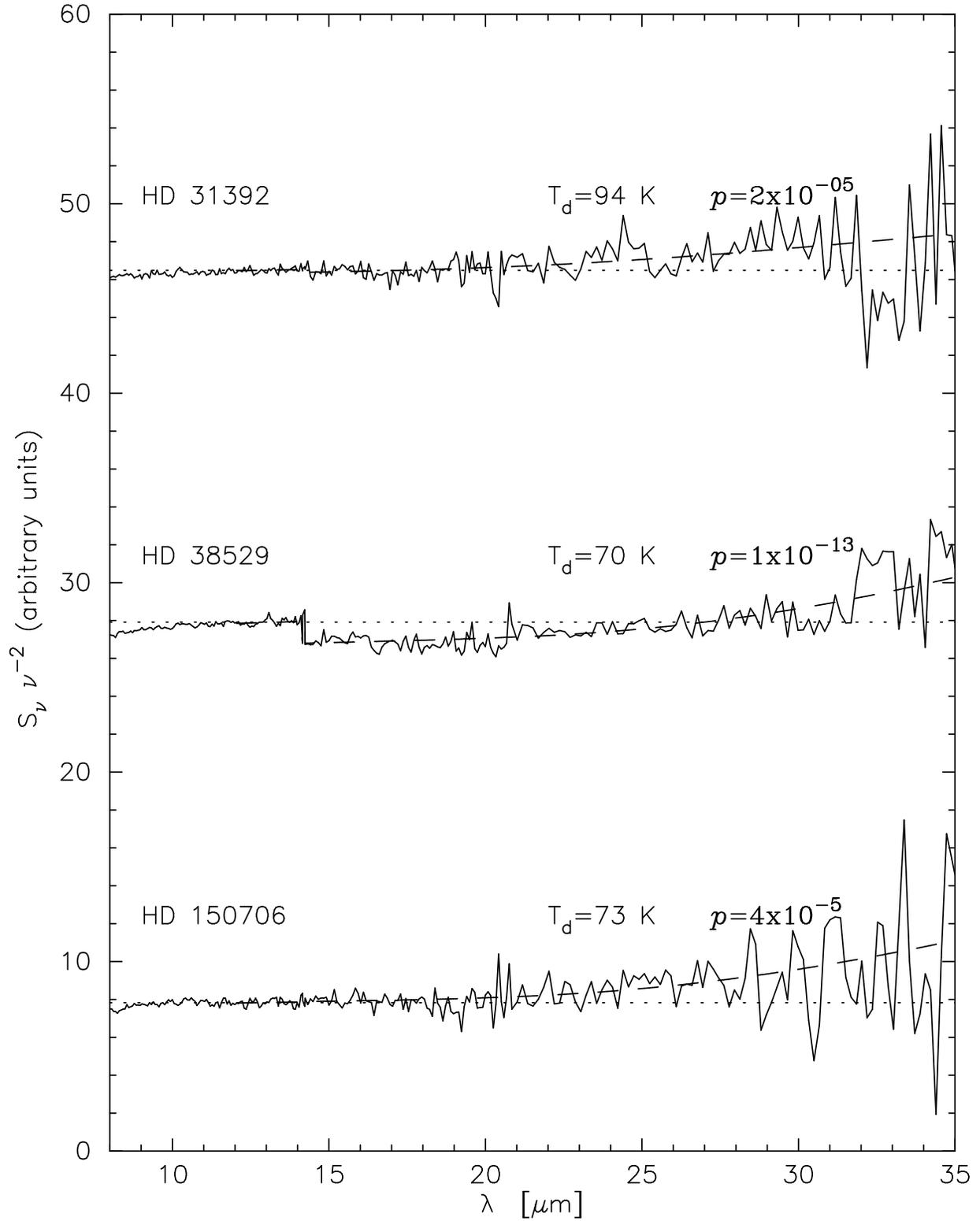}
\caption{
  \label{fig:mips70irs}
  Same as Figure.~\ref{fig:mips24irs}, but for 3 sources with MIPS~70\micron\
  and IRS excesses and no MIPS~24\micron\ excess.
}
\end{center}
\end{figure}

The reliability of sources with 70\micron\ excesses can be assessed in a
similar manner. Of the 21 sources with $\ge 3\sigma$ excess at 70\micron, 17
also have a 24\micron\ and IRS excess. Three of the remaining four 70\micron\
excess sources have an IRS excess but not a 24\micron\ excess; spectra for these
sources are presented in Figure~\ref{fig:mips70irs}. The one 70\micron\ excess
source without a confirming 24\micron\ or IRS excess is HD~187897 ($p = 0.2$),
where the signal-to-noise ratio of the 70\micron\ excess is 6. The detected
70\micron\ sources are within 5\arcsec\ of the stellar position, and the
expected number of contaminants from extragalactic sources that are nearly
centered on the stellar source is negligible \citep{Hillenbrand08}. We
anticipate that most, if not all, of these sources are real
70\micron\ excesses associated with the stellar target. 

\begin{figure}
\begin{center}
\includegraphics[angle=0,scale=0.9]{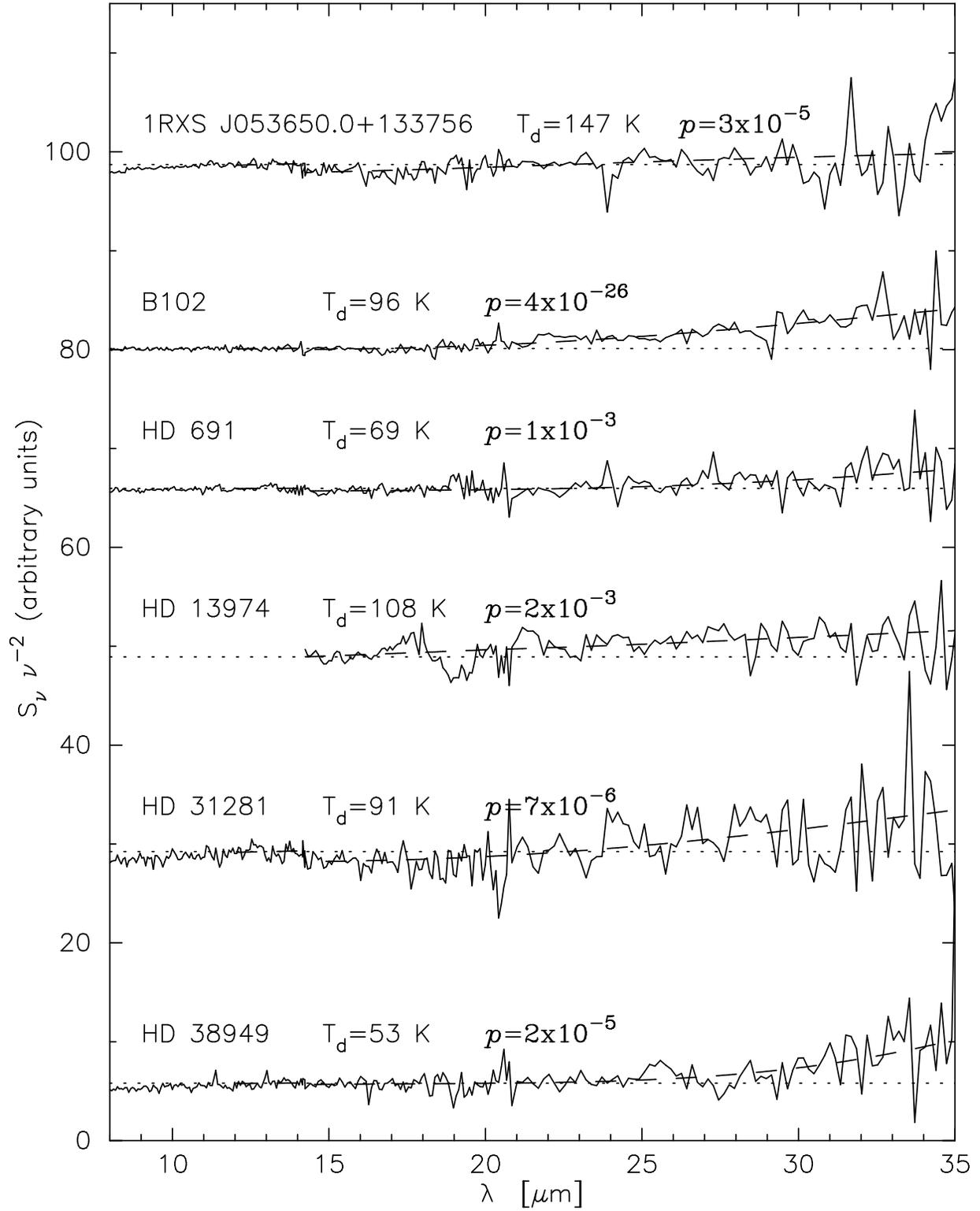}
\caption{
  \label{fig:irsonly}
  Same as Figure~\ref{fig:mips24irs}, but for 21 sources with IRS excesses
  and no detectable MIPS~24 or 70\micron\ excess.
}
\end{center}
\end{figure}

While the infrared excesses in most sources are confirmed with two or more
\Spitzer\ instruments, 23 sources have an excess detected with IRS that is not
substantiated by either 24\micron\ or 70\micron\ photometry. The apparent IRS
excesses for two sources (HD~104467 and RX~J1531.3-3329) are deemed spurious
since the inferred IRS dust temperatures are less than 14~K, and the model
70\micron\ flux density should have been readily detected ($\gg3\sigma$) if the
excess was real. For the other 21 sources, the model 70\micron\ density would
have been detected at less than 2$\sigma$ and the observations cannot rule out 
the IRS excess. Spectra for the these 21 sources are presented in
Figure~\ref{fig:irsonly}. 

\begin{figure}
\begin{center}
\includegraphics[angle=-90,scale=0.9]{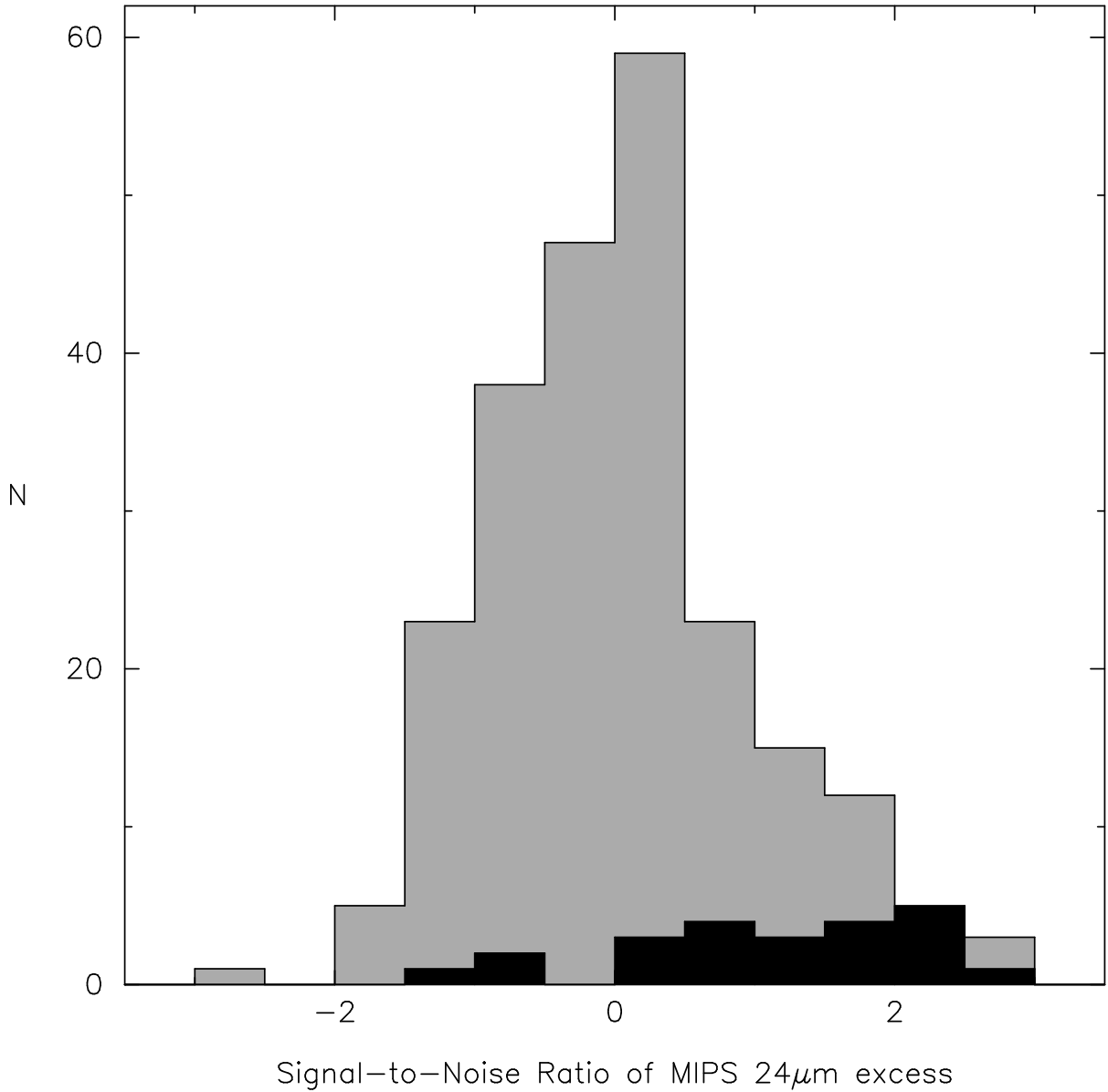}
\caption{
  \label{fig:r24_irs}
  Histograms of the MIPS~24\micron\ excess signal-to-noise ratio for subsets 
  of the FEPS sample. The black histogram represents stars that have a 
  $\ge 3\sigma$ excess from IRS spectra, but the excess is not confirmed with 
  MIPS~24 or 70\micron\ photometry. The gray histogram indicates sources
  that have less than a 3$\sigma$ excess in the IRS, MIPS 24\micron, and 
  MIPS 70\micron\ data. Sources with 
  unconfirmed IRS excesses tend to have positive signal-to-noise ratios for
  the 24\micron\ excess, suggesting that the IRS excess is indeed real for
  many of these sources.
}
\end{center}
\end{figure}

To investigate whether or not these sources likely contain real IRS excesses,
in Figure~\ref{fig:r24_irs} we compare the signal-to-noise ratio of the
MIPS~24\micron\ excess for sources with unconfirmed IRS excesses (black
histogram) and without a detectable excess in any of the \Spitzer\ instruments
(gray histogram). Sources with unconfirmed IRS excesses tend to have larger
signal-to-noise ratios for the MIPS~24\micron\ excess than sources without IRS
excesses. Comparison of the dereddened $J-K_{\rm s}$ colors for the two samples
using the Kolmogorov-Smirnoff (K-S) test indicates the two samples have
indistinguishable distributions of $J-K_{\rm s}$ colors (K-S probability =
0.24), indicating that the differences in the excess distributions is not a
result of a systematic error in the assumed intrinsic colors. By contrast, the
K-S probability that the two populations have the same signal-to-noise
distribution of 24\micron\ excesses is 10$^{-6}$, which suggests that it is
unlikely the positive bias to the 24\micron\ excesses can be attributed to
random noise. Moreover, if the IRS-only excesses were due to random noise, 
the age distribution of these sources should mimic that of the full
sample. Instead, 19 of the 21 sources with IRS-only excesses are younger than
300~Myr, while 202 of the 314 stars in the full sample are this young. The
probability that this difference in the age distributions could result by
chance is 1.6\%. These results suggest that many of the IRS-only excess sources
likely have a real infrared excess, but the 24\micron\ photometric excess is
too weak to detect.

In summary, we identified 50 stars that are considered to have a reliable 
infrared excess: 46 stars have both a MIPS~24\micron\ ($\ge 3\sigma$) and 
IRS excess ($p \le 0.003$), 3 stars have both IRS and MIPS~70\micron\ excess 
($\ge 3\sigma$) but no MIPS~24\micron\ excess, and 1 star has a 70\micron\ 
excess only. Of these 50 stars, 45 are considered ``debris'' disks and 5
are ``primordial'' disks (see \S\ref{evolution}). An additional 21 stars have
an apparent IRS excess that is unconfirmed photometrically, but many of these
sources likely have an excess based on the tendency to have positive 24\micron\
photometric excesses (but less than $3\sigma$).

\section{Temporal Evolution of Debris Disk Properties}
\label{evolution}

The results presented in Figures~\ref{fig:irac_ccd} plus
Figures~\ref{fig:mips24irs} and \ref{fig:mips70irs}
suggest a dichotomy in disk properties in that only 5 sources have excess
emission at wavelengths $\le$ 8\micron, while for the remaining
sources the excesses appear only at longer ($> 16$\micron) wavelengths. The 5
sources with 8\micron\ excesses have properties (e.g. circumstellar disk 
masses, H$\alpha$ accretion signatures, shape of the spectral energy
distribution, high fractional disk luminosities) consistent with optically 
thick accretion disks \citep{Silverstone06,Bouwman08}. We show in
\S\ref{evolution:ldust} that the remaining sources have fractional infrared
excess luminosities of $L_{IR}/L_*$ \aboutless $10^{-3}$ and are consistent
with optically thin dust emission. We assume that the thick disks represent
``primordial'' disks formed during the star formation process, and the
optically thin systems are ``debris'' disks, although the transition between
these states is not well characterized observationally. 

In this section, we establish the empirical signature of disk evolution within
the FEPS sample from this sample of debris disks. Since no debris disks were
detected at 8\micron, we investigate any evolutionary trends using MIPS
photometry and IRS spectra. We first examine if the FEPS sample of debris disks
contains any bias with respect to stellar luminosity, since dust emission will
be brighter around more luminous stars for a given dust surface area and 
orbital radius. Stars with detected debris disks have spectral types ranging 
from K3 to F5 and span an order of magnitude in stellar luminosity. 
The median stellar luminosity for stars with 24\micron\ excesses is 1.15\lsun,
compared to the median luminosity of 1.07\lsun\ for the entire sample. The K-S
test indicates a 77\% probability that the stellar luminosities for stars with
and without 24\micron\ excesses are drawn from the same parent population. A
similar probability was derived for stars with and without 70\micron\ excesses.
Thus there is no evidence for luminosity bias within the FEPS debris disk
sample, and we use these stars to investigate trends in the debris properties.

\subsection{24\micron\ excesses}
\label{evolution:excess24}

\begin{figure}
\begin{center}
\includegraphics[angle=-90,scale=0.9]{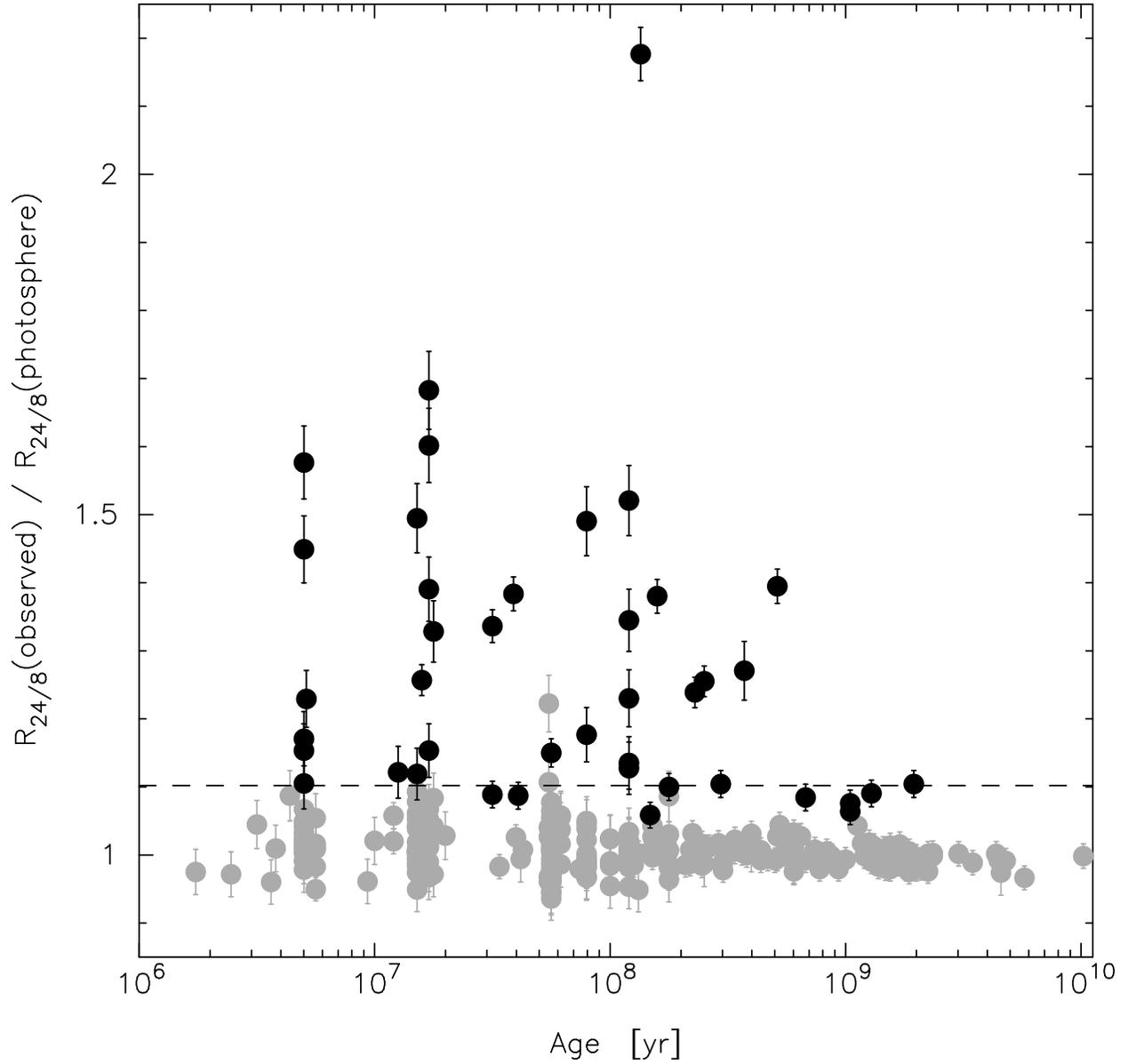}
\caption{
  \label{fig:mips24_age_fbol}
  Ratio of the observed MIPS~24\micron\ flux to 24\micron\ photospheric
  flux as a function of stellar age. Filled symbols represent the 45 sources
  that have a 24\micron\ photometric excess confirmed by a IRS spectrum. 
  The dashed line at a value of 1.102 shows the minimum 24\micron\ excess 
  that could be detected at the $\ge 3\sigma$ level over all stellar ages.
  Five sources with excesses
  from optically thick circumstellar disks are offscale on this plot.
}
\end{center}
\end{figure}

In Figure~\ref{fig:mips24_age_fbol}, we show the ratio of
observed-to-photospheric \RM\ versus stellar age to investigate the temporal
evolution of 24\micron\ excess emission. The magnitude of the 24\micron\
excesses ranges from a low of 5.9\% above the photosphere, as limited by the
accuracy of the photometric calibrations, to a high of 118\% for 
HD~61005\footnote{\citet{Hines07} present resolved scatter light images 
of the HD~61005 debris system obtained with the Hubble Space Telescope.}.
The decline in the magnitude of the excess toward older ages appears abrupt in
that the upper envelope of 24\micron\ excesses is roughly flat at \about 50\%
above the photosphere (with HD~61005 as the main outlier) for ages \aboutless
300~Myr, and \about 10\% above the photosphere for ages $>$ 500~Myr.
Based on the Kendall's rank correlation statistic, as implemented in the 
ASURV Rev 1.2 package \citep{Lavalley92}, the probability that the apparent
correlation between stellar age and the magnitude of the 24\micron\ excess 
can result from chance is $4\times10^{-5}$.

To further quantify the temporal evolution of the 24\micron\ excess, we select
a uniform sample of excess sources over all ages. While the IRS spectra are
sensitive to smaller excesses, the 24\micron\ photometry is more uniform and is
available for the full FEPS sample. Therefore, we draw a uniform sample based
upon the 24\micron\ photometry. The precision of \RM\ is poorest for sources
with $S_\nu < 100$~mJy, where $\sigma$(\RM)/\RM~=~3.4\% (see 
\S\ref{excess:mips24}). The minimum 24\micron\ excess that can be
detected over all stellar ages at $\ge 3\sigma$ is 10.2\%. A total of 38
sources have a 24\micron\ photometric excess greater than this limit, of which
5 are primordial disks and 33 are debris disks.

\begin{figure}
\begin{center}
\includegraphics[angle=-90,scale=0.9]{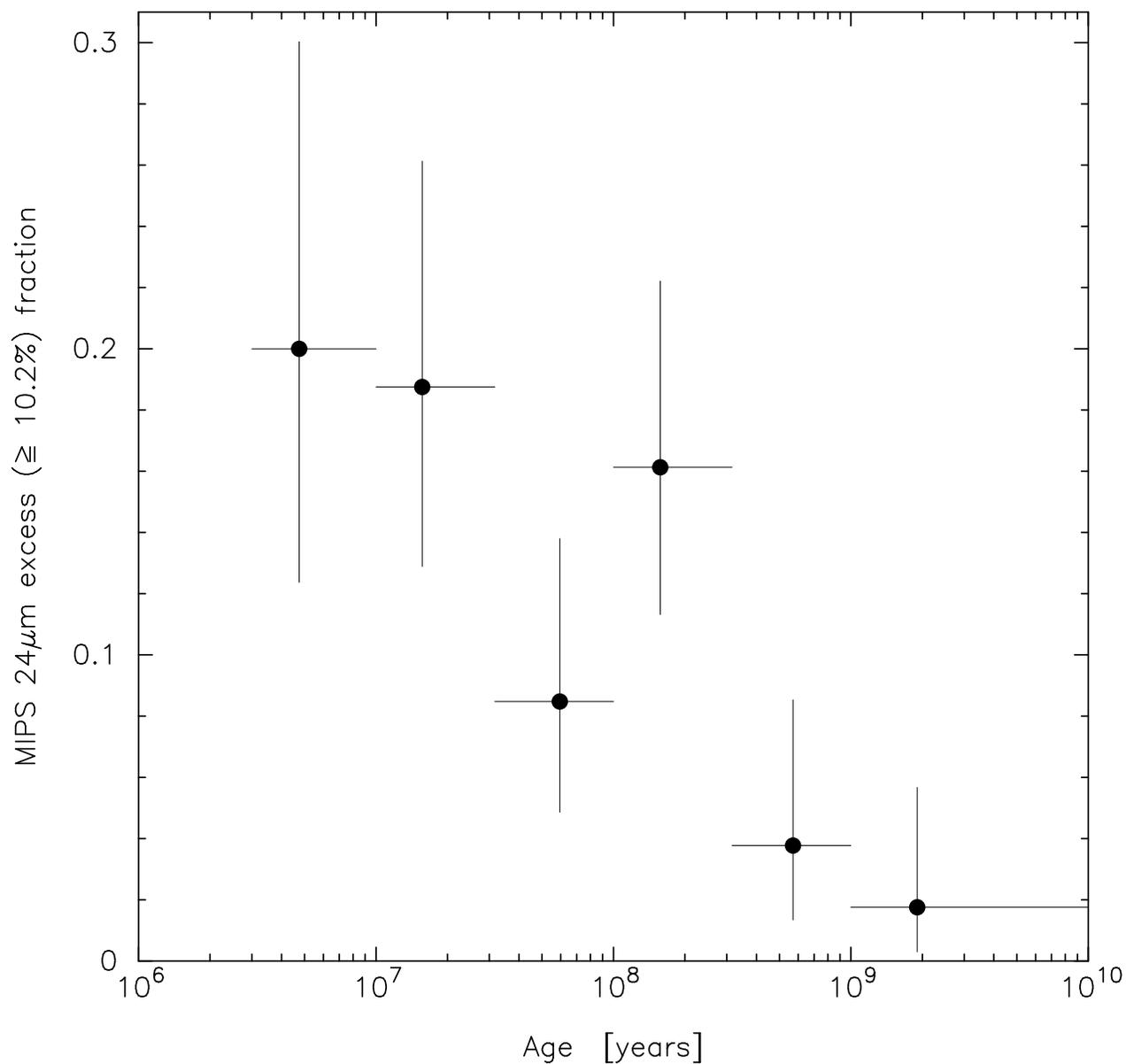}
\caption{
  \label{fig:mips24_age_fraction}
  Fraction of sources with a MIPS~24\micron\ excess greater than 10.2\% of the
  photosphere, and verified with the IRS spectrum, as a function of age. 
  Five sources with optically thick disks at ages $<12$~Myr are not included 
  to show only the temporal evolution from 
  optically thin debris dust. Vertical error bars represent the 1$\sigma$ 
  uncertainties 
  computed from binomial statistics using the tables in \citet{Gehrels86}. 
  Horizontal ``error'' bars are the range of stellar ages in the corresponding 
  bin, where the points are placed at the mean age in the bin.
}
\end{center}
\end{figure}

In Figure~\ref{fig:mips24_age_fraction}, we show the fraction of debris disks
that have a 24\micron\ excess greater than 10.2\% versus stellar age. The age
bins, selected to span a factor of three in age, are younger than 10~Myr,
between 10 and 30~Myr, 30-100~Myr, 100-300~Myr, 300-1000~Myr, and older than
1~Gyr. Preliminary results of this analysis were presented in \citet{Meyer08},
but we use the combined MIPS and IRS data to select a more reliable sample of
sources with 24\micron\ excesses. The results presented in
Figure~\ref{fig:mips24_age_fraction} and tabulated in Table~\ref{tbl:excess}
are consistent with the 24\micron\ excess fraction remaining constant to within
the uncertainties at a mean of 15\% for ages less than 300~Myr. For older ages,
the excess fraction declines to a mean of 2.7\%. Using the two-tailed Fisher's
exact test to compare ratios, the probability that the 24\micron\ excess
fraction is the same for ages younger and older than 300~Myr is 0.04\%. This
result indicates a decline in the 24\micron\ excess fraction with stellar age.
However, the stellar age uncertainties prevent us from determining the
form of the decline.  As noted in \citet{Meyer08}, age uncertainties will tend 
to soften the decline perhaps indicating that it is starker than observed here. 

Because the Fisher's exact test considers only the binomial distribution of the
excess fraction and does not account age uncertainties, we
conducted a Monte Carlo experiment where the estimated stellar ages were
modified randomly within a Gaussian distribution, and the fraction of stars
with 24\micron\ excesses younger and older than 300~Myr was re-evaluated. For
stars that are members of associations or clusters, we assumed an age
uncertainty of $\sigma[{\rm log}(age)] = 0.15$.  For field stars and members of
the Corona Australis association, we assumed uncertainty $\sigma[{\rm
log}(age)] = 0.3$. A larger uncertainty for Corona Australis was adopted since
it represents our youngest association where not only is the fractional error
in age likely larger than for the older associations and clusters, but a true
range of stellar ages may indeed be present as perhaps indicated by the mixture
of stars without any disks and with optically thick disks. Our adopted age
uncertainties are {\it ad hoc}, but we believe that they represent reasonable
estimates. In 10,000 simulations, the number of stars younger than 300~Myr with
24\micron\ excesses exceeded the number of such excess sources older than
300~Myr by at least a factor of two in 99\% of the trials. Even if the age
uncertainty for all stars was as large as $\sigma[{\rm log}(age)] = 0.5$, the
excess fraction is larger than older stars in 99.6\% of the trials. We
conclude that the temporal decline in the fraction of stars with
24\micron\ excesses is robust to plausible, random age uncertainties.

The sample of stars younger than 300~Myr contain a mixture of stars in clusters
(20\%), associations (28\%), and the field (52\%), while 80\% of the sample
older than 300~Myr are field stars. Clusters are typically considered to have 
stable dynamical times longer than 100~Myr while associations disperse into 
the field on timescales less than this \citep[e.g.][]{Lada03}.
Environment, either in the form of high initial stellar
density in clusters or the high radiation field from any OB stars in clusters
or associations, could impact the lifetime of primordial disks and ultimately
the formation of debris systems. We examine then whether or not the declining
24\micron\ excess fraction with age is a result of differences in the excess
properties in cluster, association and field-star populations\footnote{Many of
the field stars could have formed in a cluster or association that has since
dispersed. However, we have no means to distinguish dispersed cluster or
association stars from sources that actually formed in relative isolation.}.

For stars younger than 300~Myr, the fraction of stars that have a 24\micron\
excess $\ge 10.2\%$ above the photosphere is 7/38 (18\%), 11/56 (20\%),
and 12/103 (12\%) for clusters, associations, and field stars, respectively.
Given the similar percentages for clusters and associations, we combine those
samples into one. While the excess fraction for clusters/associations is nearly
twice as high as for field stars, the probability that the field star excess
fraction is drawn from the same parent population as cluster/associations is
17\% based on the Fisher's exact test. Therefore we are unable to determine
definitively if a significant difference exists in the excess fraction between
the two populations. We also considered the evolution of cluster/association
stars and field stars separately. The probability that the excess fraction of
cluster/association members younger than 300~Myr is the same as for older
cluster stars (0/22) is 2\%. Similarly, there is a 6\% chance that the excesses
fraction in young and old field stars was drawn from the same excess parent
population. The fact that the decline in the excess fraction is suggestive for
both clusters/associations and field stars, albeit at weak confidence for the
individual samples, suggests that the overall decline in the excess fraction
with age results from temporal evolution and not a change in the mix of
clusters/association and field star populations.

Our results can be compared with other \Spitzer\ surveys of debris disks.
Since each survey adopts different detection thresholds and sample selection,
this comparison is qualitative in nature. \citet{Siegler07} compiled various
\Spitzer\ 24\micron\ surveys of FGK stars and found a decline in the 
24\micron\ excess fraction with age similar to that observed in the FEPS 
sample. However, they inferred a higher excess fraction (36\% on average) in 
clusters younger than 50~Myr compared to the FEPS sample despite having a 
higher detection threshold for a 24\micron\ excess (15\%). The frequency
of debris disks found around A-stars declines on similar time scales as
solar-type stars, although the debris disk frequency is higher \citep{Rieke05}.
One should keep in mind that debris disks identified towards A-type stars as 
a relative fraction of the stellar photospheric emission possess greater dust 
masses than similarly selected debris disks around G~stars. Further, the dust 
detected at a given temperature generating 24\micron\ excess is located at a 
greater orbital radius around A~stars compared to G~stars.
\citet{Currie08a} suggest a more complicated evolutionary picture where the 
magnitude of the 24\micron\ excess for A- and F-type stars increases from 
5 to 10~Myr, peaks at ages of 10-15~Myr, and then declines toward older ages. 
Based on the Kendall's rank correlation statistic and visual inspection of 
Figure~\ref{fig:mips24_age_fbol}, we do not find a similar trend for 
solar-type stars.

\subsection{70\micron\ excesses}
\label{evolution:excess70}

\begin{figure}
\begin{center}
\includegraphics[angle=-90,scale=0.9]{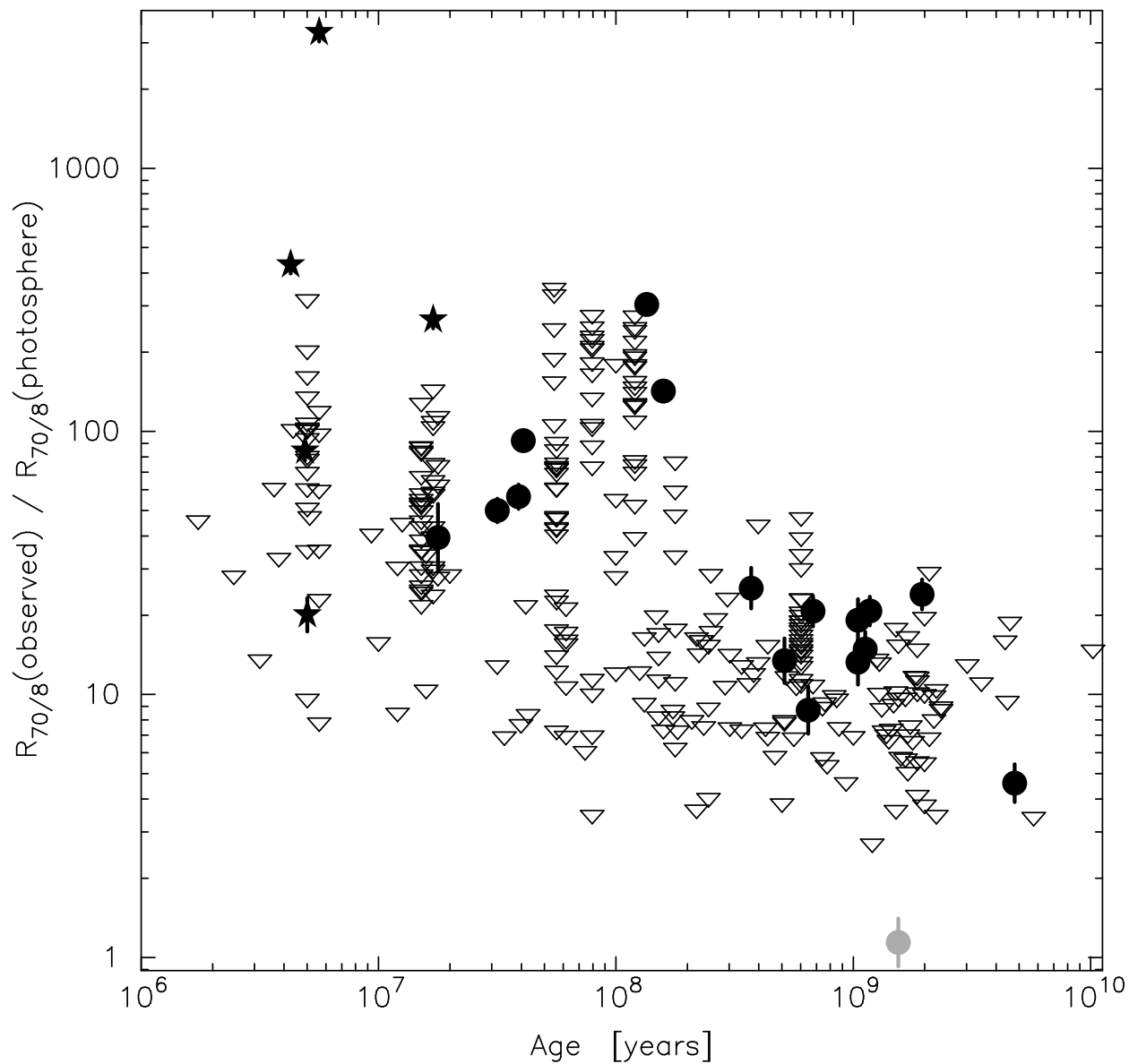}
\caption{
  \label{fig:mips70_age_fbol}
  Ratio of the observed MIPS~70\micron\ flux density to the 70\micron\ 
  photospheric value as a function of the stellar age. Filled symbols 
  represent sources that have a $\ge 3\sigma$ 70\micron\ detection, with 
  stars representing optically thick disks, black circles indicating
  optically thin debris disks, and the gray circle marking the source with 
  a 70\micron\ detection consistent with photospheric emission. Open triangles 
  represent the 3$\sigma$ upper limits for sources not detected at 70\micron.
}
\end{center}
\end{figure}

We present in Figure~\ref{fig:mips70_age_fbol} the ratio of the
observed-to-photospheric 70\micron\ flux density as a function of stellar age.
For sources that were not detected at 70\micron, upper limits to the observed
flux density were computed as 3$\sigma$ if $S_\nu(70\micron) \le 0$, or as
$S_\nu(70\micron) + 3\sigma$ if $S_\nu(70\micron) > 0$. The upper limits
include both internal and calibration uncertainties. The upper limits to the
70\micron\ emission are typically 3-30 times the photosphere for stars older
than \about 300~Myr, and 10-200 times for the younger, more distant stars. The
detected 70\micron\ sources have excesses up to 300$\times$ and 3000$\times$
the photosphere for the debris and primordial disks, respectively. 

The maximum fractional 70\micron\ excess occurs between ages of \about 30 and
200~Myr. Of the 105 stars in this age range, at least 5 sources have a
70\micron\ fractional excess more than 50 times the photosphere. By contrast,
of the 126 stars older than 200~Myr, none have this large of an excess and the
upper limits to the 70\micron\ fractional excess are all below 50. The
probability that the luminous excess sources in the two ages samples are drawn
from the same parent population is 1.8\% according to the Fisher's exact test. 
However, the significance of this comparison is sensitive to the age range 
chosen. If we compare all stars younger than 300~Myr to the older stars, the 
probability that the luminous excess are down from the same parent population
is 8\%. We also used the Kendall's rank correlation statistic to evaluate if 
the 70\micron\ excess depends on stellar age. Considering both detections and
upper limits to the 70\micron\ excess, the probability of a trend of 70\micron\
excess emission with age is 46\%. 

The results presented in Figure~\ref{fig:mips70_age_fbol} possibly suggests
that relatively few luminous 70\micron\ excess sources indicative of {\it
debris} dust are found among the youngest sources. Considering only sources
younger than 200~Myr, the Kendall's rank correlation statistic indicates 
a 84\% probability that a correlation is present the 70\micron\ excess with
age. We therefore do not find any evidence for evolution in the debris 
luminosity between 3 and 200~Myr.

In summary, we find weak evidence for a decline in the magnitude of the
peak 70\micron\ excess from intermediate-age stars (30-200~Myr) to older
ages. We do not find any systematic temporal change in the overall 70\micron\
excess, or if the amount of 70\micron\ emission evolves between stellar ages of
3 and 200~Myr. Observations of additional stars in this age range combined
with more sensitive flux density limits are needed to make more definitive
conclusions. Compared to debris disks around A-star, \citet{Su06} found that
the decay time of the 70\micron\ emission is \about 400~Myr. Qualitatively that
is consistent with the FEPS data in that the upper envelope of 70\micron\
emission appears to decline between 100 and 300~Myr.

\subsection{Dust Temperature}
\label{evolution:tdust}

Assuming the dust grains are in thermal equilibrium with the stellar radiation
field, the dust temperature is a first order indicator of the orbital radius.
For grains that are efficient absorbers of stellar radiation, the orbital
radius, $R$, is proportional to $R \propto L_*^{0.5}\ T_{\rm
dust}^{-{4+\beta\over2}}$, where $L_*$ is the stellar luminosity and $T_{\rm
dust}$ the dust temperature. In practice, the orbital radius cannot be derived
uniquely from spectral energy distributions without knowledge of the grain
size, composition, and porosity. Nonetheless, variations in the dust
temperature may translate into a range of orbital radii if the grain properties
are similar amongst the debris disks.

\begin{figure}
\begin{center}
\includegraphics[angle=0,scale=0.7]{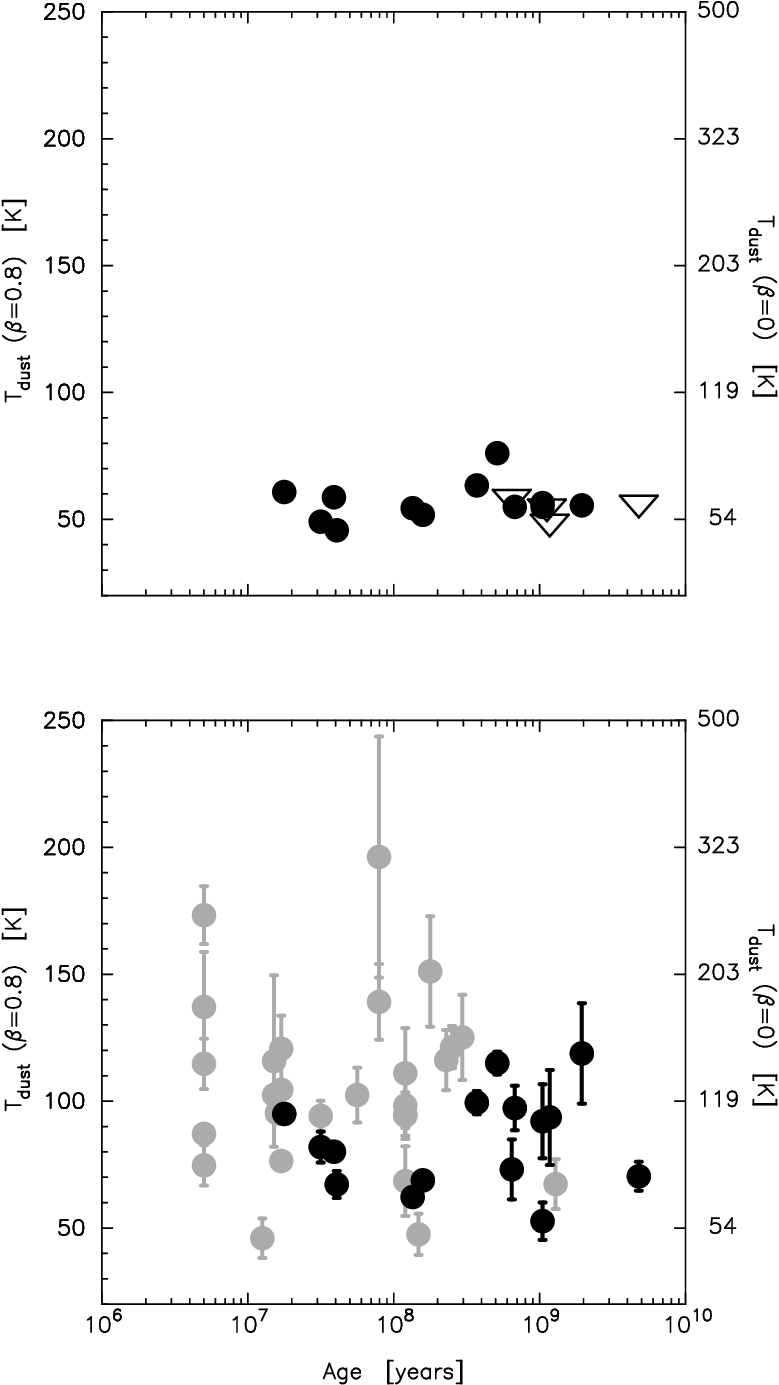}
\caption{
  \label{fig:tdust}
  Dust temperatures as a function of stellar age for debris disks in the 
  FEPS sample. The top panel shows temperatures derived from the MIPS 24 and
  70\micron\ photometry for 16 stars with 70\micron\ excesses. The bottom 
  panel shows temperatures for inferred from IRS spectra for 43 debris disks 
  with an IRS excess and either a MIPS 24 or 70\micron\ excess.
  The IRS dust temperatures were derived assuming the excess can be 
  approximated by a modified blackbody with $\beta=0.8$.
  Equivalent blackbody temperatures are shown on the right axis. Black
  and gray circles indicate sources with and without a detected 70\micron\ 
  excess, respectively. 
}
\end{center}
\end{figure}

Dust temperatures derived by fitting a Kurucz model plus a modified blackbody
($\beta=0.8$) to the IRS spectra between 12 and 35\micron\ (see
\S\ref{excess:irs}) are plotted versus stellar age in the bottom panel in
Figure~\ref{fig:tdust} for debris disks with an IRS~excess and either a MIPS 24
or 70\micron\ excess. Different colored symbols are shown for sources with
(black circles) and without (gray circles) 70\micron\ excesses. A greater
fraction of the older sources tend to have 70\micron\ excesses, which most
likely reflects that older stars tend to be closer in distance than the younger
objects, and smaller excesses can be detected. The dust temperatures range from
$46\pm7$~K (HD~281691) to $196\pm48$~K (HE~750) with a median of 112~K. The
distribution of dust temperatures overlaps for sources with and without
70\micron\ detections, although the median dust temperature is higher for
sources without 70\micron\ detections (102~K vs. 81~K). The top panel in
Figure~\ref{fig:tdust} shows the dust temperature derived from the MIPS 24 and
70\micron\ photometry. The temperatures derived from the MIPS photometry are
lower than that inferred from the IRS spectra, suggesting the presence of
cooler dust (see \S\ref{evolution:ldust}). No strong trend between dust
temperature and stellar age is evident.

The observed scatter in the ratio $L_*^{0.5}\ T_{\rm dust}^{-2}$, which is
proportional to the orbital radius assuming isothermal dust radiating as a
blackbody, is three times larger than the expected scatter if the orbital
radius was the same for all sources. These results suggest variations are
present in the orbital location of dust. Variations in dust properties may 
also contribute to the scatter, but we consider here only the range of orbital 
radii implied by the observations. The minimum orbital radius of the dust 
grains can be estimated assuming the grains radiate like blackbodies. The 
implied orbital radius is $R =
(L_* / L_\odot) (278~{\rm K} / T_{\rm dust})^2~{\rm AU}$. The range of
blackbody temperatures derived from the IRS spectra, 
$50\pm9$~K to $283\pm127$~K with a median of 113~K,
corresponds to orbital radii of 31~AU and 1~AU with a median of 6~AU.
(We excluded 1RXS~J051111.1$+$281353 since the IRS excess is poorly fitted with
 a modified blackbody.) If most of the surface area is from smaller grains
which do not emit as blackbodies, the corresponding radii will be larger. 

\subsection{Fractional Dust Luminosity}
\label{evolution:ldust}

The amount of dust emission radiated in the 24\micron\ and 70\micron\ 
bandpasses depends on both the surface area and temperature of the dust 
grains. The fractional dust luminosity 
($f_{\rm dust} = L_{\rm dust} / L_{\rm *}$), or equivalently the
fractional dust bolometric flux ($f_{\rm dust} = F_{\rm dust} / F_{\rm *}$),
accounts for variations in dust cross-sectional surface area and temperatures 
to reflect the total amount of stellar emission absorbed and 
re-radiated by dust grains. 

The luminosity of ``warm'' dust emission between 12 and 35\micron\ is
constrained by the IRS spectra. The modified-blackbody fits provide estimates
of both the dust temperature and surface area of dust grains, and extrapolation
of these fits yields the bolometric flux of warm dust ($F_{\rm warm}$) over
all wavelengths. For sources without a detectable IRS excess, upper limits to 
$F_{\rm warm}$ were computed by first integrating the excess emission in the
IRS spectrum after subtracting the model photosphere and computing the formal
uncertainty in the integrated flux. The 3$\sigma$ upper limits were then
computed in a manner similar to that for the 70\micron\ upper limits as described in
\S\ref{evolution:excess70}. The warm dust bolometric flux, 
$F_{\rm warm}$ was normalized by the photospheric bolometric flux ($F_*$)
estimated as 
\begin{equation} 
    F_{*} = {\sigma_{\rm SB}\ T_{\rm eff}^4\ 
    S_\nu(3.6\micron) \over \pi B_\nu(3.6\micron, T_{\rm eff})}, 
\end{equation} 
where $\sigma_{\rm SB}$ is the Stefan-Boltzmann constant, $T_{\rm eff}$ is the
photospheric temperature estimated from the observed stellar colors or spectral
type (see Paper~I), $S_\nu(3.6\micron)$ is the observed IRAC 3.6\micron\ flux
density, and $B_\nu(\lambda,T_{\rm eff})$ is the Planck function.

The ``cool" dust luminosity radiated at wavelengths longer than 35\micron\
is not as well determined since we have a single observation at 70\micron\ and 
most of our measurements are upper limits. Moreover, \citet{Hillenbrand08} 
found that the color temperature inferred from the observed 33 to 24\micron\ 
flux densities is often higher than that derived from the 70 to 
33\micron\ flux density ratio, although the significance of the temperature 
difference is marginal for any single star. The modified blackbody fits to the 
IRS spectra provide a more accurate assessment of the dust temperature since 
the shape of the entire emission spectrum is used to estimate the dust 
temperature and luminosity. We find that the 70\micron\ flux density from
the modified blackbody fits underestimates the observed 
70\micron\ excess flux density by 3-11$\sigma$ for 11 of the 16 debris disks
with a 70\micron\ detection
even if $\beta=0$ is assumed to maximize the predicted 70\micron\ emission. 
For the 5 remaining debris disks, the observed flux density exceeds the 
projected value, but by less than 3$\sigma$.
These results suggests that an additional cool dust component is present in 
many sources which contributes significantly to the 70\micron\ emission but 
not at IRS wavelengths. The dust luminosities estimated from the IRS 
spectra then underestimate the total luminosity.

Given that the distribution of dust temperatures is poorly constrained by the 
70\micron\ observations, we adopt a two component model to estimate the 
bolometric dust luminosity. The predicted 70\micron\ flux density from the 
warm component was subtracted from the observed 70\micron\ flux density to 
yield the 70\micron\ emission from cooler dust ($S_\nu^{\rm cool}$). The 
bolometric flux of cool dust was estimated assuming a single dust temperature,
$T_{\rm cool}$, as
\begin{equation}
    F_{\rm cool} = {S_\nu^{\rm cool}(\nu_{70}) 
       \int\nu^\beta\ B_\nu(T_{\rm cool}, \nu)\ d\nu
          \over 
       \nu_{70}^\beta\ B_\nu(T_{\rm cool}, \nu_{70})},
\end{equation}
where $\nu_{70}$ is the frequency corresponding to the mean wavelength of the 
MIPS~70\micron\ bandpass. We adopt $T_{\rm cool} = 60$~K to conform to the 
temperature frequently inferred from mid-infrared observations of debris disks 
around solar-type stars \citep{Zuckerman04,Hillenbrand08}. 
Upper limits to the cool luminosity were computed using $T_{\rm cool}$ and 
the upper limit to $F_{\rm cool}$. The upper limits do not include 
the uncertainty in $T_{\rm cool}$ nor the arbitrary amount of luminosity that 
could be generated by adding even cooler dust that does not radiate 
prominently at 70\micron.

\begin{figure}
\begin{center}
\includegraphics[angle=0,scale=0.5]{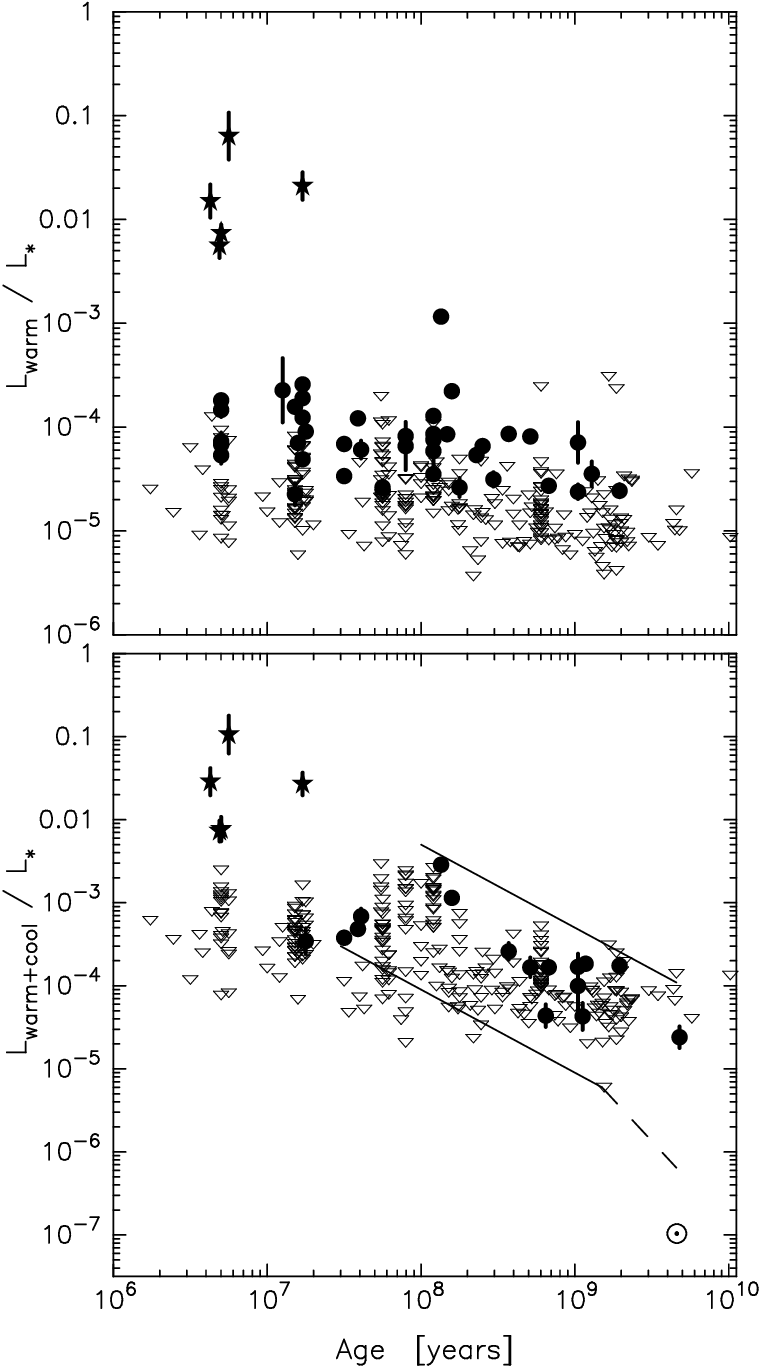}
\caption{
  \label{fig:fbol_age}
  Ratio of the infrared-excess to stellar luminosity for ``warm'' dust emission
  (upper panel) and the ``warm'' plus ``cool'' dust emission (bottom panel). 
  The warm dust luminosity was computed from modified blackbody fits to the
  observed IRS spectra, with upper limits computed assuming a dust
  temperature of 100~K. The warm plus cool dust luminosity was estimated from 
  the IRS
  spectra and the 70\micron\ flux assuming a cool dust temperature of 60~K
  (see text).
  Filled symbols represent sources that have a $\ge 3\sigma$ 70\micron\
  photometric detection above the stellar photosphere, with stars and circles
  representing optically thick and thin disks, respectively. Open triangles 
  represent the 3$\sigma$ upper limits for sources not detected at 70\micron.
  The solid lines in the bottom panel show $t^{-1}$ evolutionary curves 
  appropriate for debris systems dominated by collisions; the curves are
  normalized to the brightest and faintest detected debris disks and are not
  a fit to the data. The dashed
  curve shows the expected evolution when the system becomes dominated by
  Poynting-Robertson drag (see text). The solar symbol indicates the luminosity
  of the Kuiper Belt \citep{Backman95}.
} 
\end{center}
\end{figure}

The fractional dust luminosity as a function of stellar age is presented in
Figure~\ref{fig:fbol_age} for the warm dust, and the sum of the warm and cool
dust components. The trends apparent in this figure are similar to those
concluded based on the MIPS~24 and 70\micron\ photometry considered alone. The
fractional warm luminosity from the young (optically thick) disks is typically
an order of magnitude greater than that from older (debris) disks, although the
fractional luminosity from the strongest debris disk (HD~61005) is lower by
only a factor of \about 5. For sources with 70\micron\ detections, the debris
disk sources with the highest fractional dust luminosity are at ages younger
than 200~Myr with a decline in the peak fractional luminosity toward old ages.
The fractional luminosities of the detected debris disks in the FEPS sample
are similar to the solar-type stars of comparable age observed by 
\citet{Trilling08}. A-type stars show a similar decline in the fractional
luminosity on timescales of 100-300~Myr \citep{Su06} as observed in 
the FEPS solar-star sample.

\section{Debris Disk Models}
\label{models}

The dynamics of dust grains in optically thin systems are dominated by
radiative and collisional processes when the effects of gas drag are negligible
\citep{Takeuchi01}. \citet{Pascucci06} used \Spitzer\ high resolution 
spectroscopy to place an upper limit of 0.04~M$_{\rm Jupiter}$ to the gas mass 
in the inner disk for 15 debris systems within the FEPS program having ages 
between 5 and 400~Myr. While these limits are still too high to establish
whether gas drag is a marginal process in the evolution of dust grains
(gas--to--dust ratio $<$ 0.1), we assume here that is in fact the case. 

For debris disks detectable with current instrumentation, \citet{Dominik03} 
demonstrated that collisional processes combined with radiation pressure 
dominate over Poynting-Robertson drag in removing dust grains from the system
\citep[see also][]{Wyatt05}. \citet{Hillenbrand08} confirmed that collisions 
are likely the dominant process for the disks detected at 70\micron\ in
the FEPS sample. Stellar mass loss also produces a drag on the orbiting dust in
a manner analogous to the Poynting-Robertson effect, and for the
Sun, the current rate of dust mass loss from corpuscular drag is about 
0.3 that of radiation drag \citep{Gustafson94}. At younger ages 
the stellar mass loss rates are likely higher, and wind drag 
may be more important than Poynting-Robertson drag and even collisions 
for young debris disks \citep{Jura04,Chen05}.  However, given 
the uncertainties on the scaling of mass loss rates with stellar age 
\citep{Wood02, Wood05}, we do not consider wind drag effects.

Assuming that collisional processes are the dominant forces influencing dust
dynamics, the location of the dust grains should trace the spatial distribution
of the planetesimals. Therefore, we adopt a model in which the planetesimals
are co-spatial with the debris dust. The following section describes the model
calculations, and in subsequent sections we apply this model to infer the
properties and evolution of the planetesimals.

\subsection{Model Description}

The model planetesimal belt extends between an inner orbital radius 
$R_{\rm in}$ and an outer radius $R_{\rm out}$, and contains particles with 
radii between
$a_{\rm min}$ and $a_{\rm max}$. The particle size distribution as a function 
of orbital radius ($R$) and particle radius ($a$) is 
\begin{equation}
   \label{eq:nar}
   N(a,R) = K a^{-3.5} R^{\alpha}
\end{equation}
such that $N(a,R)\ da\ dR$ is the number of particles between $a$ and $a+da$, 
and between $R$ and $R+dR$. Since the slope of the particle size distribution
is the same over all orbital radii in this model, the particle surface area 
and mass surface density will also scale as $R^{\alpha}$. The particle size 
distribution of $a^{-3.5}$ is appropriate for a collisional cascade without 
boundary conditions on the minimum and maximum particle size 
\citep{Dohnanyi69}. 
In practice, radiation pressure imposes limits on the minimum particle size, 
and numerical simulations predict a ``wavy'' pattern of particle sizes since 
the smallest grains are not available to break up particles near the cutoff 
limit \citep{Campo94,Thebault07}. For simplicity, however, we consider only a 
power-law distribution of particle sizes. With this adopted size distribution,
most of the grain surface area is contained in the smallest particles such
that $a_{\rm min}$ is the critical parameter, while the larger particles
contain most of the disk mass (M$_{\rm disk} \propto \sqrt{a_{\rm max}}$). 

The flux density from the disk assuming optically thin emission is 
\begin{equation}
   \label{eq:fnu}
   S_\nu = \int_{R_{\rm in}}^{R_{\rm out}} 
           \int_{a_{\rm min}}^{a_{\rm max}} 
           B_\nu[T_{\rm d}(a,R)]\ Q_{\rm abs}(\nu,a)\ 
       N(a,R)\ {\pi a^2\over d^2}\ 2\pi R\ da\ dR,
\end{equation}
where $T_{\rm d}(a,R)$ is the dust temperature as a function of particle size
and orbital radius, $Q_{\rm abs}(\nu,a)$ is the absorption coefficient,
and $d$ is the distance to the star. For particle sizes smaller than 3~mm in 
radius, we computed $Q_{\rm abs}(\nu,a)$ using the procedure described in 
\citet{Wolf03} and the optical constants for ``astronomical silicates'' 
\citep{Weingartner01}.
For larger particles, we adopted $Q_{\rm abs}(\nu)$ = 1, 
which is valid for the wavelength range considered here 
($\lambda \le 70$\micron). Dust temperatures were computed assuming the disk is
optically thin and in thermal equilibrium with the stellar radiation field.

\begin{figure}
\begin{center}
\includegraphics[angle=-90,scale=0.75]{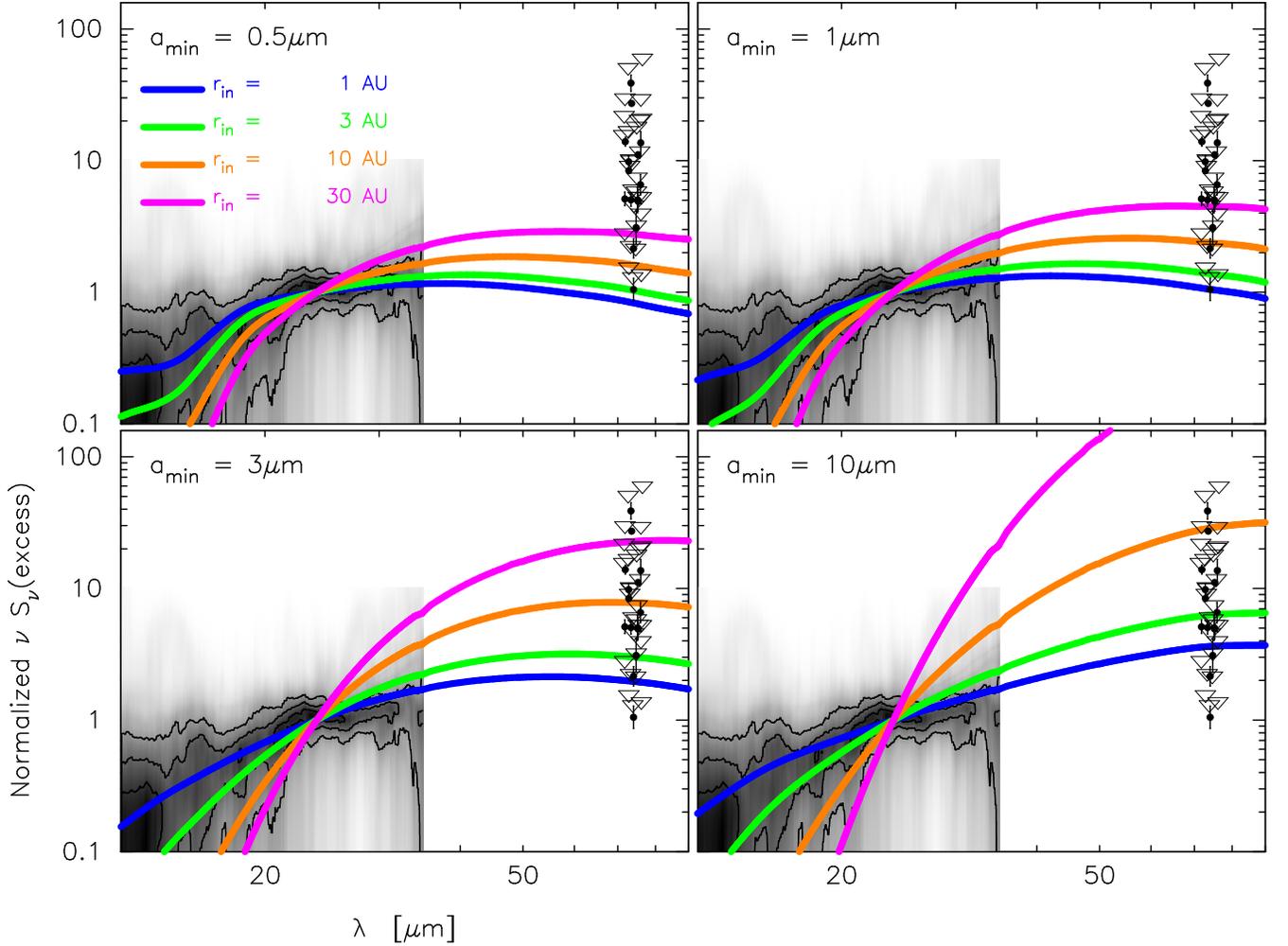}
\caption{
  \label{fig:models_belts}
  Comparison of dust disk models with the observed spectra for 45 debris disks
  with 70\micron\ excess or both 24\micron\ and IRS excess. The contours and 
  gray scale represent the average IRS emission excess above the photosphere 
  normalized to the 24\micron\ excess and converted to a density plot. 
  Contours are at 0.4, 0.6, and 0.8 of the peak density and represent the typical
  shape of the excess emission. MIPS~70\micron\ detections (filled circles) and upper limits
  (triangles) are indicated discretely. Solid curves represent model dust emission from 
  silicate grains for a constant-surface density disk ($\alpha=0$) with an outer radius 
  $R_{\rm out}$ = 50~AU
  surrounding a solar-type star with $M_*= 1$\msun, $L_*=1$\lsun, and $T_*=5780$~K. 
  The grain radii follow a power-law distribution 
  ($N(a) \propto a^{-3.5}$) with $a_{\rm max}=1$~km. Model calculations
  are shown for different minimum grain radii $a_{\rm min}$  = 0.5, 1, 3, and 
  10\micron\ in each panel. Within a panel, the four curves indicate inner disk 
  radii $R_{\rm in}$ = 1, 3, 10, and 30~AU. 
  A small wavelength offset has been added to the 70 photometry for clarity.
}
\end{center}
\end{figure}

In Figure~\ref{fig:models_belts}, we present calculations that illustrate
how various model parameters affect the resulting spectra. 
The model disk surrounds a solar type star and contains minimum grain sizes 
that span the range from the approximate radiation blow-out size 
\citep{Burns79} to the smallest
grain size that radiates like a blackbody at 70\micron. The inner disk radii
range from the terrestrial planet zone in our Solar System to the Kuiper Belt.
The models are compared to sources with IRS excesses that are confirmed by
MIPS~24\micron\ or 70\micron\ photometry.

The main features of the data that need to be explained are the shape of the
IRS spectra, and the intensity of the MIPS~70\micron\ photometry relative to
the IRS excess. The density contours allows one to assess the median
behavior and the deviation around a ``typical" debris disk.  Models with small
grains (\aboutless 1\micron) and small inner disk radii (\aboutless 10~AU) tend
to reproduce the general shape of the IRS spectra, but underestimate the
70\micron\ flux densities for those stars detected at this longer wavelength
due to the falloff in small grain radiative efficiency. If the minimum grain
size is \aboutmore 3\micron, the models can explain the level of 70\micron\
emission, but then cannot produce the tail of warm excess emission observed 
between 20 and 30\micron. When 70\micron\ excess emission is not detected, 
then a broad combination of model parameters can reproduce the observations.
From these illustrative models of debris dust located over a range of orbital 
distances 
from a central star and having a range of grain sizes, we conclude that some of
the features typical of the observations can be explained, but that a more
systematic parameter study is needed. In the following subsections we first
examine which of the disk parameters can be constrained robustly, and then
study the time evolution of the plausible debris belts.

\subsection{Constraints on the Inner Disk Radius}
\label{models:belts}

To place constraints on the disk and grain properties, we used a Bayesian
approach to infer the likelihood of model parameters given the observational
data \citep[see, e.g.,][]{Lay97} for the debris disks that have reliable
infrared excesses (see \S\ref{excess_sum}). 
We constructed a grid of models spanning the
three dimensional parameter space of inner disk radius,  disk width ($\Delta R
= R_{\rm out} - R_{\rm in}$), and the power-law slope of the mass surface 
density. Plausible ranges for each of the parameters were considered. The
inner disk radius was varied between 0.1 and 1000~AU to encompass the size
scales inferred for protoplanetary disks. The disk width was varied between 0.1
and 1000~AU to allow for narrow rings and wide belts. Models were computed for
surface density power-laws (see Eq.~\ref{eq:nar}) between $-$1.5 to 0 in steps
of 0.5, encompassing the mass surface density of the current Solar System
\citep[$\alpha=-1.5$;][]{Weidenschilling77}, of that commonly inferred for 
protoplanetary disks \citep[$\alpha=-1$;][]{Beckwith90, Kitamura02, Andrews05},
and a constant surface density ($\alpha = 0$).

Spectra over the Spitzer wavelength range were computed for each combination 
of the above parameters assuming an optically thin disk with particle size
distribution extending from the radiation blowout size to a 1000~km radius.
The spectral types for stars with debris disks span between K3 and F5,
or approximately an order of magnitude in stellar luminosity. The 
corresponding radiation blowout size varies between 0.2-0.7\micron\
\citep{Burns79}. Nonetheless, we did not compute a grid of models for each
source, but adopted stellar parameters for a canonical solar type star and
fixed the radiation blowout size at $a = 0.5$\micron. The maximum particle
radius of 1000~km corresponds to the size of the bodies needed to excite the
collisional cascade in the debris disk models by \citet{Kenyon04}. The maximum
particle size has negligible influence on the model calculations for the
observed emission since large bodies contain a relatively small surface area
for the adopted particle size distribution.

The model at a given grid point was fitted to the observations with a
normalization constant as a free parameter that is proportional to the total 
particle cross-sectional area for optically thin emission. The photospheric
contribution was determined using the fits described in \S\ref{excess}.
The relative probability of the model, given the data, is proportional to
$e^{-\chi^2/2}$, where $\chi^2$ is the sum of the squares of the difference
between the model and the data normalized by the observational uncertainties. 
The probability distribution of a single model parameter is determined by
summing the probabilities over the other two model parameters, and normalizing
the integrated probability to unity. The parameter constraints from the
ensemble of sources is then computed as the sum of the probability
distributions for each source, and re-normalizing the sum to unity.

The probability distributions are only meaningful if the model provides a
reasonable fit to the observations, which can be judged from the minimum
$\chi^2$ values across the grid. For model fits to the IRS spectra between 12
and 35\micron, 35 of the debris disks have probabilities $> 0.01$ that the
$\chi^2$ residuals are consistent with noise. For fits to both the IRS spectra
and the MIPS~70\micron\ photometry (detections and non-detections), 
27 sources have probabilities $> 0.01$. In general, sources with 70\micron\ 
detections have higher $\chi^2$ values that reflect the difficulty in finding 
models that can fit the IRS spectral shape and the 70\micron\ photometric 
excess simultaneously (as illustrated in Figure~\ref{fig:models_belts}).

\begin{figure}
\begin{center}
\includegraphics[angle=-90,scale=0.75]{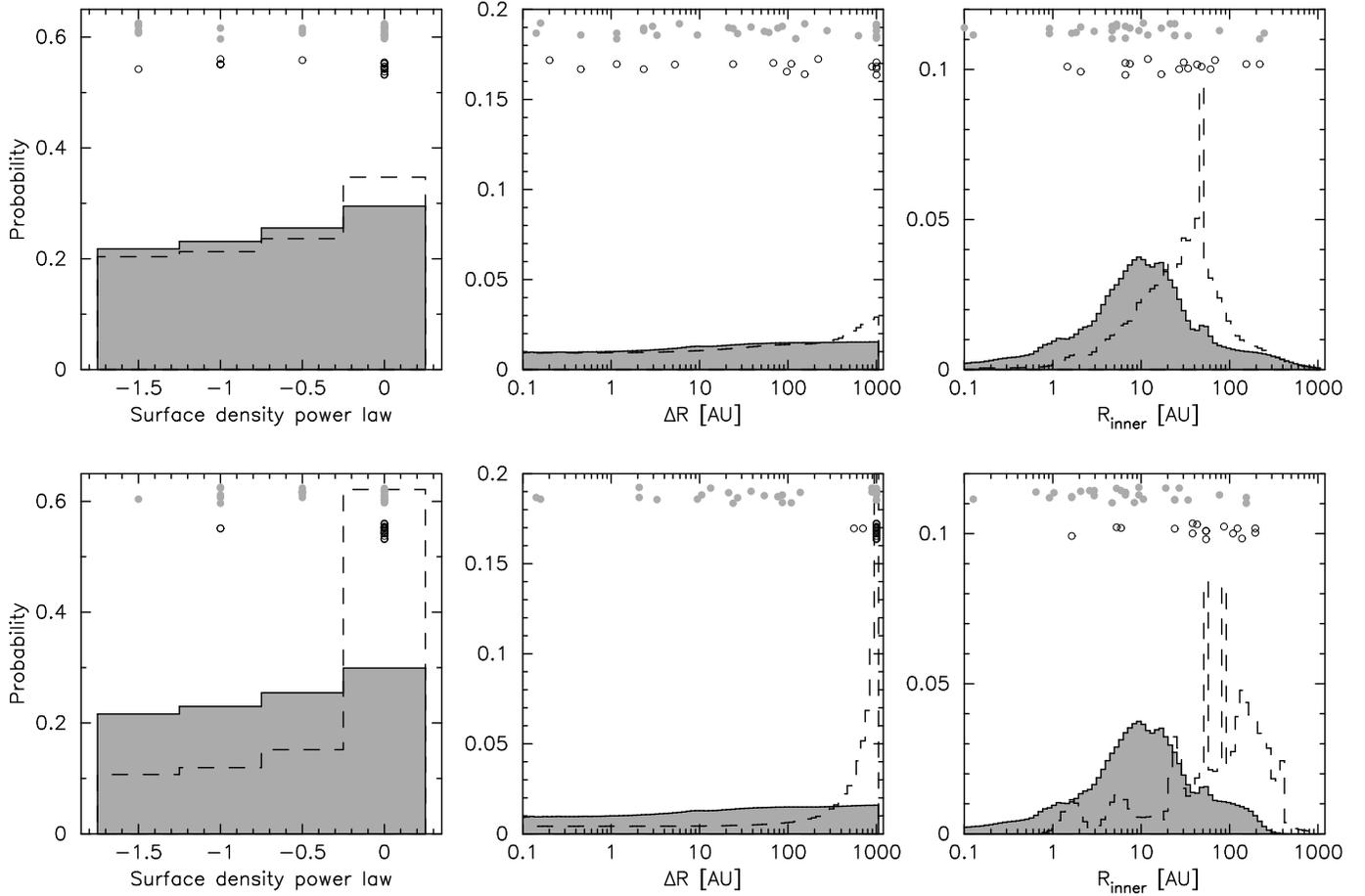}
\caption{
  \label{fig:models_bays}
  Probability distributions for model disk parameters (see text) as 
  constrained by debris disks sources with a IRS and/or MIPS~70\micron\ excess.
  The star 1RXS~J051111.1+281353 is not shown (see Fig.~\ref{fig:sed_1rxs}).
  Top panels show the probability distribution derived from fitting the
  IRS spectra between 12 and 35\micron, and the bottom panels for fitting the
  IRS spectra and the MIPS~70\micron\ photometry (for both detections and
  non-detections).
  Gray histograms represent the probability distribution for sources
  not detected at 70\micron, and dashed histograms for sources with
  70\micron\ detections. The integrated probability for each histogram has 
  been normalized to unity. Parameters for the best fit models are indicated
  by the open and filled circles for sources with and without 70\micron\
  detections, respectively.
}
\end{center}
\end{figure}

Probability distributions for infrared excess sources with and without
MIPS~70\micron\ detections are presented in the top panels of
Figure~\ref{fig:models_bays} for fits to the IRS spectra only, and in the
bottom panels for fits
to both the IRS spectra and 70\micron\ photometry. Best fit model parameters
for each source are represented by circles. We do not find any correlation
of best-fit model parameters with spectral type, suggesting that adopting a
single stellar model did not significantly impact the results.
These histograms indicate the
constraints on the model parameters for the ensemble of observed debris disks.
For two of the parameters, surface density power law and disk width, the
probability distributions are relatively flat.  The one input parameter having
moderate constraint is the inner disk radius.  This is most apparent in the
probability distribution that considers only the IRS spectra (which are most
diagnostic of warm inner disk material). Most debris disks appear to have 
inner radii between 3-40 AU, with somewhat larger inner radii for stars
with 70\micron\ detections. We explore the implications of this result 
starting in \S\ref{models:terrestrial}.

For the sources with 70\micron\ detections, adding the 70\micron\ flux 
densities as a 
constraint in the model fits accentuates many of the trends described above
(see Fig.~\ref{fig:models_bays}). Models with flatter surface density profiles
are generally preferred.  The inner disk radius increases to
\about 100-200~AU. There is a strong tendency for a large disk width;
indeed, since the most probable disk width is also the maximum value 
in the model grid (1000~AU), these model calculations have not established the 
upper 
bound of the outer disk radius. Disks this wide, however, are inconsistent 
with the other observations of debris disks around solar-type stars in that
scattered-light images images and resolved submillimeter images typically find 
radii less than 200~AU \citep{Schneider06,Ardila04,Greaves98}. Furthermore, 
the reduced $\chi^2$ values for the model fits tend to be high for sources with 
70\micron\ detections, indicating as we illustrated earlier that
the model has difficulty fitting both the 
IRS spectral shape and the 70\micron\ photometry. Better fits with smaller
outer disk radii could be obtained by modifying the model assumptions to
decrease the relative number of smaller grains. This can be accomplished, for
example, by adopting for a less steep grain size distribution than $N(a)
\propto a^{-3.5}$, or adopting a larger minimum grain size. Also,
improved model fits could be obtained by allowing for a population of
small grains that are decoupled spatially from the larger grains, as was
found for the Vega debris disk \citep{Su06}. Finally, adopting a different
grain composition (e.g. water ice for the outer disk particles), may modify
the fits as well.

\subsection{Evolution of the Planetesimal Belts}
\label{models:evolution}

The analysis presented in \S\ref{models:belts} was intended to constrain
the properties of the debris systems, which indicate the presence of 
planetesimal belts. The most strongly constrained parameter is the inner belt 
radius, which we find is typically 3-40~AU.  We now extend our
analysis, adopting these parameters, in order to study temporal evolution 
of the debris belt properties.  The main observational characteristics 
of the observed debris disks related to temporal evolution 
(Figures~\ref{fig:mips24_age_fbol}, ~\ref{fig:mips24_age_fraction},
~\ref{fig:mips70_age_fbol}, and \ref{fig:fbol_age}) are:
(1) the lack of debris emission at wavelengths shorter than 16\micron\
    for ages older than 3~Myr,
(2) a flat distribution to the upper-envelope of 24\micron\ excesses for
    age \aboutless 300~Myr with a decline toward older ages,
and
(3) the apparent decline in the maximum 70\micron\ excess emission for stars 
    older than \about 300~Myr. 
In this section, we explore if temporal collisional models can explain these 
aspects of the observational data.

We adopt the analytic model developed by \citet{Dominik03} and extended
by \citet{Wyatt07a} to describe the collisional evolution of a planetesimal 
belt in quasi-steady state equilibrium. The model posits that small grains 
are continually produced by collisional grinding of massive bodies, and mass 
is removed by radiation blowout of the smallest grains. For a system in
collisional equilibrium, the mass surface density of solid particles 
in a narrow annulus of a planetesimal belt ($\Delta R \ll R$) as a function 
of time is 
\begin{equation}
    \label{eq:nt}
    \Sigma(t) = {\Sigma_{\rm o} \over 1 + t/t_{\rm co}},
\end{equation}
where $\Sigma_{\rm o}$ is the initial mass surface density of particles and 
$t_{\rm co}$ is the 
collisional timescale for the largest planetesimals when the cascade begins
at $t=0$. A similar expression is valid for the number of particles and the 
cross-sectional surface area. In the calculations described below, we use the 
formula for $t_{\rm co}$ presented in Equation~13 in \citet{Wyatt07a}. 
However, it 
is instructive to view an approximate formula for $t_{\rm co}$ to understand 
how the collisional evolution depends on 
disk parameters. We make three assumptions: 
{\it i)} a particle size distribution of $N(a) \propto a^{-3.5}$ 
         \citep{Dohnanyi69},
{\it ii)} the orbital eccentricity ($\epsilon$) is approximately equal to 
          the orbital inclination, 
and 
{\it iii)} the size of the planetesimals that can destroy the largest
    planetesimals is much smaller than $a_{\rm max}$.
The collisional time at $t=0$ for a narrow planetesimal belt at orbital 
radius $R$ is then
\begin{equation}
\scriptsize
    \label{eq:tc}
    t_{\rm co}(R) \approx 0.09~{\rm Myr }\ 
             {\rho \over 2.7\ \rm{g\ cm^{-3}}}\ 
             {a_{\rm max} \over 1000\ \rm{km}}\ 
             \Bigl({R \over 1\ \rm{AU}}\Bigr)^{7/3}
             \Bigl({Q_{\rm D} \over 2\times10^6\ \rm{ergs\ g^{-1}}}\Bigr)^{5/6}
             \Bigl({\Sigma_{\rm o} \over 1\ \rm{g\ cm^{-2}}}\Bigr)^{-1}
             \Bigl({\epsilon \over 0.067}\Bigr)^{-5/3},
\end{equation}
where $\rho$ is the particle density and $Q_{\rm D}$ the specific incident
energy required to destroy a planetesimal \citep{Wyatt07a}. In this model,
the collisional timescale is inversely proportional to the surface density.
For $t \ll t_{\rm co}$, the mass surface density of the disk at
radius $R$ is constant in time since the largest particles in the system have
undergone few collisions. The spread in debris disk luminosities at young ages
then depends on the cross-sectional surface area at $t=0$ and the 
distance of the dust from the star. However,
for $t \gg t_{\rm co}$, the mass surface density,
and hence the emission from the disk, is independent of the initial disk 
surface density \citep[see][]{Dominik03,Wyatt07a}. 

To compute the emission from a debris disk as a function of time, the disk was
divided into annuli of width $\Delta R = 2 \epsilon R$ that corresponds to the
range of orbital radii for a particle. As an approximation to the different
evolution timescales at different radii in the disk, Equations~\ref{eq:nt} and
\ref{eq:tc} were applied to each annulus separately. The emission from
individual annuli was computed using Equation~\ref{eq:fnu} and then summed to
get to the total disk emission. The model assumes that the collisional cascade
begins at a stellar age of 0~Myr. We note that primordial disks survive for
1--10 Myr \citep[see, e.g.,][]{Haisch01}, and the collisional cascade in the
planetesimal belt will be delayed relative to the stellar age by at least the
minimum time to form the planetesimals.

\subsubsection{Gas-Giant and Kuiper Belt Zones}
\label{models:gasgiant}

We first attempt to explain the excess characteristics at 8, 16, and 24\micron.
The 70\micron\ data are not used since the disk properties needed to produce 
this emission are poorly constrained by the data (see \S\ref{models:belts} and 
Fig.~\ref{fig:models_bays}). The model planetesimal belt
has a fiducial inner radius of 10~AU in accordance with the disk properties inferred in \S\ref{models:belts}.  In the following section we relax this
assumption and explore the signatures of planetesimals in the inner 10~AU. 
For the other disk parameters we explore a range of properties. The disk 
surface density at $t=0$ is parameterized as a constant surface density disk
of $\Sigma(R) = \Sigma_{\rm o}$; models were computed for 
$\Sigma_{\rm o}$ ranging from 0.03 to 3~g~cm$^{-2}$. 
We consider disks with an outer radius 
of 15 and 100~AU to emulate a narrow ring and a wide belt, respectively.
Extending the disk beyond 100~AU will change the 24\micron\ flux density by
less than 10\% 
since the dust at these radii is too cool 
to radiate prominently at $\lambda < 24$\micron. 

\begin{figure}
\begin{center}
\includegraphics[angle=-90,scale=0.75]{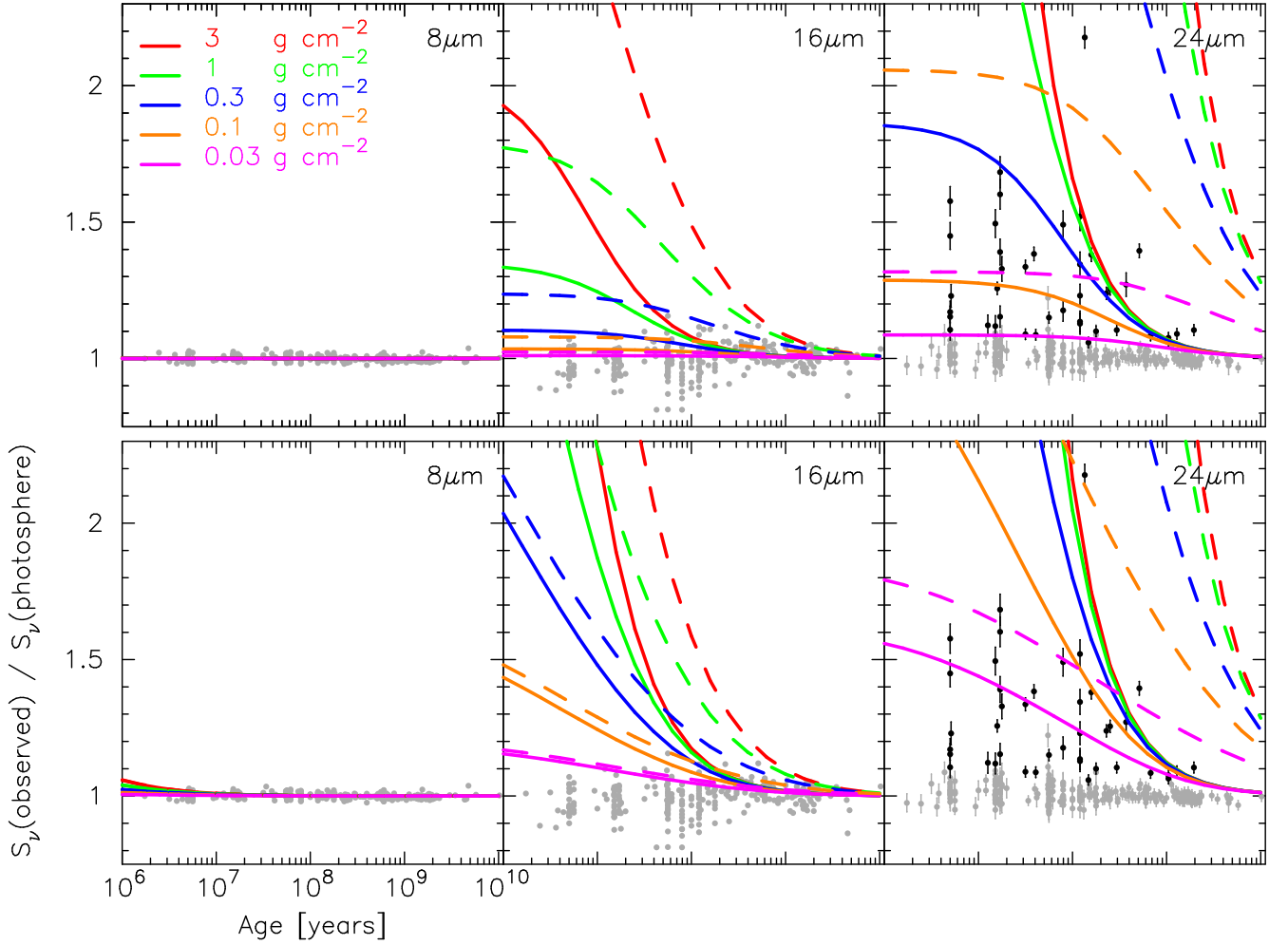}
\caption{
  \label{fig:models_time}
  Ratio of the observed-to-photospheric flux density at 8, 16, and 24\micron\
  as a function of age for a model disk having uniform surface density and
  an inner radius of 10~AU (top panels) or 0.5~AU (bottom panels) surrounding 
  a solar-luminosity star. The 
  emission evolves as a function of time based on collision depletion of the 
  disk (see text). 
  The curves represent various assumed values of the initial surface density
  from 0.03 to 3~g~cm$^{-2}$. Solid and dashed curves are models with an
  outer disk radius of 15~AU and 100~AU, respectively. Filled circles 
  represent the FEPS data, where the black circles in the right panel are
  sources with a 24\micron\ excess from a debris disk. Five sources with 
  optically thick disks are not shown.
} 
\end{center}
\end{figure}

In Figure~\ref{fig:models_time}, we present model calculations for the
temporal evolution of the 8, 16, and 24\micron\ emission.
In the top panels, different
colored curves represent distinct values of $\Sigma_{\rm o}$ for an outer 
radius of 15~AU (solid curves) and 100~AU (dashed curves). 
The IRAC~8\micron, 16\micron, and MIPS~24\micron\ observations discussed in 
\S\ref{excess} are also shown, with black circles representing 
sources with infrared excesses at the wavelength. 

The model curves shown in the top row of panels of 
Figure~\ref{fig:models_time} shows a flat emission
profiles for younger ages, followed by a $t^{-1}$ fall off in the intensity
toward older ages. This general shape is expected based on the adopted density
distribution with time (Eq.~\ref{eq:nt}). Models with an outer radius of 100~AU
have brighter emission than models with a 15~AU outer radius since the larger
disk contains more dust surface area. 

The models predict that the 8\micron\ excess is $<0.7$\% at an age of 1~Myr for
all surface densities considered here. This excess level is consistent with the
observed 3$\sigma$ upper limit of 3\% found toward individual stars. The lack
of 8\micron\ emission in the models is expected since the inner 10~AU was
assumed to be evacuated of dust. 

At 16 and 24\micron, the models cannot provide a unique interpretation of the
results since the disk size and power-law surface density are degenerate
parameters. Nonetheless, we show in Figure~\ref{fig:models_time} that a
class of models exists that can account for many of the observed
characteristics. For example, these models predict 16\micron\ emission of less
than 20\% for surface densities \aboutless 0.3~g~cm$^{-2}$. At 24\micron, the
emission is relatively constant for ages \aboutless 100~Myr if $\Sigma_{\rm o}
< 0.1$~g~cm$^{-2}$ and $R_{\rm out}$=100~AU, or for $\Sigma_{\rm o} <
0.3$~g~cm$^{-2}$ and $R_{\rm out}$=15~AU. The $R_{\rm out}$=15~AU models 
agrees well with the observations in that the emission falls sharply with age 
at \about 300~Myr. However, for models with $R_{\rm out}$=100~AU, the 
24\micron\ emission persists for as long as 10~Gyr, which is much longer than
observed. The time constant of the emission could be decreased for the $R_{\rm
out}$=100~AU models to better match the data by decreasing the maximum
planetesimal size, having a steeper radial density profile, or increasing the
particle eccentricity.

While we have not done an exhaustive parameter study, we conclude that basic 
characteristics of the 8, 16, and 24\micron\ excess emission can be explained 
by a planetesimal belt with an inner radius of 10~AU in quasi-steady state 
collisional evolution. This model can account for the magnitude of the 
excesses, and the relatively flat distribution of 24\micron\ excess for young 
ages, and the decline in the 24\micron\ excess for ages older than 
\about 100-300~Myr.

\subsubsection{Depletion of the Inner Disk}
\label{models:terrestrial}

The models presented in the top panel of Figure~\ref{fig:models_time} assume an
inner disk radius of 10~AU, which is a typical radius inferred from analysis of
the IRS spectra (\S\ref{models:belts}). By contrast, the inferred inner disk
radius in primordial disks is \aboutless 0.3~AU \citep[e.g.][]{Akeson05}, and
at least in our solar system, the terrestrial planets must have formed from
planetesimals well within 3 AU. Collisions in the inner planetesimal belt would
presumably produce warm debris that will emit at wavelengths shorter than
24\micron\ and could be traced by IRS spectra and IRAC photometry.

One difficulty in detecting warm debris disks is related to the speed at
which they are expected to evolve. For a disk with a power-law
surface density distribution (Eq.~\ref{eq:nar}), the timescale for
collisional growth of planetesimals varies with orbital radius as
$R^{3/2-\alpha}$ \citep{Lissauer87}\footnote{The radial exponent for
planetesimal growth adopted here differs from that for planetesimal destruction
(see Eq.~\ref{eq:tc}) because of additional assumptions about how physical
parameters vary with radius in the latter model \citep[see][]{Wyatt07a}.}. We
thus expect that the collisional cascade will also proceed on faster timescales
at smaller orbital radii for $\alpha < 1.5$, and that the inner disk will be
depleted more quickly of dust mass via radiation blowout of the smallest grains
than the outer disk. We consider then if a planetesimal belt that
originally extended to small radii will deplete on timescales fast enough to
remain consistent with the lack of observed excess emission at 8 and 16\micron,
or if another mechanism is needed to increase the dissipation timescale. For
this exercise, we consider a disk model with an inner radius of 0.5~AU, and
otherwise adopt the disk parameters assumed in \S\ref{models:gasgiant}. Model
calculations that show the temporal variations of the emission are presented 
in bottom row of panels in Figure~\ref{fig:models_time}.

As expected, changing the inner disk radius from 10~AU to 0.5~AU increases the
amount of short-wavelength emission. For $t$ \aboutless 3~Myr, the model 
8\micron\ disk emission is $\ge $ 3.0\% above the photosphere and detectable 
in individual sources with the FEPS IRAC observations in disk models with 
$\alpha \le -1$ and $\Sigma_{\rm o} \ge 3$\,gm~cm$^{-2}$. These models are
consistent with the observations in that no optically thin 8\micron\ emission 
was detected in FEPS sample, which covers stellar ages older than 3~Myr. 

Based on the FEPS observations, \about 12\% of stars are surrounded by
optically thick disks between ages of 3-10~Myr. If we assume that the
collisional cascade begins primarily after the cessation of accretion and all
optically thick disks pass through a debris phase, then 12\% of stars 
with ages of 3-10~Myr may contain recent production of 8\micron\ debris 
emission. Since 8\micron\ emission will be detectable
for up to 3~Myr after the start of the collisional cascade for an inner disk
radius of 0.5~AU, we expect to detect \about 1-2 stars with optically thin
8\micron\ emission. This is consistent with the null detections of optically
thin emission in the FEPS sample. We conclude that the absence of 8\micron\
detections in the FEPS sample does not meaningfully constrain the
presence or lack thereof of the initial planetesimal belts at a radius of
\about 0.5~AU, and that observations for a much larger sample of young stars
are needed to probe this rapid evolutionary stage \citep[see][]{Currie07}.

The IRS and MIPS~24\micron\ observations provide more meaningful constraints on
the temporal evolution of the inner planetesimal belts. With the smaller inner
disk radius, the collisional time scale is smaller for the $R_{\rm in}=0.5$~AU
model than the 10~AU model considered in \S\ref{models:gasgiant} for the same
surface density (see Eq.~\ref{eq:tc}). The emission curves are
largely on the $t^{-1}$ evolution phase and the predicted 16 and 24\micron\
emission decline rapidly with time for most of the surfaced densities
considered here (see Figure~\ref{fig:models_time}). Yet the observed emission
is much less than predicted by many of the models explored here.

Because the inner disk radius is smaller than considered in
\S\ref{models:gasgiant}, this class of models requires lower overall surface
densities to match the observations. Model with surface densities of
$\Sigma_{\rm o} > 0.03$ ~g~cm$^{-2}$ produce higher 16 and 24\micron\ flux
densities and a sharp decline in the flux densities with increasing age that is
inconsistent with the observations. Qualitative agreement between the models
and the observations is found for $\Sigma_{\rm o} < 0.03$~g~cm$^{-2}$ in that
the 16\micron\ excess is less than 20\%, and that the magnitude of 24\micron\
is consistent with the observations. As was found in \S\ref{models:gasgiant},
models with $R_{\rm out}=15$~AU provide a better match to the temporal
decline than models with $R_{\rm out}=100$~AU for the assumed parameters.
We conclude that if the inner planetesimal belt initially extended to 0.5~AU,
the belt can deplete fast enough to explain the observations. However,
bright 16\micron\ excess emission from debris dust would be expected to 
found around young ($< 3$~Myr) stars.

\subsection{Timescale of the Debris Phase}

The fraction of stars in the FEPS sample with 24\micron\ excesses greater than 
10.2\% above the photosphere is constant to within the uncertainties at 
\about 15\% for ages less than 300~Myr. By contrast, while some primordial
disks dissipate within 1~Myr \citep{Padgett06,Cieza07}, about half of solar 
mass stars at age of 1-3~Myr remain surrounded by primordial disks 
whether traced by emission at 3-8\micron\ \citep{Haisch01,Hernandez07}, 
10\micron\ \citep{Mamajek04}, or 24\micron\ \citep[e.g.][]{Lada06,Damjanov07}. 
We consider then why the fraction of 1~Myr stars with primordial disks, and
therefore have at least the potential to form planets, is much higher than the
stars detected with debris dust.

As discussed in \citet{Meyer08}, the observed fraction of stars with 24\micron\
excesses as a function of age (see Fig.~\ref{fig:mips24_age_fbol}) may reflect
one of two scenarios. One possibility is that only \about 15\% of disks around
solar-type stars form planets or large planetesimals, and the resulting debris
disks emit at 24\micron\ for \about 100-300~Myr. The results presented in
\S\ref{models:gasgiant} indicate that a quasi-steady state collisional cascade
can explain basic characteristics of the excess emission observed in debris
disks  under certain assumptions \citep[see also][]{Dominik03,Wyatt07a},
although this model is not a unique interpretation of the data. Another
possibility is that the debris phase is more common, but for any given star it
is short lived and so for any given age interval, only 15\% of stars exhibit
the debris phenomenon. In this interpretation, more than 60\% of FEPS stars
pass through a debris phase at some point in their lifetime, suggesting that
most solar-type stars form planetary systems.

The critical distinction between the above two scenarios is the lifetime of the
debris phase and whether the dust is continuously replenished over tens of
million of years in a quasi-steady state, or if the dust is a transient
phenomenon. We reconsider the arguments in \citet{Meyer08} using the full
complement of \Spitzer\ data from the FEPS program.
We assume that the
planetesimal belt extends from inner radius $R_{\rm in}$ to outer radius
$R_{\rm out} = R_{\rm in} + \Delta R$. The lower limit to the debris disk
lifetime is set by the timescale for the collisional cascade to propagate
through the disk. For a power-law radial surface density of $\Sigma(R) \propto
\Sigma_{\rm o} R^{\alpha}$, the timescale for collisional growth is 
proportional to $R_{\rm in}^{3/2-\alpha} \Sigma_{\rm o}^{-1}$. 
If the debris phase is a transient phenomenon, 
the duration of the debris phase, defined as $\Delta t$, is much
less than the age intervals used to create Figure~\ref{fig:mips24_age_fraction},
or $t / \Delta t < 3$. The width of the planetesimal belt ($\Delta R$) to limit
the debris phase to $\Delta t$ is then given by $\Delta R \ll 2
(1.5-\alpha)^{-1} R_{\rm in}$. For a typical inner radius of \about 10~AU and
$\alpha$ between 0 and 1.5, the width of the planetesimal belts responsible
fo the 24\micron\ emission must be $\ll 7-13$~AU to allow for a short 
debris phase.

We consider the likelihood that most of the debris disks have narrow
planetesimal belts given the observations. The model calculations presented in
\S\ref{models:belts} suggest that the grains producing the 70\micron\ emission,
and by inference the corresponding planetesimal belt, extend over tens of
astronomical units. By the above arguments, we expect the debris phase for
these systems to persist over tens of millions of years. Whether or not the
emission is detectable at 24 or 70\micron\ over this time period depends on the
characteristics of the planetesimal belt. 

As demonstrated in
\S\ref{models:gasgiant}, large planetesimal belts in a quasi-steady state can
produce 24\micron\ emission above the FEPS detection limit for \about 100~Myr.
This result is model dependent, however, and we consider the case where
24\micron\ emission is only produced when a ``wave'' of debris production
propagates through the disk \citep{Kenyon02,Kenyon05}. In this scenario,
stars have narrow planetesimal belts at various radii, and we observe
24\micron\ emission only when the collisional cascade is initiated at a given
radius. The expectation is that debris in planetesimal belts at large radii
will be produced at later times, such that the observed 100-300~Myr lifetime of
the 24\micron\ emission would translate to a factor of 6--10 variation in radii
for $\alpha=-1$. The temperature of the debris emission would decrease by a
factor of \about $2.5-3$ over this range of radius for blackbody grains. While
the observed range of dust temperatures is of this magnitude (see
Figure~\ref{fig:tdust}), there is no evidence for a systematic change in
the dust temperatures with age to support this scenario. 

Alternatively, the
100-300~Myr lifetime of strong 24\micron\ emission could primarily reflect
variations in the disk surface density, where the collisional cascade is
initiated at later times in lower density disks. Since 24\micron\ excesses are
observed over a factor of 10-30 in age, we expect a corresponding variation in
the magnitude of the 24\micron\ excess, with smaller excesses at older ages. 
As show in Figure~\ref{fig:mips24_age_fbol}, there is no evidence for a 
systematic decline of this magnitude for stars younger than 300~Myr. 

Another possibility is that the debris emission is produced by catastrophic
events that produce copious amounts of debris dusts for brief periods
\citep{Rieke05}. \citet{Wyatt07a} evaluated the probability of such collisions
and found that the frequency of such collisions will vary with time as $t^{-2}$
at a given orbital radius. Assuming the orbital radii of planetesimal belts are
approximately the same, one would expect a two order of magnitude decrease in
the fraction of disks exhibiting a catastrophic collision between 10 and
100~Myr. This contrasts with the observed flat distribution of fractional
excesses with age for ages \aboutless 300~Myr (see
Figure~\ref{fig:mips24_age_fraction}). We therefore suggest the 15\%
detection rate of debris disks younger than 300~Myr primarily reflects that
only \about 15\% of solar-type stars exhibit 24\micron\ excesses (greater than
10.2\% above the photosphere) over their lifetime.

If the fraction of stars that pass through a relatively long-lived,
luminous 24\micron\ debris phase is about 15\%, we return now to the question
of why this percentage is so low given that most young stars are surrounded by
primordial disks. The spread in lifetimes of primordial disks is \aboutless
10~Myr. If planetesimal production from primordial disks is the same 
regardless of the lifetime of the primordial phase, then a delay of 10~Myr in
the offset of debris production will have little influence on the duration
(\about 100-300~Myr) of the 24\micron\ debris phase. However, we can 
anticipate a range of planetesimal growth timescales related to the observed
dispersion in primordial disk masses, as well as diversity in dynamical
processes that sculpt planetary systems.

It is tempting to compare this frequency of detected debris ($>$ 15\%) to the
occurrence of gas giant planets. \citet{Cumming08} report the frequency of
sun-like stars with gas giants $>$ 0.1~M$_{\rm Jupiter}$ of up to 20\% when 
extrapolated out
to 20 AU. \citet{MoroMartin07a} were unable to confirm a correlation
between debris disk phenomenon and the presence of close in gas giant planets
\citep[see however][for a notable example]{MoroMartin07b}.
\citet{Apai08} did not find evidence of massive gas
giants in debris systems with large inner holes. It remains unclear to whether
these debris disks represent those with inner gas giants or are another
population representing a diverse planetary system architecture. Perhaps these
large inner holes are evidence of rapid planet formation in the terrestrial
planet zone. The question remains whether those sources lacking obvious debris
represent dynamically full planetary systems that rapidly cleared all potential
sources of debris, or primordial disks that dissipated leaving nothing behind."

Could we have missed a significant number of debris disks in our survey?
Debris disks could not have been detected if the dust is colder
or less luminous than the sample detected here. In the context of the
collisional planetesimal model, on timescales shorter than the collisional
time scale, the most massive and larger planetesimal belts will produce more
luminous 24\micron\ emission. Similarly, increasing the inner disk radius would
decrease the 24\micron\ disk emission. Evidence for additional debris disks
is present within the FEPS data. Analysis of the IRS spectra suggest
a possible additional 21 disk candidates that do not exhibit a 24\micron\
excess, and most of the sources appear younger than 300~Myr. If these infrared
excesses are real, the debris disk percentage would increase
from \about 15\% to 24\%. However, this is still much less than the primordial
disk fraction found around stars, and it remains unclear what primordial
disk properties produced this small percentage of debris disk detections.

\section{Comparison to the Solar System}
\label{comparison}

The basic geometry of the planetesimal disks inferred from debris dust
surrounding the FEPS sample is a cleared inner disk out to \about 10~AU
with optically thin dust beyond these radii. For the sources with 70\micron\
detections, the inferred radii are many tens of AU wide
\citep[see also][]{Hillenbrand08}. This
morphology is similar to that of the Kuiper Belt of our Solar System.
In this section, we make more quantitative comparison between the Solar
System and the debris disks in the FEPS sample, considering both Kuiper
Belt-like dust on tens of AU scales and the available constraints on
zodiacal dust on several AU scales.

The model results in \S\ref{models} suggest that the initial surface density 
of solids for FEPS detected debris disks is
\aboutless 3~g~cm$^{-2}$ at 1~AU; this is less than surface density of solids
expected for the ``minimum-mass-solar-nebula'' \citep[\about
7-30~g~cm$^{-2}$,][]{Weidenschilling77,Hayashi81}. However, the model surface
densities are extremely sensitive to the adopted parameters since most of the
surface area is contained in smaller particles, while the large particles
contain the majority of the mass. For example, by changing the power law slope
of the size distribution from $-3.5$ to $-3.2$ for particle sizes between
0.5\micron\ and 1000~km, the total cross-sectional surface density of particles
can be maintained if the mass surface density is increased by a factor of 1000. 
Given the poor constraints on the planetesimal belt mass, our comparisons to
the solar system will focus on the fractional dust luminosity which is more
closely tied to the observations.

We first consider the sensitivity of the \Spitzer\ observations to the present
day Solar System zodiacal dust \citep[see also][]{Mamajek04}. Following
\citet{Gaidos99}, the luminosity emitted by zodiacal dust is $L_{\rm zodi} =
8\times10^{-8}$~\lsun\ with a characteristic temperature of 260~K
\citep{Reach96}. The implied surface area of particles to reproduce this
luminosity given the characteristic temperature is \about $10^{21}$~cm$^2$,
which \citet{Gaidos99} defined as 1~``zody'' ($\equiv$ 1\zody). 
For the median observed 8\micron\ flux density of 127~mJy, and the 3$\sigma$ 
sensitivity limit of 3\% to 8\micron\ excesses (see \S\ref{excess:irac}), the 
IRAC observations can detect 8\micron\ excesses of \about 4~mJy typically. 
At the median distance of 50~pc to stars in the FEPS sample, this limit 
corresponds to a sensitivity limit of \about 4500\zody\ for 260~K dust.
Similarly, the synthetic IRS~16\micron\ photometry is sensitive to an 
infrared excess of 16.2\% (5~mJy for the typical star), and 5.4-10.2\% 
(0.8-1.6~mJy) for MIPS~24\micron. For 260~K dust, these limits translate to 
a sensitivity limit of 1400\zody\ and 220-440\zody, respectively. 
Our sensitivity limits then do not preclude the presence of zodiacal-type dust 
present in the FEPS debris disk sample.

The MIPS~70\micron\ observations are sensitive to cooler temperature dust
(\about 70~K), and the more appropriate comparison is the Kuiper Belt. 
Emission from debris dust in the Kuiper Belt has not been detected.
\citet{Stern96} place an upper limit of $3\times10^{-6}$ to the optical depth 
at 60\micron, and models of the Kuiper Belt population suggest that the current 
luminosity emitted by dust in the Kuiper Belt is \about $10^{-7}$\lsun\
\citep{Backman95}. This disk luminosity is substantially lower than the 
luminosities derived for disks in the FEPS sample. However, the Sun, at an age 
of 4.57~Gyr \citep{Bahcall95}, is significantly older than most stars observed
here (\aboutless 3~Gyr). Therefore we must account for temporal depletion of 
small dust grains from Poynting-Robertson drag and collisional processes to 
make a proper comparison between the FEPS sample and the Kuiper Belt
\citep[e.g.][]{Meyer07}. Including the effects of the dynamical rearrangements
of the outer planets indicated in the Nice Model \citep[e.g.]{Gomes05} is vital
for a complete comparison to the solar system, but is beyond the scope of this
work.

We use the quasi-steady state collisional model developed by \citet{Dominik03}
and described in \S\ref{models} to project forward the evolution of the FEPS
debris disks. We can identify three regimes for the anticipated evolution of
the disks. When the age of the debris disks is shorter than the initial
collisional timescale ($t_{\rm co}$) of the largest particles in the cascade,
the debris emission will be approximately constant in time. When system ages is
$> t_{\rm co}$, but collisions still dominate, 
the debris luminosity will vary in time as $t^{-1}$. Finally, when the
Poynting-Robertson (PR) drag time scale becomes shorter than the collisional
timescale, the number of particles and the emission will vary in time as
$t^{-2}$. 

For the debris disks detected by FEPS, the collisional timescale is likely
shorter than the PR timescale for all particle sizes 
\citep{Dominik03,Hillenbrand08}. However, it is not possible to determine if 
the debris disk luminosity is on the constant, $t^{0}$, or the $t^{-1}$ phase
of collisional evolution. We assume that each debris disk is currently in the 
$t^{-1}$ phase to achieve the fastest dissipation time. The debris system 
luminosity evolves as $t^{-1}$ until the 
Poynting-Robertson drag is shorter than the instantaneous collisional 
timescale. From \citet{Backman93} the collisional time scale is 
\begin{equation}
    t_{\rm c} = \Bigl({R\over{\rm AU}}\Bigr)^{1.5}\ 
                \sqrt{{M_*\over{\rm M_\odot}}}\ 
                \Bigl({8 \sigma(R)}\Bigr)^{-1}\ {\rm yr},
\end{equation}
where $\sigma(R)$ is the fraction of the geometric surface area that is
covered with particles. For a uniform surface density disk, 
$\sigma(R) = {2 f / {\rm ln}(R_{\rm out} / R_{\rm in})}$, where $f$ is the 
fractional dust luminosity \citep{Backman04}. Also from \citet{Backman93} the 
Poynting-Robertson time scale is given by
\begin{equation}
    t_{\rm PR} = 1900\ {\rho\over 2.7~{\rm g~cm^{-3}}}\ 
                 {a \over \mu{\rm m}}\ 
                 \Bigl({R\over{\rm AU}}\Bigr)^2\ 
                 \Bigl({L_*\over{\rm L_\odot}}\Bigr)^{-1}\ 
                 {\rm yr},
\end{equation}
where $a$ is the particle size. Adopting an inner disk radius of 10~AU and an
outer radius of 100~AU, the PR time scale for the smallest grains (0.5 $\mu$m)
   will be shorter than the collisional time scale for $f < 6 \times 10^{-6}$. 

The debris disks detected by FEPS at ages of 20-200~Myr have fractional
luminosities between $2\times10^{-4}$ and $3\times10^{-3}$. If these
luminosities are projected forward in time using the above prescription for a
disk that extends between 10 and 100~AU, by the age of the solar system the
expected luminosity is between $3\times10^{-8}$ and $10^{-4}$ 
(see Figure~\ref{fig:fbol_age}).   Between ages of 400 and 1200~Myr the median 
fractional luminosity observed for the FEPS detected 
debris disks is $2\times10^{-4}$. These systems project forward to luminosities
of \about $2\times10^{-5}$ at 4.5~Gyr. Clearly any such projections are
extremely uncertain, though we conclude that with the possible exception of the
youngest debris disks, the projected luminosities are 1-2 orders of magnitude
brighter than the estimated luminosity of the Kuiper Belt.  These data are 
consistent with the solar system debris luminosity being ``typical'' 
\citep[c.f.][]{Bryden06}.

\section{Summary}
\label{summary}

We have completed a \Spitzer\ photometric (3.6, 4.5, 8, 24, and 70\micron) and 
spectroscopic (7-35\micron) survey of 314 FGK solar-type stars that span ages 
between \about 3~Myr and 3~Gyr. These data were used to identify sources that 
have infrared emission above the stellar photosphere that is diagnostic of
circumstellar dust. This study complements previous Spitzer studies which have
surveyed either more massive A- and  F- type stars \citep{Rieke05, Su06},
lower mass M-type stars \citep{Gautier07},  solar-mass field stars
\citep{Trilling08,Beichman06b,Bryden06}, or stars in open clusters of well
defined ages \citep{Siegler07,Gorlova07,Cieza08,Currie08b}. Moreover, the
extensive photometric and spectroscopic coverage enables a thorough
investigation of the circumstellar dust properties.

The multi-wavelength dataset was utilized to select a reliable sample of 
sources with infrared excesses. Five sources display infrared excesses in
the IRAC, MIPS, and IRS data, and have characteristics of optically thick,
gas-rich, primordial disks. The other sources identified with infrared excesses
have properties more akin to debris disks in that the excess emission is
detected at wavelengths longer than 16\micron\ and the fractional luminosity is
$L_{\rm IR} / L_* < 10^{-3}$. Physical properties of the putative planetesimal 
belts that produce the debris were inferred by fitting the spectral energy
distribution emitted from an optically thin debris disk containing a power-law
distribution of silicate particles ($N(a) \propto a^{-3.5}$). The results
suggest that the inner disk is typically cleared out to 3-40~AU, and
for sources with detected 70\micron\ excesses, the debris extends over tens
of astronomical units. 

An average 15\% of the stars younger than 300~Myr have a 24\micron\ excess more
than 10.2\% above the photosphere, and this fraction declines to 2\% for older
stars. The maximum 70\micron\ excess exhibits decline over the same age range.
The temporal properties were modeled with a planetesimal belt in quasi-steady
state collisional equilibrium where mass is removed from the system by
radiation blowout of the smallest particles. Such a model can account for the
lack of excess emission shortwards of 16\micron\ and the relatively flat
distribution of 24\micron\ excesses with age. These results suggest it is not
necessary to invoke transient collisional events to explain the emission
characteristics though they cannot be ruled out. Another possibility is that
they were cleared out to 10 AU by dynamical processes associated with inner
planet formation. We can only speculate whether those sources lacking evidence
for debris in the present survey will never (or have never) exhibited such
evidence, or whether they represent systems with the maximum (or minimum)
numbers of planets expected from primordial disks.

The properties of the debris disks in the FEPS sample were compared to that of
the Solar System zodiacal dust and Kuiper Belt. The FEPS observations are
sensitive to $> 220$ times the luminosity of the zodiacal disk and cannot rule
out the presence of a Solar System type debris disk. The luminosity of the
debris systems was compared with a simple model for the evolution of the debris
dust that neglects dynamical events suggested in the Nice model of the 
solar system \citep{Gomes05}. Comparing the expected evolution of the younger, 
luminous debris disks detected in our survey, as well as our upper limits, we 
cannot rule out that cold outer debris disks comparable to our own solar 
system are the rule rather than the exception.

\acknowledgements

JMC thanks Dave Frayer and the \Spitzer\ Science Center staff for patiently 
answering numerous questions regarding \Spitzer\ data. We are grateful to the 
anonymous referee for their extensive comments on the paper. We would like
to also thank all members of the FEPS team for their contributions over the
years. This work is based on observations made with the {\it Spitzer Space
Telescope}, which is operated by JPL/Caltech under a contract with NASA. The
program made use of data and resources from the FEPS project, which receives
support from NASA contracts 1224768, 1224634, and 1224566 administered through
JPL. This research made use of the SIMBAD database, operated at CDS,
Strasbourg, France, and data products from the Two Micron All Sky Survey, which
is a joint project of the U. Massachusetts and the Infrared Processing and
Analysis Center/Caltech, funded by NASA and the NSF. S.W. was supported by the
German Research Foundation (DFG) through the Emmy Noether grant WO 857/2.
MRM is grateful for support from the NASA Astrobiology Institute. 

{}

\clearpage



\begin{deluxetable}{lccccc}
\tablewidth{0pt}
\tablecaption{Summary of Sample by Stellar Association\label{tbl:sample_associations}}
\tablehead{
  \colhead{Group}  &
  \colhead{Age Range}     &
  \colhead{Number Stars} &
  \multicolumn{3}{c}{Spectral type} \\
  \cline{4-6}
  \colhead{}    &
  \colhead{(Myr)} &
  \colhead{}    &
  \colhead{F} &
  \colhead{G} &
  \colhead{K} \\
}
\startdata
Corona Australis        &   3-10 &   7 &   0 &   2 &   5\\
Upper Scorpius          &      5 &  16 &   0 &   8 &   8\\
Upper Centaurus Lupus   &     15 &  23 &   1 &  13 &   9\\
Lower Centaurus Crux    &     17 &  14 &   0 &   9 &   5\\
IC 2602                 &     55 &   5 &   0 &   4 &   1\\
Alpha Per               &     80 &  13 &   6 &   6 &   1\\
Pleiades                &    120 &  20 &   5 &  14 &   1\\
Hyades                  &    600 &  22 &   5 &  10 &   7\\
Field                   & 3-3000 & 194 &  22 & 131 &  41\\
\enddata
\end{deluxetable}




\begin{deluxetable}{ccc@{\extracolsep{10pt}}ccc@{\extracolsep{10pt}}ccc}
\tablewidth{0pt}
\tablecaption{Summary of Sample by Stellar Age\label{tbl:sample_age}}
\tablehead{
  \colhead{Age Range}   &
  \multicolumn{3}{c}{Number}  &
  \multicolumn{2}{c}{Percentage} &
  \multicolumn{3}{c}{Spectral Type}\\
  \cline{2-4}
  \cline{5-6}
  \cline{7-9}
  \colhead{(Myr)}       &
  \colhead{Total}       &
  \colhead{Associations}&
  \colhead{Field}       &
  \colhead{Associations}&
  \colhead{Field} &
  \colhead{F} &
  \colhead{G} &
  \colhead{K}\\
}
\startdata
$<$10      &     34 &     21 &     13 &     62 &     38 &   1 &  15 &  18\\
10-30      &     49 &     39 &     10 &     80 &     20 &   2 &  26 &  21\\
30-100     &     59 &     18 &     41 &     31 &     69 &  10 &  37 &  12\\
100-300    &     60 &     20 &     40 &     33 &     67 &  11 &  38 &  11\\
300-1000   &     55 &     22 &     33 &     40 &     60 &   6 &  38 &  11\\
$\ge$1000  &     57 &      0 &     57 &      0 &    100 &  10 &  42 &   5\\
All        &    314 &    120 &    194 &     38 &     62 &  40 & 196 &  78\\
\enddata
\end{deluxetable}




\begin{deluxetable}{llccccr@{ $\pm$ }rrccccc}
\rotate
\tablewidth{0pt}
\tabletypesize{\scriptsize}
\tablecaption{Source List\label{tbl:sources}}
\tablehead{
  \colhead{Name}       &
  \colhead{Stellar}       &
  \colhead{log (age/years)}        &
  \colhead{$(J-K_{\rm s})_{\rm o}$\tablenotemark{2}}   &
  \multicolumn{4}{c}{Normalized ratios\tablenotemark{3}}   &
  \colhead{log $p$\tablenotemark{4}}   &
  \multicolumn{5}{c}{Excess?}\\
  \cline{5-8}
  \cline{10-14}
  \colhead{}       &
  \colhead{Group\tablenotemark{1}}       &
  \colhead{}       &
  \colhead{(mag)}       &
  \colhead{\RII}   &
  \colhead{\RM}   &
  \multicolumn{2}{c}{\RMM}   &
  \colhead{}       &
  \colhead{8\micron} &
  \colhead{16\micron} &
  \colhead{IRS} &
  \colhead{24\micron} &
  \colhead{70\micron}
}
\startdata
1E 0307.4+1424               & Field &      7.8 &     0.40 &     0.92 $\pm$     0.05 &     1.06 $\pm$     0.03 &    -41.9 &     21.0 &    -0.73\ &      &      &      &      &     \\
1E 0324.1-2012               & Field &      7.8 &     0.37 &     1.16 $\pm$     0.05 &     1.08 $\pm$     0.03 &     -7.1 &     15.0 &    -2.52\ &      &      &      &      &     \\
1RXS J025216.9+361658        & Field &      7.8 &     0.69 &     0.81 $\pm$     0.05 &     0.94 $\pm$     0.03 &     -0.6 &      4.3 &    -1.37\ &      &      &      &      &     \\
1RXS J025751.8+115759        & Field &      7.8 &     0.49 &     0.95 $\pm$     0.05 &     0.97 $\pm$     0.03 &      3.6 &     15.0 &    -0.16\ &      &      &      &      &     \\
1RXS J030759.1+302032        & Field &      8.2 &     0.43 &     0.95 $\pm$     0.05 &     0.96 $\pm$     0.03 &      2.2 &      5.4 &    -1.18\ &      &      &      &      &     \\
1RXS J031644.0+192259        & Field &      7.7 &     0.43 &     1.09 $\pm$     0.05 &     1.03 $\pm$     0.03 &    -57.8 &     36.9 &    -0.68\ &      &      &      &      &     \\
1RXS J031907.4+393418        & Field &      7.8 &     0.49 &     0.98 $\pm$     0.05 &     1.02 $\pm$     0.03 &    -22.1 &     31.6 &    -0.23\ &      &      &      &      &     \\
1RXS J034423.3+281224        & Field &      7.8 &     0.38 &     1.01 $\pm$     0.05 &     0.96 $\pm$     0.03 &     -0.4 &      4.9 &    -1.44\ &      &      &      &      &     \\
\enddata
\tablenotetext{1}{$\alpha$\,Per: Alpha Perseus; CrA: Corona Australis; LCC: Lower Centaurus Crux; UCL: Upper Centaurus Crux; USco: Upper Scorpius}
\tablenotetext{2}{Dereddened $J-K_s$ color}
\tablenotetext{3}{Ratio of the observed flux density ratio to the photospheric value.}
\tablenotetext{4}{Probability statistic computed from $F$-test for fitting model photospheres and modified blackbodies to IRS spectra.}
\end{deluxetable}




\begin{deluxetable}{ccccc}
\tablewidth{0pt}
\tablecaption{24\micron\ Excess Fraction vs. Age\tablenotemark{1}\label{tbl:excess}}
\tablehead{
  \colhead{Age Range}       &
  \colhead{Mean Age}        &
  \multicolumn{2}{c}{Number of Stars} &
  \colhead{Excess Fraction} \\
  \cline{3-4}
  \colhead{(Myr)}  &
  \colhead{(Myr)}  &
  \colhead{Excess} &
  \colhead{Total}
}
\startdata
    $<$ 10 &          5 &          6 &         30 &       0.20$^{+0.10}_{-0.08}$\\\\
     10-30 &         16 &          9 &         48 &       0.19$^{+0.07}_{-0.06}$\\\\
    30-100 &         59 &          5 &         59 &      0.085$^{+0.05}_{-0.04}$\\\\
   100-300 &        150 &         10 &         60 &       0.17$^{+0.06}_{-0.05}$\\\\
  300-1000 &        560 &          2 &         55 &      0.036$^{+0.046}_{-0.024}$\\\\
$\ge$ 1000 &       1800 &          1 &         57 &      0.018$^{+0.039}_{-0.015}$\\\\
\enddata
\tablenotetext{1}{Values computed for a 10.2\% excess above the photosphere and exclude 5 sources with optically thick disks.}
\end{deluxetable}


\clearpage

\clearpage

\clearpage

\clearpage

\clearpage

\clearpage

\clearpage

\clearpage
\begin{center}
\includegraphics[angle=0,scale=0.9]{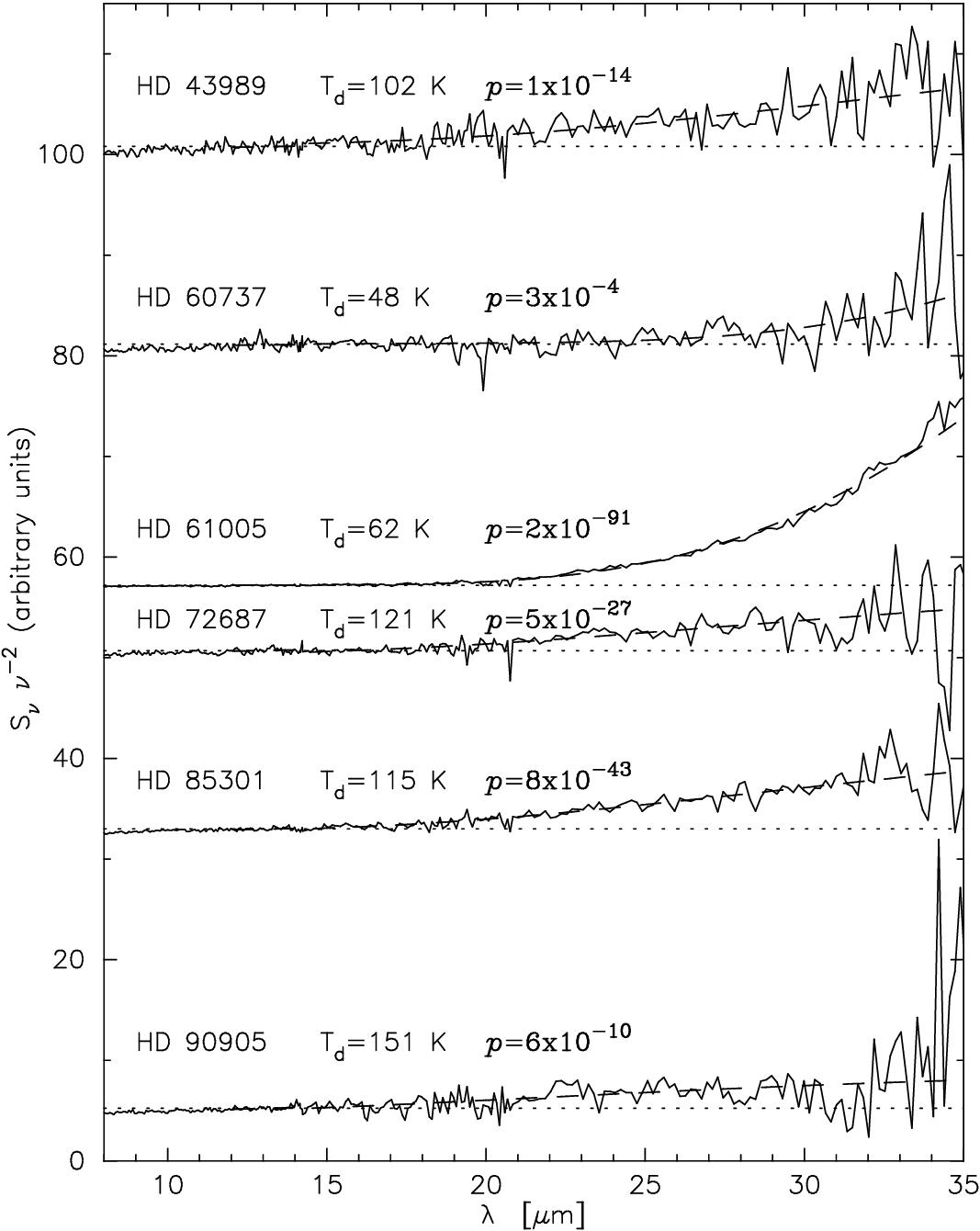}
\clearpage
\includegraphics[angle=0,scale=0.9]{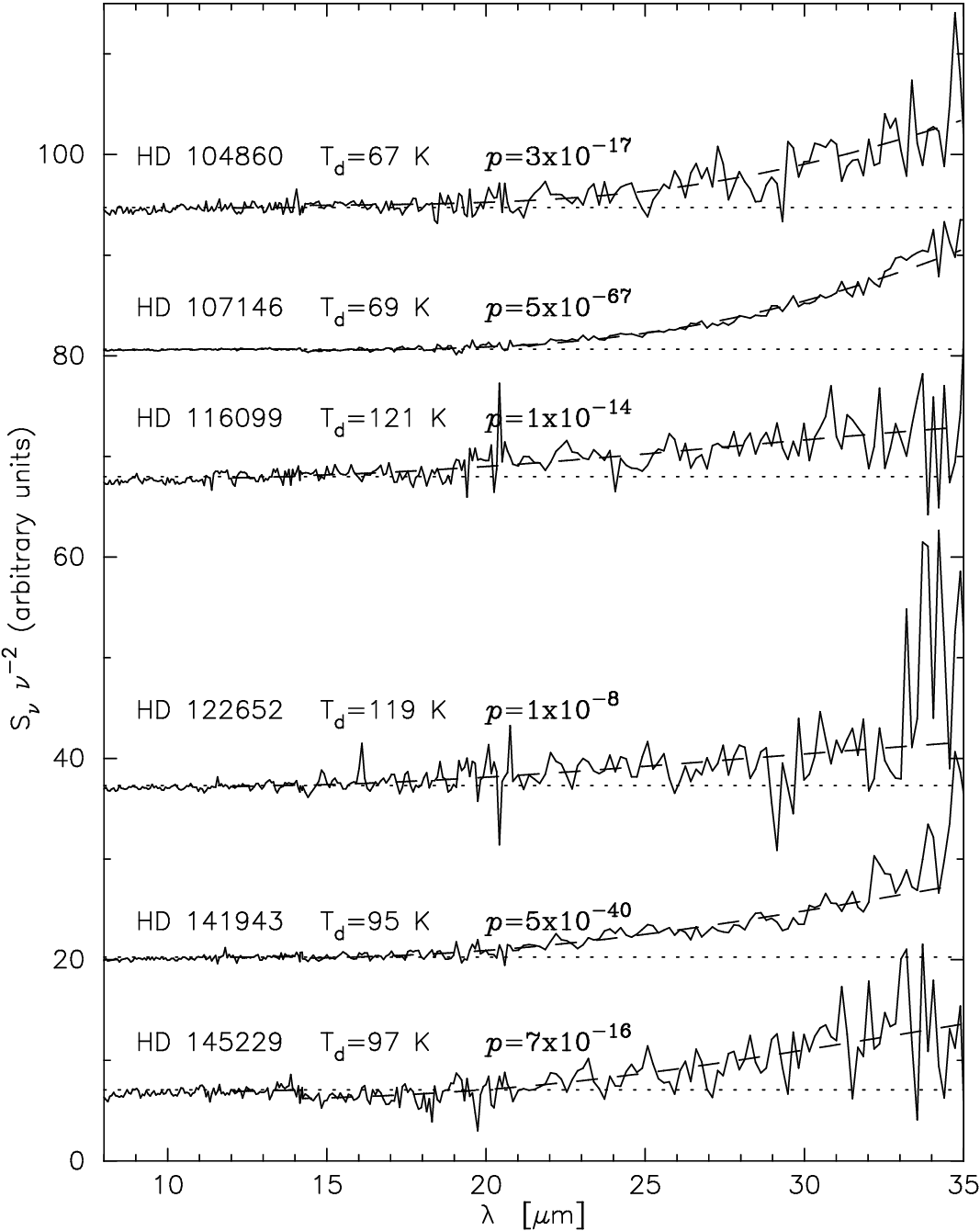}
\clearpage
\includegraphics[angle=0,scale=0.9]{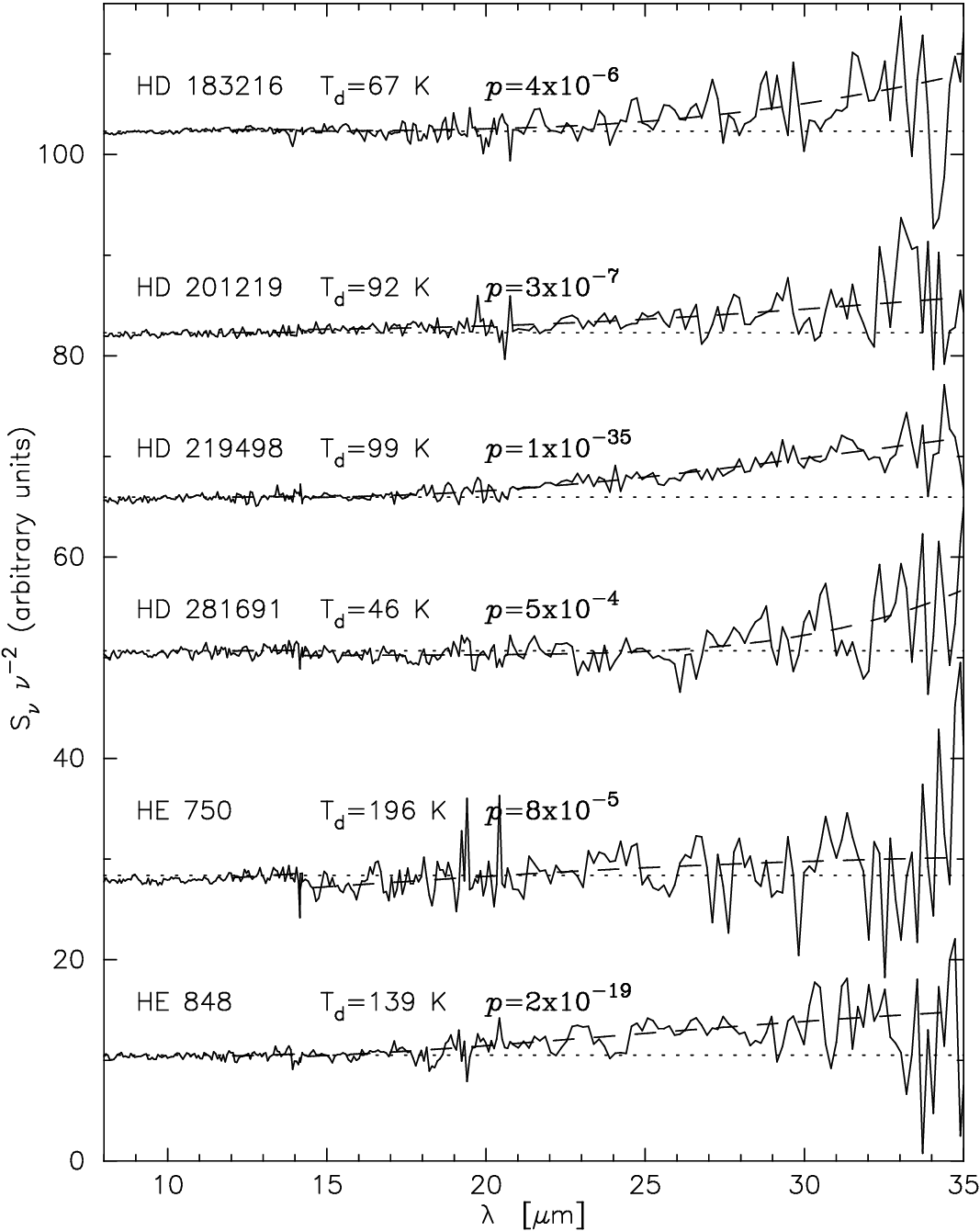}
\clearpage
\includegraphics[angle=0,scale=0.9]{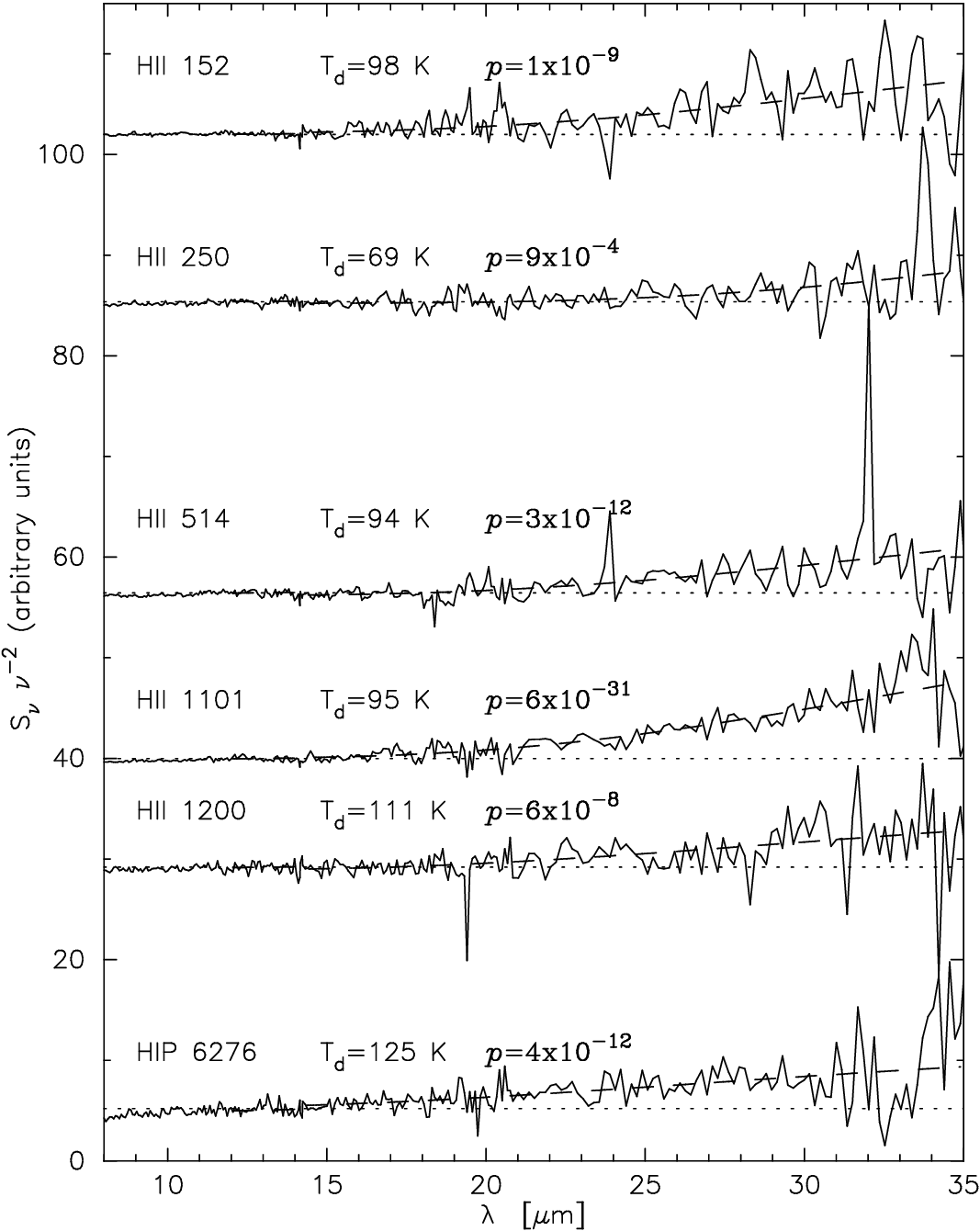}
\clearpage
\includegraphics[angle=0,scale=0.9]{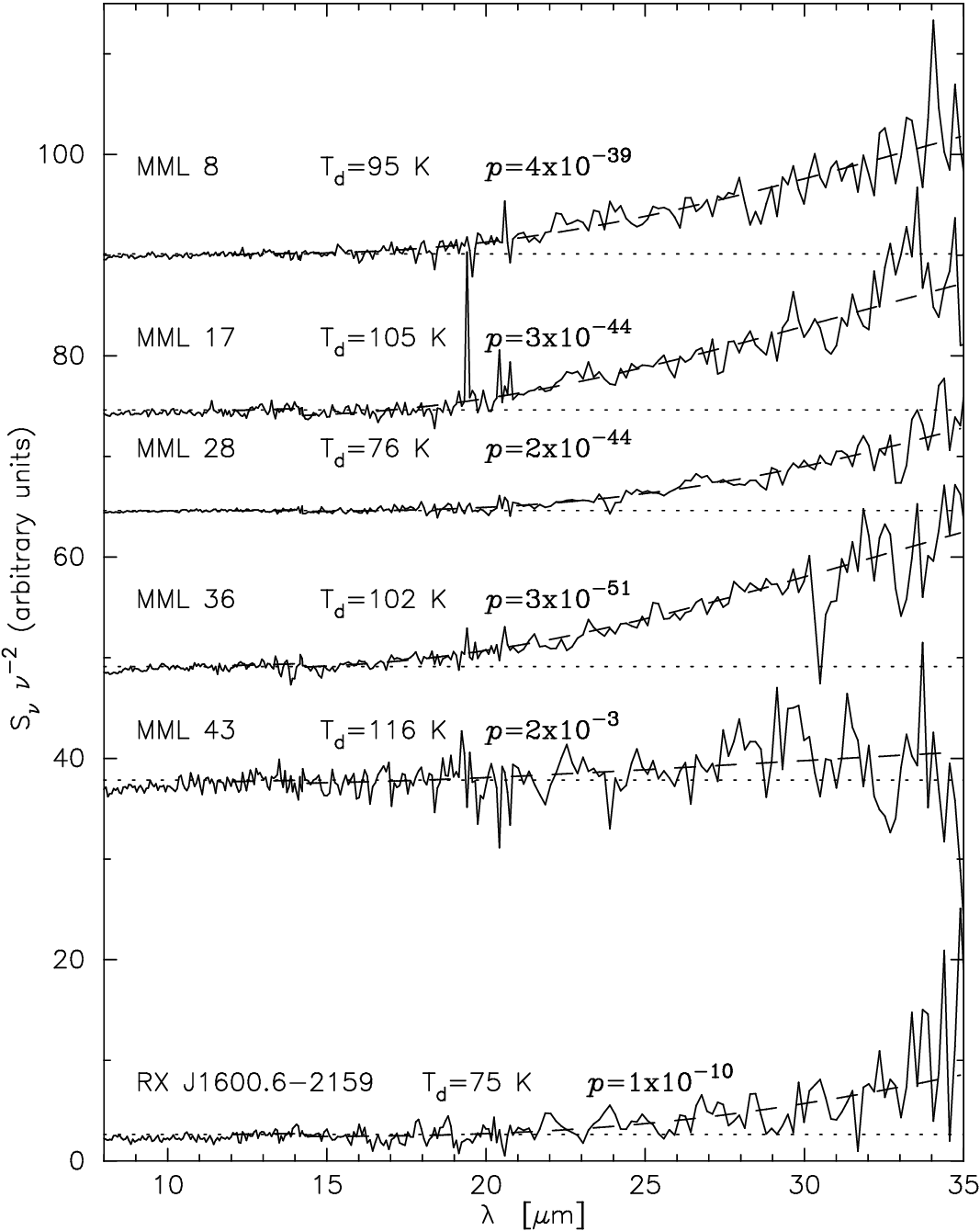}
\clearpage
\includegraphics[angle=0,scale=0.9]{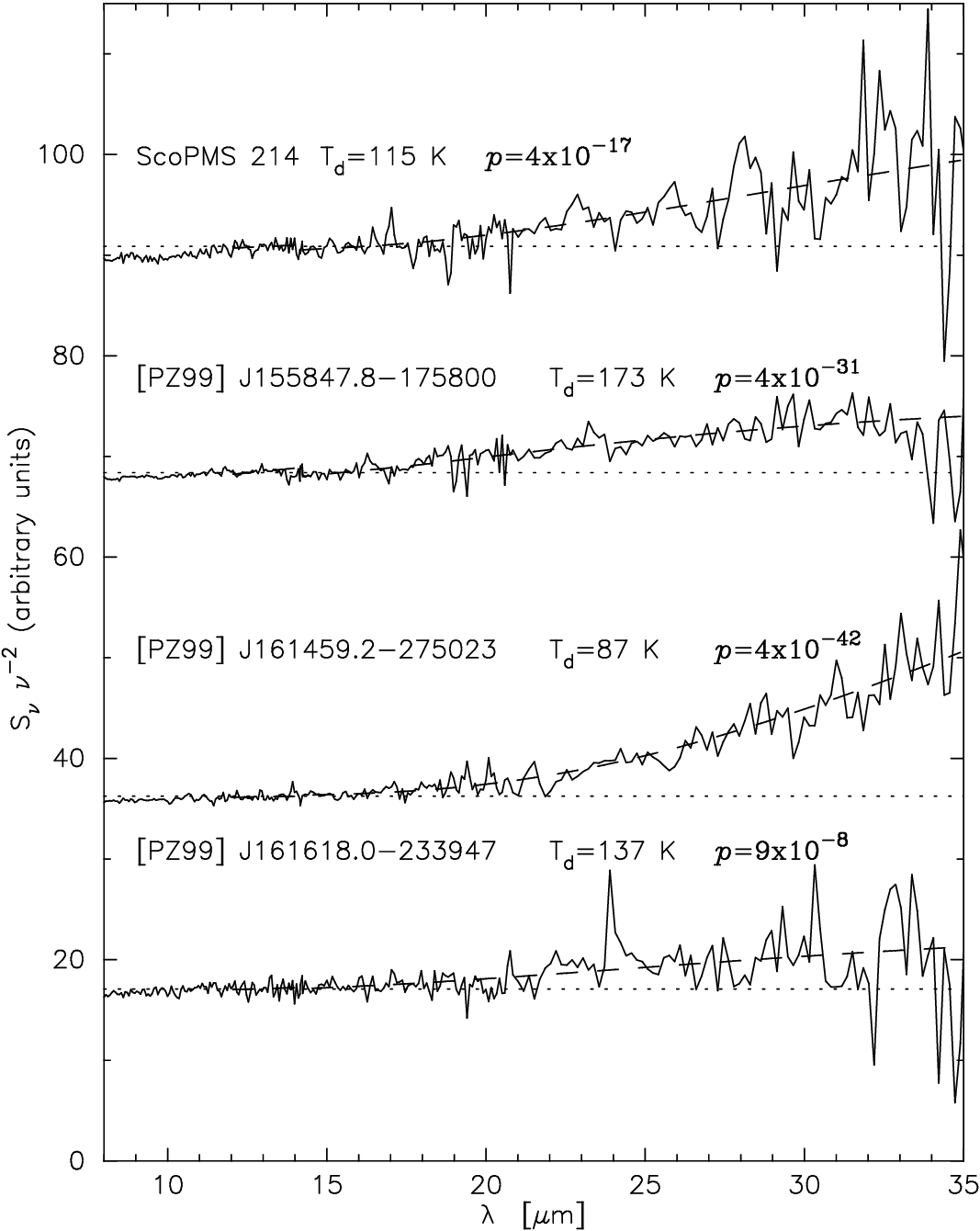}
\end{center}
\clearpage

\clearpage

\clearpage

\clearpage
\begin{center}
\includegraphics[angle=0,scale=0.9]{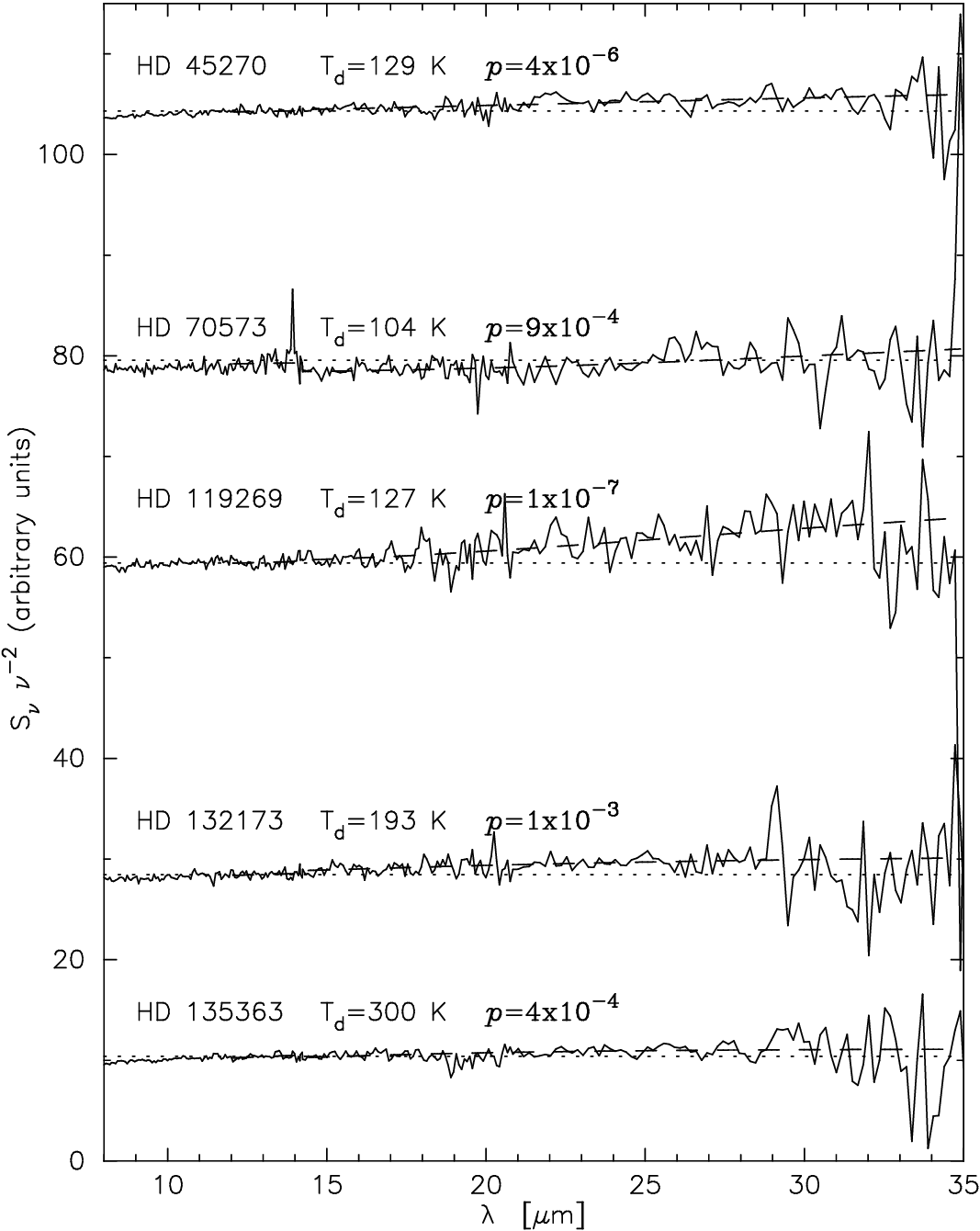}
\clearpage
\includegraphics[angle=0,scale=0.9]{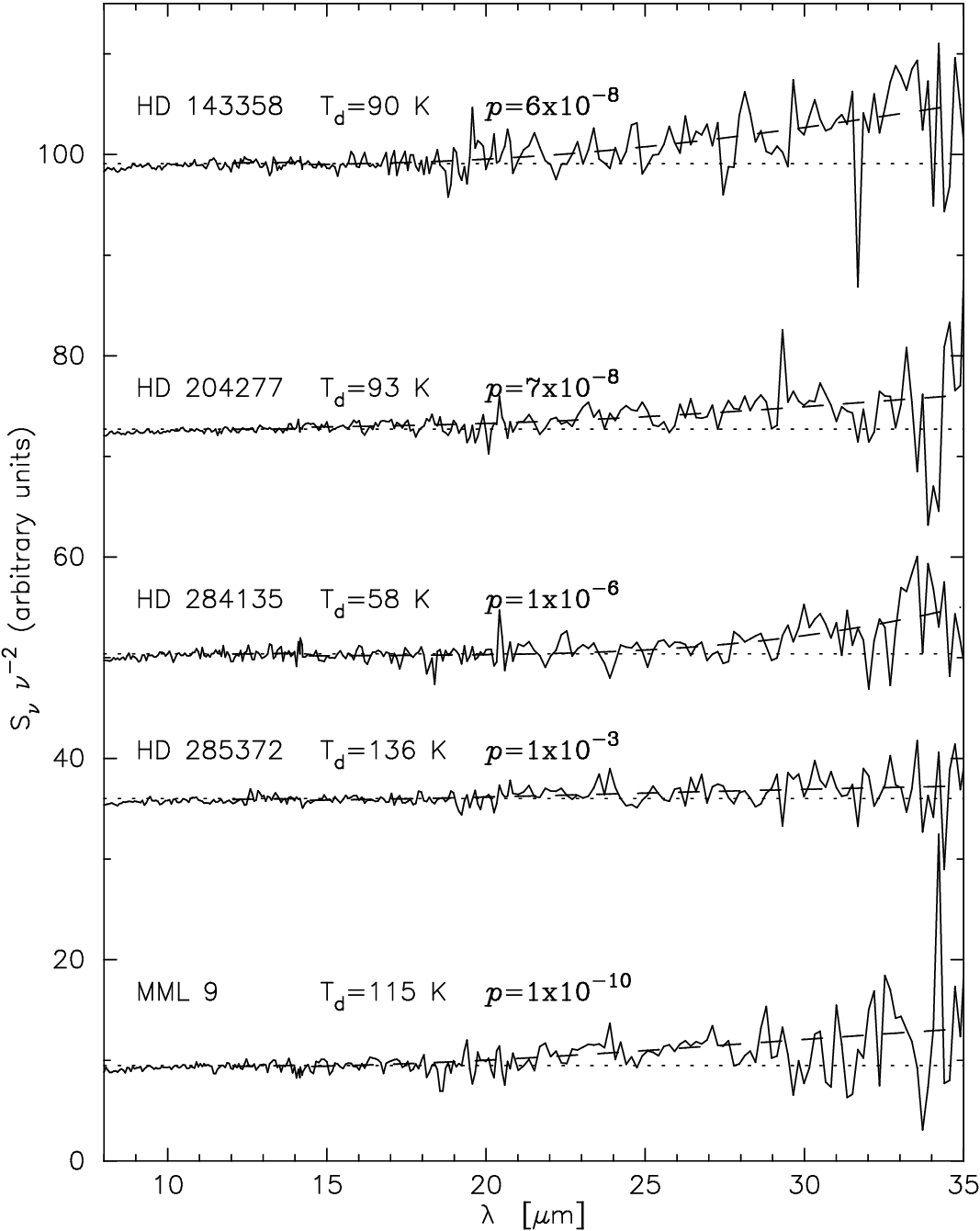}
\clearpage
\includegraphics[angle=0,scale=0.9]{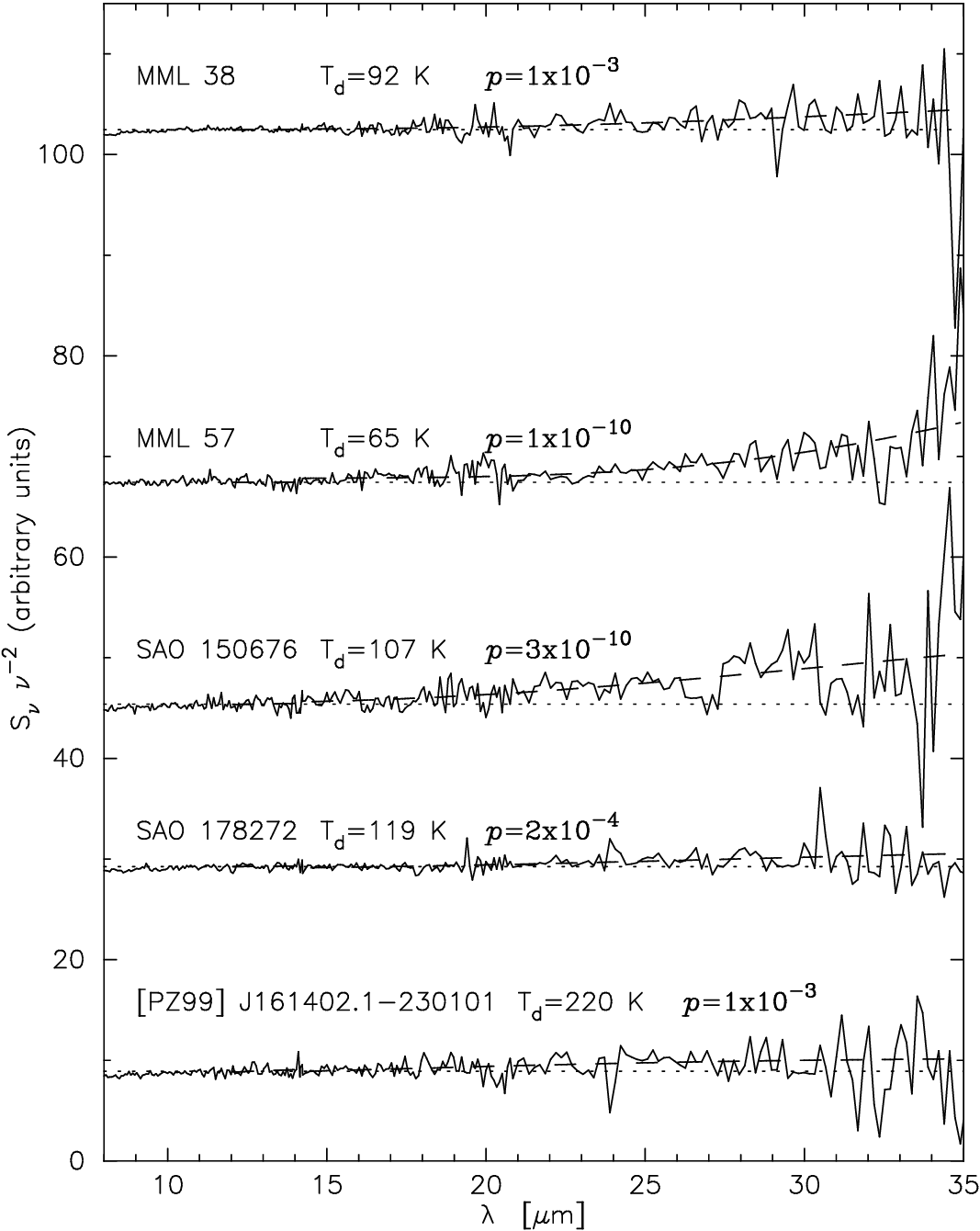}
\end{center}
\clearpage

\clearpage

\clearpage

\clearpage

\clearpage

\clearpage
 
\end{document}